%% file: main.tex
\documentclass[a4paper,twoside,openany,11pt]{memoir} 
\pdfoutput=1

\def\fullheadfoot{0} 
\input{PreRamble}

\newcommand{\braces}[1]{\left\lbrace #1 \right\rbrace}

\newcommand{\Tr}{\mathop{\mathrm{Tr}}}
\newcommand{\tr}{\mathop{\mathrm{tr}} }
\newcommand{\eminus}{\vcenter{\hbox{\scalebox{0.6}[1]{$ - $}}}}	
\newcommand{\ord}[1]{\mathcal{O}( #1 )}
\newcommand{\commutator}[2]{\big[#1, \, #2 \big]}

\newcommand{\dd}{\mathop{}\!\mathrm{d}}
\newcommand{\ud}[2]{\phantom{}^{#1}\phantom{}_{#2}}

\newcommand{\transpose}{^\intercal}

\newcommand{\ineqgraphics}[1]{\vcenter{\hbox{\includegraphics[]{#1}}}}
\newcommand{\sscript}[1]{{\scriptscriptstyle \mathrm{#1}}}

\newcommand{\U}{\mathrm{U}}

\newcommand{\UV}{\sscript{UV}}
\newcommand{\EFT}{\sscript{EFT}}

\newcommand{\cL}{\mathcal{L}}
\newcommand{\cO}{\mathcal{O}}

\newcommand{\cQ}{{\mathcal Q}}

\newcommand{\cI}{{\mathcal I}}
\newcommand{\cJ}{{\mathcal J}}

\newcommand{\lzm}{\left(}
\newcommand{\dzm}{\right)}
\newcommand{\lzs}{\left[}
\newcommand{\dzs}{\right]}

\newcommand{\msbar}{$ \overline{\text{\small MS}} $\xspace}

\newcommand{\ber}{\mathrm{Ber}}

\begin{document}

\thispagestyle{empty}
\renewcommand*{\thefootnote}{\fnsymbol{footnote}}

\begin{center}
    {\sffamily \bfseries \fontsize{16.}{20}\selectfont \mathversion{boldsans}
    A Guide to Functional Methods Beyond One-Loop Order\\[-.5em]
    \textcolor{blue!80!black}{\rule{\textwidth}{2pt}}\\
    \vspace{.05\textheight}}
    {\sffamily \mathversion{sans} \Large 
    Javier Fuentes-Mart\'in,$^{1}$\footnote{javier.fuentes@ugr.es} 
    Adri\'an Moreno-S\'anchez,$^{1}$\footnote{adri@ugr.es} 
    Ajdin Palavri\'c,$^{2}$\footnote{ajdin.palavric@unibas.ch} 
    and \\[5pt]
    Anders Eller Thomsen$^{3}$\footnote{anders.thomsen@unibe.ch}
    }\\[1.25em]
    { \small \sffamily \mathversion{sans} 
    $^{1}\,$Departamento de Física Teórica y del Cosmos, Universidad de Granada,\\
    Campus de Fuentenueva, E–18071 Granada, Spain\\[5pt]
    $^{2}\,$Department of Physics, University of Basel, Klingelbergstrasse 82, CH-4056 Basel, Switzerland\\[5pt]
    $^{3}\,$Albert Einstein Center for Fundamental Physics, Institute for Theoretical Physics, University of Bern, CH-3012 Bern, Switzerland
    }
    \\[.005\textheight]{\itshape \sffamily \today}
    \\[.03\textheight]
\end{center}
\setcounter{footnote}{0}
\renewcommand*{\thefootnote}{\arabic{footnote}}%
\suppressfloats	

\begin{abstract}\vspace{+.01\textheight}
Functional methods can be applied to the quantum effective action to efficiently determine counterterms and matching conditions for effective field theories. We extend the toolbox to two-loop order and beyond and show how to evaluate the expansion of the path integral in a manifestly gauge-covariant manner. We also generalize the method to theories with mixed spin statistics and prove the validity of the hard-region matching formula to all loop orders. The methods are exemplified with a two-loop matching calculation of the Euler--Heisenberg Lagrangian resulting from decoupling the electron in QED. 
\end{abstract}

\newpage
\section*{Table of Contents}
\toc
\newpage


\section{Introduction}

The absence of direct evidence for New Physics (NP) states suggests a pronounced scale separation between the electroweak and any possible new scales. This has prompted an adjustment of the paradigm and strategy in the experimental searches, transitioning from an emphasis on direct searches to a renewed focus on uncovering hints of NP through precision measurements. This methodology was validated during the foundational years of what is now known as the Standard Model (SM) of particle physics. Precision studies, for instance, provided clear indications of the scale of electroweak symmetry breaking and the existence of a third quark generation well before their discoveries. From a theoretical perspective, it is essential to carry out relevant higher-order calculations within both the SM and beyond-the-SM (BSM) theories, as many phenomena cannot be adequately captured at leading order. Examples of these within BSM context include Renormalization Group (RG) effects~\cite{Feruglio:2016gvd,Aoude:2020dwv,Machado:2022ozb,Allwicher:2023shc,Greljo:2023bdy,Palavric:2024gvu,Allwicher:2024sso,Boughezal:2024zqa,Grojean:2024tcw}, dipoles~\cite{Ardu:2021koz,Kley:2021yhn,Fajfer:2023gie,Bonnefoy:2024gca}, finite matching contributions~\cite{Hurth:2019ula,Gherardi:2020qhc,Mantzaropoulos:2024vpe,Loisa:2024xuk}, and many more. 

The effects of heavy NP can be captured systematically in the framework of low-energy Effective Field Theories (EFTs). For BSM analyses, the current state-of-the-art is to include one-loop Standard Model Effective Theory (SMEFT)~\cite{Jenkins:2013zja,Jenkins:2013wua,Alonso:2013hga,Machado:2022ozb} and Low-Energy Effective Theory (LEFT)~\cite{Jenkins:2017dyc,Naterop:2023dek} running effects with one-loop matching between the two theories~\cite{Dekens:2019ept}.\footnote{There is also a recent effort to include dimension-8 one-loop running in the SMEFT~\cite{Chala:2021pll,DasBakshi:2022mwk,DasBakshi:2023htx,Chala:2023xjy,Bakshi:2024wzz,Liao:2024xel,Boughezal:2024zqa}.} While these calculations are somewhat universal, the matching conditions between ultraviolet (UV) models and the SMEFT have to be determined on a case-by-case basis. A comprehensive overview of the UV mediators and the corresponding tree-level SMEFT matching relations and phenomenological implications can be found in~\cite{deBlas:2017xtg,Greljo:2023adz}. At the one-loop level, the process of matching has recently been automated~\cite{Carmona:2021xtq,Fuentes-Martin:2022jrf,Guedes:2023azv,Gargalionis:2024jaw}, with the exception of theories with heavy vectors. While many relevant effects can be accurately captured at this order, there are numerous examples for which this degree of precision proves insufficient. Additionally, scheme-independence of a calculation demands one-loop matching to go with two-loop running, see~\cite{Ciuchini:1993ks,Ciuchini:1993fk} and references therein. First partial results beyond the one-loop order in the SMEFT and the LEFT have recently appeared~\cite{Jenkins:2023bls,DiNoi:2024ajj,Haisch:2024wnw,Born:2024mgz}. However, the complexity of the calculations makes it desirable to explore methods that could render higher-order RG and matching calculations more tractable.

Loop calculations have most commonly been performed in the familiar language of Feynman diagrams, a proven technique with all the benefits of decades of development and refinement within the field. That being said, functional methods~\cite{Aitchison:1984ys,Fraser:1984zb,Aitchison:1985pp,Gaillard:1985uh,Cheyette:1985ue,Chan:1985ny,Chan:1986jq,Cheyette:1987qz,Dittmaier:1995cr} have garnered renewed interest over the last decade as a viable alternative when it comes to EFT matching~\cite{Henning:2014wua,Drozd:2015rsp,delAguila:2016zcb,Boggia:2016asg,Ellis:2016enq,Fuentes-Martin:2016uol,Zhang:2016pja,Ellis:2017jns,Summ:2018oko,Cohen:2019btp,Kramer:2019fwz,Angelescu:2020yzf,Ellis:2020ivx,Cohen:2020fcu,Dedes:2021abc,Dittmaier:2021fls,Larue:2023uyv,Banerjee:2023xak,Li:2024ciy}, counterterm calculations~\cite{Henning:2016lyp,Fuentes-Martin:2022vvu}, anomaly calculations~\cite{Quevillon:2021sfz,Filoche:2022dxl,Cohen:2023gap,Cohen:2023hmq}, and (semi-)automated matching tools~\cite{DasBakshi:2018vni,Cohen:2020qvb,Fuentes-Martin:2020udw,Fuentes-Martin:2022jrf}. Functional methods are also the long-standing choice for calculations of the effective potential~\cite{Coleman:1973jx,Jackiw:1974cv}. In contrast to diagrammatic methods, the functional approach has the advantage of keeping background fields (the sources of the generating functionals) at intermediate steps, facilitating a very direct computation of the EFT or counterterm action. It also ensures that calculations can be performed with manifest gauge invariance through all intermediate steps and that there is no need to account for permutations of external legs. All the same, calculations can be cast in terms of loop-momentum vacuum integrals, allowing practitioners to draw on known results and methods for evaluating these.

Until recently~\cite{Fuentes-Martin:2023ljp,Born:2024mgz}, functional methods have been limited to the domain of one-loop calculations, with the exception of heat-kernel calculations~\cite{Jack:1982hf,Bijnens:1999hw,vonGersdorff:2022kwj,Banerjee:2024rbc} and higher-order computations of the effective potential~\cite{Ford:1992pn,Martin:2013gka,Martin:2017lqn}. This work aims to give a detailed account of how to use functional methods at two-loop order (and beyond) to perform EFT matching and RG calculations. Expanding upon our previous work~\cite{Fuentes-Martin:2023ljp}, which introduced the functional approach for higher-order computations, we extend the functional formalism to incorporate fermions and gauge symmetries. We demonstrate how the expressions for the relevant two-loop topologies generalize to include the fermionic degrees of freedom and manifest background-gauge symmetry at all stages. This generalization recently allowed~\cite{Born:2024mgz} to perform a functional two-loop calculation of the RG equations in the bosonic SMEFT. Establishing a hard-region matching formula, is a crucial step for adapting functional methods for EFT matching calculations. Although this formula has tacitly been assumed to hold (see e.g.~\cite{Manohar:1997qy}) across many matching calculations, it has only been formalized to one- and two-loop orders~\cite{Fuentes-Martin:2016uol,Zhang:2016pja,Fuentes-Martin:2023ljp} more recently. Thus, we have devoted a part of this work to demonstrate the validity of the hard-region matching formula to all loop orders. Putting this result on a firm basis is also relevant for diagrammatic matching calculations, including a recent approach based on on-shell amplitudes~\cite{Chala:2024llp}, where it provides important simplifications. 

This paper is structured in the following manner: In Section~\ref{sec:Effective_action}, we present the general derivation of the vacuum functional and effective action, directing particular attention to the functionals with mixed statistics, i.e., with both bosonic and Grassmannian variables present. Section~\ref{sec:gauge_inv_evaluation} describes how to evaluate the topologies from the effective action in a manifestly gauge-invariant manner. We derive the analytical expressions for each contributing topology, which can be readily applied to concrete theories. We further provide generic rules to derive expressions for arbitrary loop orders based on diagrammatic representations in terms of vacuum (super)graphs. Section~\ref{sec:matching} is devoted to the more formal analysis of matching and running computations, discussing how to apply the functional methods to these problems. We also give an all-order proof to the hard-region matching formula. In Section~\ref{sec:EH}, we demonstrate the application of the developed techniques to a concrete example, namely the Euler--Heisenberg Lagrangian, for which we derive the two-loop matching and running contributions. Lastly, we conclude in Section~\ref{sec:conclusion} by providing a summary and pointing towards promising directions for future work. Extensive appendices provide additional mathematical details for the use of superfields in the path integral (\ref{app:superfields}) and gauge covariant delta functions (\ref{app:pdp}); an alternative covariant formula for the evaluation of the sunset graph (\ref{app:alt_sunset}); and an all-orders expansion by regions for loop integrals with the heavy mass expansion~(\ref{app:regions}).

\section{Perturbative Expansion of the Generating Functionals}
\label{sec:Effective_action}

There is a long tradition in Quantum Field Theory (QFT) for using and manipulating generating formulas for Green's functions based on the path integral~\cite{Iliopoulos:1974ur,Bijnens:1999hw,Fuentes-Martin:2023ljp}. These quantities are instrumental, as they contain all the quantum information of a theory, and it is possible to recover renormalization and matching quantities from them.
To cover all realistic models, we aim to recover a perturbative expansion of the path integral fully generalized to include fields with mixed spin statistics. 
We consider a generic theory described by the action $ S[\eta] $, where $ \eta_I $ denotes a collection of fields---scalars, spinors, and gauge bosons.  
The \emph{vacuum functional} is the generating functional for all connected $n$-point functions and contains all physical information of the theory. It is given by
    \begin{equation} \label{eq:vac_func_def}
    e^{i \hbar^{\eminus 1} W[\mathcal{J}]} = \int \,[\mathcal D \eta] \exp\!\left[ \dfrac{i}{\hbar}  \big( S[\eta] + \mathcal{J}_I \eta_I \big) \right]\,, \qquad 
    S = \sum_{\ell=0}^{\infty} \hbar^\ell S^{(\ell)}\,,
    \end{equation} 
having employed DeWitt notation where the capital indices $ I= (x, a) $ take values in both spacetime and internal degrees of freedom. Thus, $ \mathcal{J}_I \eta_I = \int_x\ \mathcal{J}_a(x) \,\eta_a(x) $, with the shorthand notation $ \int_x = \int \, \dd^d x $. The spacetime dimension $ d=4 - 2\epsilon $ is used as a regulator, and the vacuum functional is regularized perturbatively by including counterterms $ S^{(\ell)}$ with $ \ell > 0$ to the tree-level action $ S^{(0)} $. Factors of $ \hbar $ are kept manifest for loop-counting purposes but are otherwise irrelevant.

To reflect the presence of fields with mixed spin statistics, it is convenient to arrange them into \emph{superfields} with schematic form $\eta=(\phi,\chi)$, where $\phi$ and $\chi$ denote the collections of bosonic and fermionic fields, respectively. The signature $ \zeta_{IJ} $ is then introduced to keep track of Grassmannian signs:
\begin{align} \label{eq:field_commutation}
    \eta_I \eta_J =\zeta_{IJ}\, \eta_J \eta_I\,,\qquad \zeta=\begin{pmatrix}1 & 1\\ 1 & \eminus 1\end{pmatrix}\,,
\end{align}
with the blocks of the indices $I,J$ indicating whether the variable is bosonic or Grassmannian.\footnote{The source $ \mathcal{J}$ naturally inherits the mixed statistics of the fields and behaves identically under commutation.} The indices of $\zeta_{IJ}$ are in some sense diagonal meaning that they are \emph{not} counted when determining if Einstein summation takes place. The reader will find a short primer on supervectors in Appendix~\ref{app:Grassmann}.

The perturbative loop expansion of the vacuum functional coincides with the saddlepoint approximation to the path integral around the classical background, which is defined implicitly as the solution to the equation of motion (EOM)
     \begin{equation} \label{eq:bkg_field_EOM}
    0 = \dfrac{\delta S^{(0)}}{\delta \eta_I}[\overline{\eta}] + \zeta_{II} \mathcal{J}_I\,.
    \end{equation}
The background configuration $ \overline{\eta}[\mathcal{J}] $, thus, depends on the source $ \mathcal{J} $. We parameterize the expansion of the action around the classical field as
    \begin{align} 
    \label{eq:action_expansion}
    S[\overline{\eta} + \hbar^{1/2}\eta] = \overline{S} 
    + \sum_{\ell=0}^{\infty} \hbar^\ell &\Big[ \hbar^{1/2}\,\eta_I \overline{\mathcal{V}}^{(\ell)}_I + \frac{\hbar}{2}\, \eta_I \overline{\mathcal{V}}^{(\ell)}_{IJ}\eta_J+\sum_{n=3}^\infty \frac{\hbar^{n/2}}{n!}\,\eta_{I_1}\dots\eta_{I_n}\overline{\mathcal{V}}^{(\ell)}_{I_1\dots I_n} \Big]\,, 
    \end{align}
where the bar is shorthand for dependence on $ \overline{\eta} $; that is, $ \overline{S} = S[\overline{\eta}] $ and
\begin{align}
\overline{\mathcal{V}}_{IJ}^{(\ell)}&= \zeta_{JJ}\dfrac{\delta^2 S^{(\ell)}[\eta]}{\delta \eta_I\, \delta \eta_J}\bigg|_{\eta=\overline{\eta}}\,,&
\overline{\mathcal{V}}_{I_1\dots I_n}^{(\ell)}&= \dfrac{\delta^n S^{(\ell)}[\eta]}{\delta \eta_{I_n}\dots\, \delta \eta_{I_1}}\bigg|_{\eta=\overline{\eta}}\qquad (n\neq2)\,.
\end{align}
Special attention should be given to the order of functional derivation, as different orders give rise to different Grassmannian signs. Following standard conventions, we denote the inverse dressed propagator (also known as the fluctuation operator) by
    \begin{equation}
    \overline{\mathcal{Q}}_{IJ}\equiv\overline{\mathcal{V}}^{(0)}_{IJ}\,,
    \end{equation}
since it plays a special role in our derivations below. 

\begin{figure}
    \centering
    \includegraphics[width=0.9\linewidth]{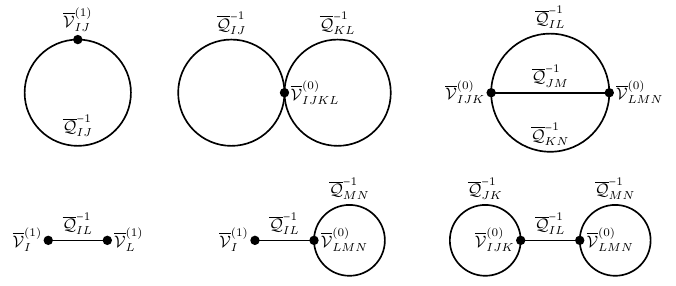}
    \caption{Two-loop contributions to the vacuum functional. All vertices and propagators are dressed with insertions of the background-field configuration $ \overline{\eta}[J] $.}
    \label{fig:vacuum_functional}
\end{figure}

By the EOM satisfied by $ \overline{\eta} $, $\overline{\mathcal{V}}^{(0)}_I = - \zeta_{II} \mathcal{J}_I$, the tree-level linear term in the quantum field vanishes from the exponential in \eqref{eq:vac_func_def}. This lets us evaluate the vacuum functional~\eqref{eq:vac_func_def} perturbatively. Shifting the integration variable $ \eta \to \overline{\eta} + \hbar^{1/2}\eta $ and performing the Gaussian integration of the quantum fields up to second order in the $\hbar$ expansion yields\footnote{In index notation, the supertrace is $ \mathrm{STr} \,\mathcal{A} = \zeta_{II} \mathcal{A}_{II}$.}
\begin{align} \label{eq:vac_func_2-loop}
    \begin{aligned}
    W[\mathcal{J}] 
    &= \overline{S}^{(0)} + \mathcal{J}_I \overline{\eta}_I + \hbar \,\overline{S}^{(1)}+ \dfrac{i\hbar}{2} \mathrm{STr} \log \overline{\mathcal{Q}} + \hbar^2 \overline{S}^{(2)}
    + \dfrac{i \hbar^2}{2} \zeta_{II} \overline{\mathcal{Q}}^{\eminus1}_{IJ} \overline{\mathcal{V}}^{(1)}_{JI} \\
    &\quad +\dfrac{\hbar^2}{12} \zeta_{IN}\zeta_{IM}\zeta_{JN}\, \overline{\mathcal{Q}}_{LI}^{\eminus1} \overline{\mathcal{Q}}_{MJ}^{\eminus1} \overline{\mathcal{Q}}_{NK}^{\eminus1} \overline{\mathcal{V}}^{(0)}_{IJK} \overline{\mathcal{V}}^{(0)}_{LMN} - \dfrac{\hbar^2}{8}  \overline{\mathcal{Q}}_{IJ}^{\eminus1} \overline{\mathcal{Q}}_{KL}^{\eminus1}\overline{\mathcal{V}}^{(0)}_{IJKL} \\
    &\quad - \dfrac{\hbar^2}{2} \zeta_{II}\bigg( \overline{\mathcal{V}}^{(1)}_I+ \dfrac{i}{2} \overline{\mathcal{Q}}_{JK}^{\eminus1} \overline{\mathcal{V}}^{(0)}_{IJK} \bigg)  \overline{\mathcal{Q}}_{IL}^{\eminus1} \bigg( \overline{\mathcal{V}}^{(1)}_L+ \dfrac{i}{2} \overline{\mathcal{Q}}_{MN}^{\eminus1} \overline{\mathcal{V}}^{(0)}_{LMN} \bigg)+ \ord{\hbar^3}\,.
    \end{aligned}
\end{align}
More details regarding the derivation can be found in Appendix~\ref{app:Details_Gen_Func}. This generalizes the expression of the vacuum functional from Ref.~\cite{Fuentes-Martin:2023ljp} with the Grassmannian signs $\zeta_{IJ}$, accounting for the mixed spin statistics.
The tensors $ \overline{\mathcal{V}}^{(0)}_{IJK} $ and $ \overline{\mathcal{V}}^{(0)}_{IJKL} $ play the role of 3- and 4-point vertices, respectively, dressed with background fields. Similarly, $ \overline{\mathcal{Q}}^{\eminus 1}_{IJ} $ is the dressed propagator of the quantum fields. 
The counterterms $ \overline{\mathcal{V}}^{(1)}_I $ and $ \overline{\mathcal{V}}^{(1)}_{IJ} $ from the renormalized action $ \overline{S}^{(1)} $ cancel all UV subdivergences of the genuine two-loop terms, reproducing the renormalization of ordinary Feynman--diagram methods. In a language more reminiscent of ordinary Feynman diagrams, the various contractions of functional tensors appearing in Eq.~\eqref{eq:vac_func_2-loop} can be viewed as connected \emph{vacuum supergraphs}, as shown in Fig.~\ref{fig:vacuum_functional}. In Section~\ref{sec:higher_order}, we introduce simple rules, analogous to Feynman rules, to relate the different tensor structures arising at any order in the loop expansion to their corresponding vacuum supergraphs.

Typically, it is preferable to work with the \emph{(quantum) effective action} for the purposes of loop calculations. The effective action is defined as the Legendre transformation of the vacuum functional, given by 
    \begin{equation} \label{eq:eff_action_def}
    \Gamma[\hat{\eta}] = W[\mathcal{J}] - \mathcal{J}_I \hat{\eta}_I\,, \qquad 
    \hat{\eta}_I \equiv \dfrac{\delta W}{\delta \mathcal{J}_I}\,,
    \end{equation}
and it is the generating functional for all 1PI diagrams. The background field $ \hat{\eta} $ is the expectation value of the field in the presence of the source $ \mathcal{J} $ and it corresponds to the solution of the quantum EOMs. 
We provide more details on the procedure for obtaining the effective action from the vacuum functional in Appendix~\ref{app:Details_Eff_Action}. In the end, one finds that the effective action can be written as
\begin{align} \label{eq:eff_action_generic_2-loop}
\begin{aligned}
\Gamma[\hat\eta]&=\widehat{S}^{(0)}+\hbar\widehat{S}^{(1)}+\frac{i\hbar}{2}\mathrm{STr}\log\widehat{\mathcal{Q}}+ \hbar^2 \widehat{S}^{(2)}+ \frac{i\hbar^2}{2}\zeta_{II}\widehat{\mathcal{Q}}^{\eminus 1}_{IJ}\widehat{\mathcal{V}}_{JI}^{(1)}-\frac{\hbar^2}{8}\widehat{\mathcal{Q}}_{IJ}^{\eminus 1}\widehat{\mathcal{Q}}_{KL}^{\eminus 1}\widehat{\mathcal{V}}_{IJKL}\\
&\quad+\frac{\hbar^2}{12}\zeta_{IM}\zeta_{JN}\zeta_{IN}\widehat{\mathcal{Q}}_{LI}^{\eminus 1}\widehat{\mathcal{Q}}_{MJ}^{\eminus 1}\widehat{\mathcal{Q}}_{NK}^{\eminus 1}\widehat{\mathcal{V}}_{IJK} \widehat{\mathcal{V}}_{LMN}+\mathcal{O}(\hbar^3)\,,
\end{aligned}
\end{align}
which is nothing but the 1PI part of the vacuum functional, sourced by the background field.  
It is well-established also at higher-loop order that $ \Gamma[\hat{\eta}] $ corresponds to all the 1PI vacuum diagrams dressed with the background fields $ \hat{\eta} $. The corresponding 1PI supergraphs are generated with the same Feynman-like rules as the vacuum functional, see Section~\ref{sec:higher_order}.

\section{Gauge-Invariant Evaluation of Functional Supergraphs}
\label{sec:gauge_inv_evaluation}

The functional formalism we are developing in this paper is more than an abstract way of expressing the generating functionals. In this section, we will show how to cast the functional supergraphs as momentum space integrals in a manner reminiscent of ordinary loop integrals, which can then form the basis for powerful practical computations. While a simple version has been presented in Ref.~\cite{Fuentes-Martin:2023ljp}, we will extend the formalism to keep the evaluation formulas manifestly gauge invariant. This gauge covariance simplifies calculations and was one of the main arguments for the development of functional methods for one-loop matching and renormalization calculations~\cite{Cheyette:1987qz,Dittmaier:1995cr,Henning:2014wua,Henning:2016lyp,Cohen:2020fcu}; however, the covariant one-loop techniques described in these references do not immediately generalize to higher-loop orders, and new methods have to be employed. The methods presented here also elucidate how gauge-invariance can be preserved through the entirety of the calculation rather than recovered at an intermediate step through clever mathematical manipulations.

\subsection{The background-field gauge}

Perturbative calculations in gauge theories are only possible once the gauge has been fixed, thereby removing unphysical singularities from the gauge-field propagator. In conventional gauges, gauge fixing breaks the symmetry of the quantum effective action down to the smaller BRST symmetry, increasing the number of independent Green's functions. An alternative is to use a class of gauges known as the \emph{background-field gauges} (BF gauges)~\cite{DeWitt:1967ub,Abbott:1980hw,Abbott:1981ke}, which fixes the gauge using the background gauge fields and has proven well-suited for RG and matching calculations. We refer the reader to Ref.~\cite{Thomsen:2024abg} for a detailed account of the use of the BF gauge in the functional formalism and merely sketch the essentials here to keep the discussion self-contained. 

We take our theory $ S[\eta] $ to possess a gauge symmetry $ G $, meaning that the corresponding gauge fields $ A_\mu \subset \eta_I $ are included among the degrees of freedom. 
The BF gauges are based on a peculiar version of the quantum effective action given by\footnote{Here, we set $\hbar=1$ for simplicity, as the loop counting question was resolved in the previous section.}
    \begin{align} 
    \label{eq:gauge-fixed_BF_vacuum_functional}
    \Gamma[\underline{\eta},\, \hat{\eta}] = -i \log \!  
    \int\, [\mathcal{D} \eta] [\mathcal{D} \boldsymbol{\omega}]\, \exp\! \left[i \!\left(\! S[\eta + \hat{\eta}] + S^G_\mathrm{fix}[\eta + \hat{\eta},\, \boldsymbol{\omega},\, \hat{\eta}] + \mathcal{J}_I (\eta_I- \underline{\eta}_I) \right) \right]\,,
    \end{align}
where the source $ \mathcal{J} $ is a function of $ \underline{\eta} $ and $ \hat{\eta} $ that is implicitly determined as the solution to\footnote{In contrast to the previous section, $ \hat{\eta} $ is technically a background field (source) rather than the expectation value of the quantum field in the presence of sources, which here is $ \underline{\eta} $. Through manipulations in the background field method, $ \hat{\eta} $ will nevertheless end up playing a similar role to the one in the previous section, hence the naming.}
    \begin{equation}
    -\zeta_{II}\mathcal{J}_I = \dfrac{\delta \Gamma[\underline{\eta},\, \hat{\eta}]}{\delta \underline{\eta}_I}\,.
    \end{equation}
The term $ S^G_\mathrm{fix}[\eta,\, \boldsymbol{\omega},\, \hat{\eta}] $ fixes the gauge of the quantum fields and is constructed with the Faddeev--Popov procedure. It involves new ghost and anti-ghost degrees of freedom collectively referred to as $ \boldsymbol{\omega} = (\overline{\omega},\, \omega) $. For instance, in case of a simple gauge group in the usual background $ R_\xi $ gauge, it reads\footnote{In this section, we normalize the gauge field such that the Yang--Mills action reads $ - \tfrac{1}{4g^2} (G_{\mu\nu}^A)^2 $. The reader can easily recover the alternative convention (with the gauge coupling in the covariant derivatives) by multiplying all occurrences of the gauge field by $ g $. } 
    \begin{equation}
    S^G_\mathrm{fix}[\eta+ \hat{\eta}, \boldsymbol{\omega}, \hat{\eta}] 
    = -\dfrac{1}{2g^2 \xi} \big( \widehat{D}^\mu A^A_\mu  \big)^2 
    - \overline{\omega}_A \widehat{D}^\mu \big(\widehat{D}_\mu \omega^A + f\ud{A}{BC} A_\mu^B \omega^C \big)\,,
    \end{equation}
where $ \widehat{D}_\mu = \partial_\mu - i \widehat{A}^A_\mu T_A $ and $ T_A $ are the generators of the relevant representation and $ f\ud{A}{BC} $ are the structure constants of the gauge group. It is common to fix the gauge parameter $ \xi = 1 $ in functional calculations (see e.g.~\cite{Henning:2014wua}) to simplify the dressed vector propagator.   

In this construction, the gauge of the quantum field $ \eta $ is fixed using the background field $ \hat{\eta} $ in a manner that the quantum effective action becomes invariant under a background-field gauge symmetry. This lets us define the \emph{gauge-invariant effective action} 
    \begin{equation}
    \Gamma[\hat{\eta}] \equiv \Gamma[0,\, \hat{\eta}]\,.
    \end{equation}
The gauge-invariant effective action can be used for matching and renormalization calculations and the manifest background-gauge invariance simplifies calculations considerably. 

The perturbative construction of the gauge-invariant effective action in the BF gauge now looks exactly like in \eqref{eq:eff_action_generic_2-loop}, except that the functional tensor building blocks differ from the original definition~\eqref{eq:action_expansion} based on the UV action. Instead, we write (with $\hbar=1$)
    \begin{align} \label{eq:BF_gaue_action_expansion}
    \begin{aligned}
    S[\eta+ \hat{\eta}] + S^G_\mathrm{fix}[\eta+ \hat{\eta}, \boldsymbol{\omega}, \hat{\eta}] = \widehat{S} 
    + \sum_{\ell=0}^{\infty} &\Big( \Omega_I \widehat{\mathcal{V}}^{(\ell)}_I + \frac{1}{2}\, \Omega_I \widehat{\mathcal{V}}^{(\ell)}_{IJ}\Omega_J + \sum_{n=3}^\infty \frac{1}{n!}\,\Omega_{I_1}\dots\Omega_{I_n}\widehat{\mathcal{V}}^{(\ell)}_{I_1\dots I_n}\Big)\,,
    \end{aligned}
    \end{align}
where $ \Omega_I = (\eta,\, \boldsymbol{\omega}) $ denotes the collection of all quantum fields. New contributions from the gauge-fixing term are added to the functional tensors compared to the un-gauged version. In contrast to the regular action, the gauge-fixing terms depend on the fields not only through the combination $ \hat{\eta} + \eta $, and we emphasize that the expansion \eqref{eq:BF_gaue_action_expansion} is performed in the quantum fields rather than in the background fields. We also observe that ghost fields do not appear as background fields; they are purely quantum, that is, integration variables.

\subsection{Gauge-covariant functional derivatives}

Let us now examine the basic building block of the functional supergraphs, the dressed kinetic operator. It can be determined as a second derivative of the gauge-fixed action:
    \begin{equation} \label{eq:dressed_kin_op}
    \widehat{\mathcal{Q}}_{IJ} \equiv \widehat{\mathcal{Q}}_{ab}(x,y) = \zeta_{bb} \dfrac{\delta^2 \big(S^{(0)}[\eta + \hat{\eta}] + S^G_\mathrm{fix}[\eta+ \hat{\eta}, \boldsymbol{\omega}, \hat{\eta}] \big)}{\delta \Omega_a(x) \delta \Omega_b(y)} \bigg|_{\eta = 0}\,.
    \end{equation}
This definition introduces a subtle problem when it comes to gauge covariance. Usually, we would define the functional derivative as $ \delta \eta_a(x)/ \delta \eta_b(y) = \delta_{ab} \, \delta(x-y) $. However, we observe that the r.h.s. does not transform as the l.h.s. under background-gauge transformations. Under a background-gauge transformation, $g(x) \in G $, we have\footnote{Even the quantum-gauge fields transform homogeneously under a background-gauge transformation. This transformation property of the quantum fields does not hold for quantum-gauge transformations. Our only concern is to maintain manifest background-gauge covariance.}
    \begin{equation}
    \dfrac{\delta \eta_a(x)}{ \delta \eta_b(y) } \longrightarrow g_{ac}(x) \dfrac{\delta \eta_c(x)}{ \delta \eta_d(y) } g^{\dagger}_{\,db}(y)\,.
    \end{equation}
Clearly, this transformation property is not manifest in the regular product of Kronecker and Dirac delta functions. The issue with the standard construction is that it does not account for the action of the background-gauge group on the field fiber bundle. 

The fields may be understood as sections of a vector bundle, and their coordinates can only be compared by parallel-transporting them with the gauge connection (the background gauge field). A fiber $ f_a(y) $ parallel-transported to a point $ x $ along a line $\gamma $ is transported as
    \begin{equation}
    f_a(x) = [U_\gamma(x,\, y)]_{a b}\, f_b(y)\,, \qquad U_\gamma(x,\,y) = \mathscr{P} \exp \! \left[ i \! \int_\gamma \dd z^\mu \widehat{A}_\mu(z) \right]\,,
    \end{equation}
such that $U_\gamma(x,\, y) $ is the Wilson line along $\gamma $, formally identified with the path-ordered integral over the background-gauge field. 
Under a background-gauge transformation, we have
    \begin{equation}
        [U_\gamma(x,\, y)]_{ab}  \to [g(x) U_\gamma  (x,\, y) g^\dagger(y)]_{ab}\,.
    \end{equation}
That is, the first index transforms as the field $\eta_a $ at position $ x$, while the second transforms in the conjugate representation at $ y$.
The parallel transport tells us how to compare field vectors at different spacetime points and we can construct a covariant functional derivative by setting
    \begin{equation} \label{eq:cov_delta_definition}
    \dfrac{\delta \eta_a(x)}{ \delta \eta_b(y) } = \delta_{ab}(x,\, y)\,, \qquad \delta_{ab}(x,\, y) \equiv [U_{\gamma} (x,\, y)]_{ab}\; \delta(x-y)\,,
    \end{equation}
where $ \delta_{ab}(x,\, y)$ is a \emph{covariant delta function}.\footnote{The covariant delta function has all the properties of the Dirac delta function. Although our arguments for its introduction are somewhat loose, we merely multiply the delta function with $ U(x,x) = \mathds{1} $ where it is non-zero. The Wilson line simply ensures that the object is covariant so that we can consistently apply functional derivatives.}
The transformation properties of the Wilson line ensure that both sides of the equality transform identically. We emphasize that
    \begin{equation}
    \dfrac{\delta\, \delta_{ab}(x,y)}{\delta \eta_c(z)} = 0 
    \implies \dfrac{\delta^2 \eta_a(x)}{ \delta \eta_c(y) \, \delta \eta_b(z)} =0\,,
    \end{equation}
seeing as $ U_\gamma(x, y)$ has no dependence on the quantum fields.

We still need to specify the curve from $ x $ to $ y $ in the covariant delta function $\delta_{ab}(x,\, y) $. A unique choice stands out for its simplicity, namely the straight line (parametrized by $ \gamma(s) = s\, x + (1-s) y $). In what follows, we will exclusively use this choice and denote the corresponding Wilson line by $U(x,y)$ (no subscript for the curve). The straight line---or a geodesic in curved space---identifies the Wilson line with the \emph{parallel displacement propagator} (PDP), which has many useful properties that facilitate practical calculations~\cite{Barvinsky:1985an,Kuzenko:2003eb}.
We will relegate the careful examination of the properties of the PDP to Appendix~\ref{app:pdp}. For now, we note that the covariant delta function with this choice of Wilson line behaves like the ordinary Dirac delta function, but with covariant derivatives. For instance, we have that
    \begin{equation} \label{eq:dev_on_cov_delta}
    P_x^\mu \delta_{ab}(x,\,y) = - P_y^\mu \delta_{ab}(x,\,y)\,,
    \end{equation}
with $ P^\mu_x = i D^\mu_x $ being the covariant momentum operator of spacetime coordinate $ x $. This is a covariant version of the familiar property of ordinary derivatives on an ordinary delta function. 

Locality of the QFT action endows a notion of locality on the superindex tensors $ \mathcal{Q}_{IJ}$, $ \mathcal{V}^{(0)}_{IJK} $, etc. For superindex matrices, it means that they can be written as local, covariant differential operators acting on covariant delta functions. These superindex matrices satisfy convenient multiplication properties. For example, let us consider two of them
    \begin{equation}
    \mathcal{A}_{IJ} = A_{ac}(x, P_x) \delta_{cb}(x,y)\,, \qquad  \mathcal{B}_{IJ} = B_{ac}(x, P_x) \delta_{cb}(x,y)\,,
    \end{equation} 
with $ A_{ab} $ and $ B_{ab} $ being differential operators and the generic assumption that $A_{ad}$ can be written as $A_{ad}(x,P_x)=\sum_{n=0}^\infty A^{\mu_1\cdots\mu_n}_{ad}(x)P_x^{\mu_1}\cdots P_x^{\mu_n}$. From recursively applying  Eq.~\eqref{eq:dev_on_cov_delta} along with integration-by-parts identities, we have the relation
    \begin{align}\label{eq:prod_func_matrices}
     \mathcal{A}_{IJ} \mathcal{B}_{JK} &= \int_y\; \sum_{n=0}^\infty A^{\mu_1\cdots\mu_n}_{ad}(x)P_x^{\mu_1}\cdots P_x^{\mu_n} \delta_{db}(x,y) B_{bf}(y, P_y) \delta_{fc}(y,z)\nonumber\\
     &=\int_y\; \sum_{n=0}^\infty (-1)^n A^{\mu_1\cdots\mu_n}_{ad}(x) [P_y^{\mu_n}\cdots P_y^{\mu_1} \delta_{db}(x,y)] B_{bf}(y, P_y) \delta_{fc}(y,z)\nonumber\\
     &=\int_y\;    \delta_{db}(x,y) \sum_{n=0}^\infty A^{\mu_1\cdots\mu_n}_{ad}(x) P_y^{\mu_1}\cdots P_y^{\mu_n}B_{bf}(y, P_y) \delta_{fc}(y,z)\nonumber\\
     &
     =A_{ab}(x, P_x) B_{bd}(x, P_x) \delta_{dc}(x,z)\,,     
\end{align}
where the brackets indicate that the differential operators only act onto the terms which are inside of them. With inductive arguments, one easily obtains the corollary that 
    \begin{equation}
    \mathcal{A}^n_{IJ} = A^n_{ac}(x, P_x) \delta_{cb}(x,y)\,, \qquad n \in \mathbb{N}\,.
    \end{equation}
We can also make sense of an inverse functional matrix, noting that the identity is given by $ \mathcal{I}_{IJ} \equiv \mathds{1}_{ac} \delta_{cb}(x,y) = \delta_{ab}(x,y) $. Setting aside issues of convergence, the inverse of a functional matrix can be expressed as the series 
    \begin{equation}\label{eq:inv_func_mat}
    \mathcal{A}^{\eminus 1}_{IJ} = \sum_{n=0}^{\infty} (\mathcal{I} - \mathcal{A})^{n}_{IJ} = \sum_{n=0}^{\infty} \big(\mathds{1} - A(x,P_x) \big)^{n}_{ac} \delta_{cb}(x,y) =  A^{\eminus 1}_{ac}(x,P_x) \delta_{cb}(x,y)\,.
    \end{equation}   
Naturally, this definition satisfies $ \mathcal{A}^{\eminus 1} \mathcal{A} = \mathcal{A} \,\mathcal{A}^{\eminus 1} = \mathcal{I} $. Using the power-series definition of a matrix logarithm, we can also give meaning to
    \begin{equation} \label{eq:log_func_mat}
    (\log \mathcal{A})_{IJ} = - \sum_{n=1}^{\infty} \dfrac{1}{n} (\mathcal{I} - \mathcal{A})^{n}_{IJ} = - \sum_{n=1}^{\infty} \dfrac{1}{n} \big(\mathds{1} - A(x,P_x) \big)^{n}_{ac} \delta_{cb}(x,y) =  \big(\log A(x,P_x) \big)_{ac} \delta_{cb}(x,y)\,.
    \end{equation}   
This is exactly the kind of logarithm that shows up in the one-loop effective action, as already indicated by Eq.~\eqref{eq:eff_action_generic_2-loop}.

\subsection{Gauge-invariant one-loop effective action}

Unlike higher-loop contributions, which are represented as sums of supergraphs, the one-loop effective action is obtained from a functional logarithm, see~\eqref{eq:eff_action_generic_2-loop}. This result makes its evaluation more subtle and distinctly different from higher-loop calculations. Building on the approach of Ref.~\cite{Kuzenko:2003eb}, we demonstrate how to covariantly evaluate the one-loop effective action using the covariant delta function introduced in the previous section.

Returning to the dressed kinetic operator~\eqref{eq:dressed_kin_op}, locality of the action ensures that it can be written as local functional matrix\footnote{The reader may apply two functional field derivatives to an action and verify that such a rearrangement is possible. Indeed, it becomes readily apparent that this is always the case.}
    \begin{equation}
    \mathcal{Q}_{IJ} = Q_{ac}(x,\,P_x) \delta_{cb}(x,y)\,. 
    \end{equation}
Both the logarithm and the inverse of the operator have to be evaluated as a series expansion. The logarithm requires particular care, since we need to extract an infinite contribution from a vacuum loop, all while the various terms in the kinetic operator do not generally commute. The usual approach is to decompose the kinetic operator into two pieces
    \begin{equation} \label{eq:Q_split}
    \mathcal{Q}_{IJ} = \boldsymbol{\Delta}_{IJ} - \mathcal{X}_{IJ}\,, \qquad \begin{dcases}
    \mathcal{X}_{IJ}=  X_{ac}(x, P_x) \delta_{cb}(x,y) \\  \boldsymbol{\Delta}_{IJ} =  \Delta_{ac}(P_x) \delta_{cb}(x,y)
    \end{dcases}\,,
    \end{equation}
where $ \Delta $ is identified with the kinetic piece of a free Lagrangian (with covariant derivatives), whereas $X$ contains the rest of the kinetic operator, which is usually referred to as the \textit{interaction terms}. The kinetic pieces encoded in $\Delta$ are diagonal for real degrees of freedom. Using the Feynman gauge for the gauge fields, we have\footnote{In these expressions, and the analogous ones for complex degrees of freedom, implicit identities for the internal degrees of freedom, such as gauge and flavor indices, are left implicit.}
\begin{align}
\Delta_{\phi\phi}&=P_x^2 - m_\phi^2\,,&
\Delta_{AA}&=-g_{\mu\nu} (P_x^2 -m_A^2)\,,
\end{align}
with $\phi$ ($A$) denoting a generic real scalar (vector) field. For complex degrees of freedom, we take their components to be $\phi=(\varphi~~\varphi^*)$, $A_\mu=(V_\mu~~V_\mu^*)$ and $\chi=(\psi~~\psi^c)$ for generic complex scalar, vector, and fermion fields, respectively.\footnote{Here, $\psi^c=C\bar \psi^\intercal$ denotes the charge-conjugated fermion, with $C$ being the charge conjugation matrix and $\psi$ a 4-component Dirac spinor.} The only non-zero entries of the kinetic piece are
\begin{align}
\begin{aligned}
\Delta_{\varphi\varphi^*}&=\Delta_{\varphi^*\varphi}=P_x^2 - m_\varphi^2\,,&
\Delta_{VV^*}&=\Delta_{V^*V}=-g_{\mu\nu} (P_x^2 -m_V^2)\,,\\
\Delta_{\psi\psi^c}&=-\Delta_{\psi^c\psi}=C(\slashed P_x - m_\psi)\,,&
\Delta_{\omega\overline{\omega}}&=-\Delta_{\overline{\omega}\omega}=P_x^2 - m_\omega^2\,,
\end{aligned}
\end{align}
where $\omega$ denotes a generic ghost field. For renormalization calculations or light fields in matching calculations (where the light masses are order parameters), it proves convenient to absorb the masses into $\mathcal{X} $.  

With the split of the kinetic operator in~\eqref{eq:Q_split}, we can decompose the genuine one-loop contribution to the effective operator into \emph{log-type} and \emph{power-type} traces according to\footnote{Even for non-commuting (functional) matrices, it holds that $ \mathrm{STr}\log(\mathcal{A}\mathcal{B})= \mathrm{STr} \log(\mathcal{A}) + \mathrm{STr} \log(\mathcal{B}) $. }  
    \begin{equation}
    \begin{split}
    \Gamma^{(1)} \supset \dfrac{i}{2} \mathrm{STr} \log \mathcal{Q} &= \dfrac{i}{2}\mathrm{STr} \log \boldsymbol{\Delta} + \dfrac{i}{2}\mathrm{STr} \log\! \big(\mathcal{I} - \boldsymbol{\Delta}^{\! \eminus 1} \mathcal{X} \big)\\
    &= \dfrac{i}{2}\mathrm{STr}\log \boldsymbol{\Delta} - \dfrac{i}{2} \sum_{n=1}^{\infty} \dfrac{1}{n} \mathrm{STr} \big(\boldsymbol{\Delta}^{\! \eminus 1} \mathcal{X} \big)^n \equiv \Gamma^{(1)}_\mathrm{log} + \Gamma^{(1)}_\mathrm{power}\,.
    \end{split}
    \end{equation}
In terms of ordinary Feynman diagrams, one may think of the log-type trace as containing all contributions from loops over a single field dressed with any number of external gauge lines. Everything else is contained in the power-type traces.  It is not a given that expansions of this type always converge. However, as discussed in Section~\ref{sec:matching}, the convergence of the series can be justified in EFT matching or renormalization calculations, where the power-counting provides a cut-off for the series.

The power-type traces are first expressed in position space by extracting the spacetime coordinates from the super indices:
    \begin{align}
    \begin{aligned}
    \Gamma^{(1)}_\mathrm{power} &= - \dfrac{i}{2} \sum_{n=1}^{\infty} \dfrac{1}{n} \mathrm{STr} \Big[ \big( \Delta^{\!\eminus 1}(P_x) X(x, P_x)\big)_{ac}^n \delta_{cb}(x,y) \Big] \\
    &= - \dfrac{i}{2} \sum_{n=1}^{\infty} \dfrac{1}{n} \int_{x,y} \zeta_{bb}\,\delta_{ba}(y,x) \big(\Delta^{\!\eminus 1}(P_x) X(x, P_x)\big)_{ac}^n \delta_{cb}(x,y)\,.
    \end{aligned}
    \end{align}
The trace is understood as a double integral constrained with an additional covariant delta function. This avoids unnecessary confusion from setting $x= y$ prematurely. The original covariant delta function factorizes into a PDP and 
an ordinary Dirac delta function~\eqref{eq:cov_delta_definition}. The latter is written in Fourier space as $\delta(x-y) = \int_k \,e^{\eminus ik\cdot (x-y)} $, where the momentum-space integral comes with the usual prefactor: $ \int_k\, \equiv (2\pi)^{\eminus d} \int\,\dd^d k $.\footnote{The Fourier representation of the delta function is adopted to regulate the divergence of the functional, which ultimately takes the schematic form $\int_{x,y}\,\delta(x-y)^2$. Other ways of regulating this divergence are possible; however, the Fourier expansion of the delta function has the added advantage of casting the result in terms of traditional loop-momentum integrals, which allows for the application of a plethora of standard methods.} Next, moving the exponential to the left of the differential operator by noting that $ e^{ik\cdot x} \partial^\mu_x e^{-ik\cdot x} = \partial^\mu_x -i k^\mu $,  we obtain
    \begin{align}\label{eq:PowerTraces}
    \begin{aligned}
    \Gamma^{(1)}_\mathrm{power}  
    &= - \dfrac{i}{2} \sum_{n=1}^{\infty} \dfrac{1}{n} \int_{x,y} \,\int_k  e^{\eminus ik \cdot (x-y)} \,\zeta_{bb}\,\delta_{ba}(y,x) \big[\Delta^{\!\eminus 1}(k+ P_x) X(x, k+ P_x)\big]_{ac}^n U_{cb}(x,y)  \\
    &= - \dfrac{i}{2} \sum_{n=1}^{\infty} \dfrac{1}{n} \int_{x} \,\int_k \,\zeta_{aa}\big[\Delta^{\!\eminus 1}(k+ P_x) X(x, k+ P_x)\big]_{ab}^n U_{ba}(x,y) \bigg|_{y=x}\,.
    \end{aligned}
    \end{align}
For the evaluation of this expression, one can readily expand the inverse kinetic operator $\Delta^{\!\eminus 1}(x, k+ P_x)$ as a series of covariant derivatives, keeping in mind that the series will admit a truncation for RG and matching calculations (see Section~\ref{sec:matching}). For instance, for heavy real scalar fields, we have
\begin{align}
\Delta_{\phi\phi}^{\!\eminus 1}(k+ P_x)&=\frac{1}{k^2-m^2} \sum_{n=0}^\infty \,(\eminus 1)^n \left( \dfrac{2k \cdot P_x + P_x^2}{k^2 - m^2} \right)^{\!\! n} \,,
\end{align}
and analogously for other field types. 
The final result is similar to previous derivations except for the presence of the PDP, which ensures covariance of the result. We may evaluate its derivatives in the coincidence limit, $ y=x$. We will give an illustrative example of this procedure at the end of this section.

The log-type trace follows in a fashion reminiscent of the power-type traces, except for subtleties involving taking the logarithm of a differential operator---a combination known to cause endless grief unless one is exceedingly careful in its evaluation. 
Using Eq.~\eqref{eq:log_func_mat} for the logarithm of a local superindex matrix, the log-type trace is cast as 
    \begin{equation}\label{eq: log-trace}
    \Gamma^{(1)}_\mathrm{log} = \dfrac{i}{2}\mathrm{STr} \Big[ \log\! \big[ \Delta(P_x) \big]_{ac} \delta_{cb}(x,y) \Big]\,.
    \end{equation}
At this stage, there is no handle to expand the logarithm around anything. Instead, now that we have separated out the differential operator, we may proceed as with the power-type traces and perform a loop-momentum shift, obtaining
    \begin{equation}\label{eq:LogTraces}
    \Gamma^{(1)}_\mathrm{log} =\dfrac{i}{2} \,\zeta_{aa} \! \int_{x}\, \int_k\; \log \! \big[\Delta(k+ P_x)\big]_{ab} U_{ba}(x,y) \bigg|_{y=x}\,.
    \end{equation}
As for the power-type traces, we can expand the logarithm as a series of covariant derivatives. It is a simple enough matter to expand the logarithm for bosons, seeing as $ k_\mu $ and $ P_x^\mu $ commutes. It follows that
    \begin{equation}\label{eq: log-trace-scalar}
    \log\!\big[(k+P_x)^2 -m^2 \big] = \log[k^2 -m^2] - \sum_{n=1}^\infty \dfrac{(\eminus 1)^n}{n} \left( \dfrac{2k \cdot P_x + P_x^2}{k^2 - m^2} \right)^{\!\! n}\,,
    \end{equation}
for scalar fields. The leading term is an infinite constant, which can simply be ignored. The same procedure can be applied to the vector fields yielding a similar expression.

The same approach cannot be used directly for fermions because $ \commutator{\slashed k}{ \slashed P_x} \neq 0$. This issue is quickly resolved with a little trick: The Dirac algebra ensures that the trace depends on the mass only through the combination $ m^2 $. Thus, $ \tr \log \big[ \slashed k + \slashed P_x -m \big] = \tr \log \big[ \slashed k + \slashed P_x + m \big] $ with the trace being over the Dirac algebra. Furthermore, the arguments of the two logarithms commute, so the addition rule for logarithms applies:
    \begin{equation}
    \tr \log \big[ \slashed k + \slashed P_x -m \big] = \dfrac{1}{2} \tr \log \big[ k^2 -m^2 + 2 k\cdot P_x  +\slashed P_x \slashed P_x \big]\,.
    \end{equation}
All terms with factors of $ P_x $ now commute with the leading $ k^2-m^2 $, facilitating the expansion\footnote{The reader may be tempted to substitute $  \slashed P_x \slashed P_x = P_x^2 - \sigma^{\mu\nu} G_{\mu \nu}$, where the field-strength tensor is in the appropriate representation. However, in practical calculations, we have found it better not to apply this substitution.}
    \begin{equation}\label{eq:logtrace_fermions}
    \tr \log\big[\slashed k+ \slashed P_x -m\big] = 2 \log[k^2 -m^2] - \dfrac{1}{2} \tr \sum_{n=1}^\infty \dfrac{(\eminus 1)^n}{n} \left( \dfrac{2k \cdot P_x + \slashed P_x \slashed P_x}{k^2 - m^2} \right)^{\!\! n}\,.
    \end{equation}
Hence, with $ \Delta_{ab} $ (anti-)diagonal in the fields, we now know how to evaluate all contributions to the log-type trace. The \texttt{Matchete}~\cite{Fuentes-Martin:2022jrf} package employs, as of version \texttt{v0.2}, the Wilson-line approach presented here.

\subsubsection{Practical example} \label{sec:PDP_example}
As an illustrative example of working with the PDP in practice, we consider the computation of the log-trace in the coincidence limit for a scalar field $\phi$ charged under a $\U(1)$ gauge symmetry. A relevant contribution from the $ n=3 $ term in the expansion~\eqref{eq: log-trace-scalar} is given by
\begin{align}\label{eq: example-coincidencelimit}
\int_x\ \int_k \, \log \,\Delta_{\phi\phi} U_{\phi\phi}\bigg|_{y=x}\supset 
\int_x\ \int_k\ \frac{4}{3}\frac{k_\mu k_\nu}{(k^2-m^2)^3}[P_x^2P_x^\mu P^\nu_x+P_x^\mu P^2_xP^\nu_x+P_x^\mu P^\nu_xP^2_x] U_{\phi\phi}\bigg|_{y=x}.
\end{align}
If the field is charged under an abelian $\U(1)$ gauge symmetry, the following identity holds (see App.~\ref{app:pdp} for details in the generic case)
\begin{align}\label{eq: dev-U1-PDP}
    P_x^\mu U_{\phi\phi}(x,y)=&
    \frac{1}{2}(x-y)_\nu F^{\nu\mu}\,U_{\phi\phi}(x,y)-\frac{1}{6}(x-y)_\rho(x-y)_\nu D^\rho F^{\nu\mu}\,U_{\phi\phi}(x,y)\nonumber\\
    &+\frac{1}{24}(x-y)_\sigma(x-y)_\rho(x-y)_\nu  D^\sigma D^\rho F^{\nu\mu}\,U_{\phi\phi}(x,y) + \mathcal{O}\big( (x-y)^4 \big),
\end{align}
where $F^{\mu\nu}$ is the field strength tensor associated with the abelian gauge symmetry. By iteratively applying Eq.~\eqref{eq: dev-U1-PDP} and neglecting total-derivative terms, one obtains
\begin{align}
    P_x^\mu P_x^\nu P_x^\rho P_x^\sigma \, U_{\phi\phi}(x,y)=\frac{1}{4}(F^{\mu\nu}F^{\rho\sigma}+F^{\mu\rho}F^{\nu\sigma}+F^{\mu\sigma}F^{\nu\rho})+ \mathcal{O}(x-y).
\end{align}
Therefore, inserting this expression into \eqref{eq: example-coincidencelimit} and taking the coincidence limit leads to 
\begin{align}
    \int_x\ \int_k \log \,\Delta(k+P_x)_{\phi\phi} U_{\phi\phi}\bigg|_{x=y}\supset -\frac{2}{3}\int_{x}\ \int_k\ \frac{k_\mu k_\nu}{(k^2-m^2)^3}F^{\mu\rho}{F_{\rho}}^\nu.
\end{align}

The last spacetime integral over $ x $ does not need to be performed in practice. The integrand is interpreted as a Lagrangian density and the $ x $ integration merely makes it into a local action functional. On the other hand, the integral over loop momentum $k$ has to be evaluated and its generic evaluation can be rather complicated. As we will discuss later its evaluation is particularly simple and essentially amounts to computing vacuum loop integrals, when it comes to counterterm and matching calculations.

\subsubsection{Comparison with previous approaches}

Many works have addressed the evaluation of the functional logarithm in a way that preserves explicit gauge covariance, either by manually commuting open covariant derivatives~\cite{Fuentes-Martin:2016uol,Zhang:2016pja} or using the Covariant Derivative Expansion (CDE)~\cite{Gaillard:1985uh,Chan:1986jq,Cheyette:1987qz,Henning:2014wua,Cohen:2020fcu,Fuentes-Martin:2020udw}. These approaches rely on rearrangements of the integrands in~\eqref{eq:PowerTraces} and~\eqref{eq:LogTraces} into sums of gauge-singlet terms, which renders the PDP trivial. For example, the CDE works by sandwiching the integrands between the operators $e^{-P_x\cdot\partial_k}$ and $e^{P_x\cdot\partial_k}$, with $k$ being the loop momentum. For a generic differential operator $A(X(x),P_x^\mu)$ acting on the PDP, the CDE gives
\begin{align}
\int_x\,\int_k\;A(X(x),k^\mu+P_x^\mu)\, U(x,y)\bigg|_{x=y}&=\int_x\,\int_k\;e^{-P_x\cdot\partial_k}\,A(X(x),k^\mu+P_x^\mu)\,e^{P_x\cdot\partial_k}\, U(x,y)\bigg|_{x=y} \nonumber\\
&=\int_x\,\int_k\;A(\tilde X(x),k^\mu+i \tilde F_{\mu\nu}\,\partial_k^\nu)\, U(x,y)\bigg|_{x=y}\,.
\end{align}
Here, the first equality follows from the PDP independence of the loop momentum and the vanishing of total derivatives. In the second step, commuting the operator $e^{P_x\cdot\partial_k}$ with $A$ produces the objects
\begin{align}\label{eq:CDE_byproducts}
\tilde{X}(x)&\equiv \sum_{n=0}^{\infty} \dfrac{(-i)^n }{n!}\, (D_{\{\alpha_1,\ldots, \alpha_n\}}  X) \,\partial^{\alpha_1}_k \cdots \partial_k^{\alpha_n},\qquad D_{\{\mu_1,\cdots\mu_n\}}\equiv \frac{1}{n!} \sum_{\sigma\in \mathcal{S}_n}D_{\mu_{\sigma(1)}}\cdots D_{\mu_{\sigma(n)}}\,, \nonumber\\
\tilde{F}_{\mu \nu} &\equiv \sum_{n=0}^{\infty} \dfrac{(-i)^n }{(n+2) n!}\, (D_{ \{\alpha_1,\ldots, \alpha_n\} }  F_{\mu \nu})\, \partial_k^{\alpha_1} \cdots \partial_k^{\alpha_n}\,.
\end{align}
Since, after applying the CDE, every individual term in $A$ is free of covariant derivatives acting on the PDP, the coincidence limit for the PDP trivially simplifies to the identity. 

In practice, many authors assume from the outset that the PDP effectively becomes this formal identity (in a suitable representation of the gauge group), whether or not the CDE is explicitly applied. Indeed, even when the CDE is not used, one commonly postulates the existence of a formal identity that the open covariant derivatives ultimately act upon. After integrating over the loop momentum, the gauge non-invariant pieces produced by the open derivatives recombine into gauge-invariant objects. This observation has also motivated alternative approaches where the open derivatives are arranged into commutators. In both cases, however, the existence of  PDP is hidden, as both methods serve to trivialize it. In fact, only by including the PDP the origin of gauge invariance is revealed; when used from the beginning, gauge invariance is manifest.

While it remains unclear whether strategies analogous to the CDE or other rearrangement techniques can systematically be extended to higher-loop calculations, such extensions appear challenging. Moreover, these methods would likely yield more cumbersome expressions, as the CDE typically leads to a significant proliferation of additional terms—an aspect evident in~\eqref{eq:CDE_byproducts}. Instead, we adopt an alternative approach: apply all open derivatives until they act directly on the PDP. As demonstrated in Section~\ref{sec:PDP_example}, the derivatives acting on the PDP can readily be evaluated in the coincidence limit, thereby automating the process of commuting and closing the open covariant derivatives.

\subsection{Gauge-invariant two-loop effective action}
\label{sec:GaugeCov_Evaluation}

In contrast to the one-loop case, covariant evaluation at multi-loop order seems to have been the exclusive remit of heat-kernel~\cite{Jack:1982hf,vonGersdorff:2022kwj} and diagrammatic~\cite{Kuzenko:2003eb} calculations. Using the PDP techniques, we can now derive manifestly covariant evaluation formulas for the two-loop effective action within the functional framework, building on the non-gauged approach of Ref.~\cite{Fuentes-Martin:2023ljp}. Several topologies appear in the two-loop effective action~\eqref{eq:eff_action_generic_2-loop}, and we treat them one by one. 
We parametrize the contributions to the two-loop effective action~\eqref{eq:eff_action_generic_2-loop} as
    \begin{equation} 
    \Gamma^{(2)} =  S^{(2)} + \dfrac{i}{2} G_\mathrm{ct.}
    + \dfrac{1}{12} G_\mathrm{ss.} - \dfrac{1}{8} G_\mathrm{f8.}\,.
    \end{equation}
For notational simplicity, we omit the Grassmann signs in what follows, keeping in mind that their inclusion is straightforward and essentially amounts to multiplying with the corresponding $\zeta_{ab}$ factors.
It may strike the reader as intimidating to extend the presented program beyond two-loop order given that each topology requires unique attention. However, we will show how to generalize the techniques in Section~\ref{sec:higher_order} with a minimal set of rules.

\subsubsection{Counterterm topology}

We begin with the counterterm contribution, $  G_\mathrm{ct.} =  \mathcal{Q}^{\eminus 1}_{IJ} \mathcal{V}_{JI}^{(1)}$, which is in essence nothing but a one-loop topology with a one-loop vertex. As such, no new techniques are required beyond what is used for the one-loop power-type traces. 
In analogy with the fluctuation operator, the one-loop counterterm super-index matrix is local and is parametrized by
    \begin{equation}
    \mathcal{V}_{IJ}^{(1)} \equiv V_{ac}^{(1)}(x,\,P_x) \delta_{cb}(x,y)\,.
    \end{equation}
With the contraction rule~\eqref{eq:prod_func_matrices}, we can write
    \begin{equation}
    G_\mathrm{ct.}= \mathcal{Q}^{\eminus 1}_{IJ} \mathcal{V}_{JI}^{(1)} 
    = \int_{x,y} \delta_{ba}(y,x) Q^{\eminus 1}_{ac}(x,\, P_x) V^{(1)}_{cd}(x,\, P_x) \delta_{db}(x,y) \,. 
    \end{equation}
We reiterate that the coincidence limit of the covariant delta function can be taken only after one has acted with all open covariant derivatives. Similarly to the one-loop calculation, the Dirac delta is extracted from the covariant delta, cast as a momentum-space integral, and moved to the left before vanishing in the coincidence limit. Thus,
    \begin{equation}
    G_\mathrm{ct.}=  \int_{x}\, \int_k Q^{\eminus 1}_{ab}(x,\, P_x +k) V^{(1)}_{bc}(x,\, P_x+k) U_{ca}(x,y) \Big|_{y=x}.
    \end{equation}
The dressed propagator $Q^{\eminus 1}_{ab}(x,\, P_x +k)$ appears in all two-loop contributions and it needs to be expanded around $k$ in order to evaluate the loop integrals. This expansion takes the following form:
\begin{align}
\begin{aligned}
Q^{\eminus 1}(x,P_x+k)&=[\Delta^{\!\eminus 1}(\mathds{1}-X \Delta^{\eminus 1})^{\eminus 1}](x,P_x+k)\\
&=\Delta^{\!\eminus 1}(P_x+k)\sum_{n=0}^\infty \big[X(x,P_x+k)\Delta^{\!\eminus 1}(P_x+k)\big]^n,
\end{aligned}
\end{align}
after which the inverse kinetic operator $\Delta^{\!\eminus 1}(P_x+k)$ can be further expanded as a series of covariant derivatives, as discussed in the one-loop evaluation. Once these expansions have been performed, the loop integral can easily be evaluated for each term in the series. The covariant derivatives act to their right using the chain rule until they act on the PDP. All derivatives acting on the PDP are then evaluated in the coincidence limit utilizing, e.g., the formula given by Eq.~\eqref{eq:pdp_at_coincidence}.

\subsubsection{Sunset topology}
\label{sec:sunset_evaluation}

Taking three functional derivatives of the action to obtain the three-point function, will generally produce an expression that is a sum of terms of the form 
    \begin{equation}\label{eq:vertex}
    \mathcal{V}^{(0)}_{IJK} \supset \int_v \;V^{(\underline{m}, \underline{n}, \underline{r})}_{a'b'c'}(v) P_{v}^{\underline{m}} \delta_{a'a}(v,x) P_{v}^{\underline{n}} \delta_{b'b}(v,y) P_{v}^{\underline{r}} \delta_{c'c}(v,z)\,,
    \end{equation}
where, for compactness, we employ a power-like notation with underlined superscripts for the Lorentz indices under which we denote, e.g., $A^{\underline{m}}B^{\underline{m}}\equiv A^{\mu_1\dots\mu_m} B_{\mu_1\dots\mu_m}$ for the contraction of the $ m $ Lorentz indices between the two objects. One of the momenta can always be eliminated in favor of the other two by performing integration-by-parts (IBP) relations under the integral (this is equivalent to momentum conservation). One should keep in mind that doing IBP will reverse the $ \underline{m} $ derivative indices as they are distributed on the various terms. Moreover, the derivatives on the delta functions can be exchanged for derivatives in the second coordinate:
    \begin{equation}\label{eq:momentum_cord_change}
    P^{\underline{s}}_x \delta_{ab}(x,y) = (\eminus 1)^{s} P^{\underleftarrow{s}}_y \delta_{ab}(x,y)\,,
    \end{equation}
where the left arrow under the $ s $ index is used to indicate that the order of the index set is now reversed, which follows from the repeated application of~\eqref{eq:dev_on_cov_delta}.\footnote{If $ \underline{s} = \nu_1 \ldots \nu_s $ then $ \underleftarrow{s} = \nu_{s} \ldots \nu_1 $.}
In general, we may therefore parametrize the three-point functions by
    \begin{equation}
    \mathcal{V}^{(0)}_{IJK} \equiv V_{ade}(x,\,P_y,\, P_z) \delta_{db}(x,y)\delta_{ec}(x,z)\,.
    \end{equation}
This is by no means a unique parametrization, as there is freedom to change the coordinates of the delta functions and/or the covariant derivatives, but we find this form useful for our calculations. In contrast to the dressed kinetic operators, we need to extract the covariant derivatives from these operators in the course of evaluating the supergraph and parametrize 
    \begin{equation}
    V_{abc}(x,\,P_y,\, P_z) = \sum_{m,n=0} V^{(\underline{m},\underline{n})}_{abc}(x) P_y^{\underline{m}} P_z^{\underline{n}}\,.
    \end{equation}
The three-point functions of renormalizable Lagrangians will have at most one external derivative, ensuring a quick termination of the $ m,n $ sums. While the sums might terminate later in an EFT, the truncation of the EFT expansion ensures that they are finite in all cases.  

We will now cast the sunset topology in terms of loop integrals. First, in position space we obtain
    \begin{align}
    G_\mathrm{ss.} 
    =&\, \mathcal{Q}_{LI}^{\eminus1} \mathcal{Q}_{MJ}^{\eminus1} \mathcal{Q}_{NK}^{\eminus1} \mathcal{V}^{(0)}_{IJK} \mathcal{V}^{(0)}_{LMN}  \\
    =& \int_{xyzx'y'z'}  \big[Q_{af}^{\eminus 1}(x,\,P_x) \delta_{fa'} (x,x') \big] \big[Q_{bg}^{\eminus 1}(y,\,P_y) \delta_{gb'}(y,y') \big] \big[Q_{ch}^{\eminus 1}(z,\,P_z) \delta_{hc'}(z,z') \big]\nonumber\\
    &\quad \times \big[V_{a'd'e'}(x',\,P_{y'},\, P_{z'}) \delta_{d'b'}(x',y') \delta_{e'c'}(x',z') \big]\big[V_{ade}(x,\,P_y,\, P_z) \delta_{db}(x,y) \delta_{ec}(x,z) \big]\,, \nonumber 
    \end{align}
and emphasize that \emph{all derivatives are understood to close inside the brackets}.
Next, integration by parts lets us move the open derivatives from the vertices into the propagators, after which the covariant delta functions associated with the vertices can be integrated out. We find
    \begin{equation} \label{eq:sunset_formula_ibp}
    \begin{split}
    G_\mathrm{ss.} =& \!\! \sum_{m,n,m',n'} \!\!(\eminus 1)^{m+n} \int_{xx'}  \big[Q_{ad}^{\eminus 1}(x,\,P_x) \delta_{da'}(x,x') \big] \big[P_x^{\underleftarrow{m}} Q_{be}^{\eminus 1}(x,\,P_x) P_x^{\underline{m}'} \delta_{eb'}(x,x') \big] \\
    &\qquad \qquad \qquad \qquad \times \big[P_x^{\underleftarrow{n}} Q_{cf}^{\eminus 1}(x,\,P_x) P_x^{\underline{n}'} \delta_{fc'}(x,x') \big]V^{(\underline{m}',\underline{n}')}_{a'b'c'}(x') V^{(\underline{m},\underline{n})}_{abc}(x)\,,
    \end{split}
    \end{equation}
with the left arrow under the $ m, n $ indices indicating that their orderings are reversed as a consequence of IBP. For the $m^\prime, n^\prime$ indices, this reversal is compensated by the change of coordinate through Eq.~\eqref{eq:momentum_cord_change}

As we did for the one-loop evaluation, we now use the Fourier decomposition of the delta function to regulate the divergences. However, contrary to the one-loop case, there are now several ways to proceed, as there is no unique choice for which delta functions to write in momentum space. Indeed, while we need to write at least two delta functions ($n$ in the case of an $n$-loop graph) in momentum space to regulate the divergences, we could instead do it for all of them. At this stage, we cannot be sure which approach offers the best computational performance. Here, we discuss the approach with the minimal number of delta functions in momentum space, which seems to yield simpler expressions, while the alternative approach where all delta functions are written in momentum space is discussed in Appendix~\ref{app:alt_sunset}. 
Thus, we cast the covariant delta functions in the last two propagators as loop-momentum integrals and get
    \begin{align}
    G_\mathrm{ss.} 
    = \!\! \sum_{m,n,m',n'} \!\!(\eminus 1)^{m+n} &\int_{xx'} \int_{k\ell} e^{i(k+\ell) \cdot (x'-x)} V^{(\underline{m},\underline{n})}_{abc}(x) V^{(\underline{m}',\underline{n}')}_{a'b'c'}(x') \big[Q_{ad}^{\eminus 1}(x,\,P_x) \delta_{da'}(x,x') \big] \nonumber\\
    &\times \big[ (P_x+k)^{\underleftarrow{m}} Q_{be}^{\eminus 1}(x,\,P_x +k) (P_x+k)^{\underline{m}'} U_{eb'}(x,x') \big] \nonumber\\
    &\times \big[ (P_x+\ell)^{\underleftarrow{n}} Q_{cf}^{\eminus 1}(x,\,P_x +\ell) (P_x+\ell)^{\underline{n}'} U_{fc'}(x,x') \big]\,.
    \end{align}
Ideally, we would now like to carry out the $ x' $ integration with the last delta function to ensure that all field dependence moves to the $ x $ coordinate. Changing the momentum coordinates in $Q_{ad}^{\eminus 1}(x,\,P_x)$ by means of Eq.~\eqref{eq:momentum_cord_change} and using IBP to move the differential operator from the last covariant delta function, we obtain
    \begin{align}
    G_\mathrm{ss.} 
    = \!\! \sum_{m,n,m',n'} \!\!(\eminus 1)^{m+n} &\int_{xx'} \int_{k\ell} \delta_{da'}(x,x') V^{(\underline{m},\underline{n})}_{abc}(x) Q_{ad}^{\eminus 1}(x',\,P_{x'}) e^{i(k+\ell) \cdot (x'-x)} V^{(\underline{m}',\underline{n}')}_{a'b'c'}(x')  \nonumber\\
    &\times \big[ (P_x+k)^{\underleftarrow{m}} Q_{be}^{\eminus 1}(x,\,P_x +k) (P_x+k)^{\underline{m}'} U_{eb'}(x,x') \big] \nonumber\\
    &\times \big[ (P_x+\ell)^{\underleftarrow{n}} Q_{cf}^{\eminus 1}(x,\,P_x +\ell) (P_x+\ell)^{\underline{n}'} U_{fc'}(x,x') \big]\,.
    \end{align}
The open derivatives of $ Q_{ad}^{\eminus 1}(x',\,P_{x'}) $ act on all $ x'$ coordinates including the Wilson lines at the end of the other propagators. Commuting through the Fourier factor and carrying out the integral over $ x' $ finally yields
    \begin{align} \label{eq:sunset_formula}
    \begin{aligned}
    G_\mathrm{ss.} 
    = \!\! \sum_{m,n,m',n'} \!\!(\eminus 1)^{m+n} &\int_{x}\, \int_{k\ell} V^{(\underline{m},\underline{n})}_{abc}(x) Q_{aa'}^{\eminus 1}(y,\,P_{y} -k - \ell)  V^{(\underline{m}',\underline{n}')}_{a'b'c'}(y) \\
    &\times \big[ (P_x+k)^{\underleftarrow{m}} Q_{be}^{\eminus 1}(x,\,P_x +k) (P_x+k)^{\underline{m}'} U_{eb'}(x,y) \big] \\
    &\times \big[ (P_x+\ell)^{\underleftarrow{n}} Q_{cf}^{\eminus 1}(x,\,P_x +\ell) (P_x+\ell)^{\underline{n}'} U_{fc'}(x,y) \big] \bigg|_{y=x}\,.
    \end{aligned}
    \end{align}	
In practice, we can expect this rather intimidating expression to be more manageable than might be initially feared, due to the $ m^{(\prime)}, n^{(\prime)} $ sums terminating quickly. After expanding the inverse propagator as discussed in the previous subsection, loop-momentum integration for each of the terms in the series reduces to ordinary vacuum two-loop integrals, which can readily be evaluated with the standard techniques presented in, among others, Refs.~\cite{Chetyrkin:1997fm,Martin:2016bgz}. As before, all covariant derivatives are propagated through the chain rule, and derivatives acting on the PDPs are evaluated at the coincidence limit using the results of Appendix~\ref{app:PDP_devs}.

\subsubsection{Figure-8 topology}

The figure-8 topology makes up the final piece of the two-loop effective action. It involves the local rank-four tensor, which is parametrized as\footnote{As for the three-point vertex, we have eliminated the dependence in one of the momenta by performing IBP on the vertex.}  
    \begin{equation}
    \mathcal{V}^{(0)}_{IJKL} \equiv V_{aefg}(x,\,P_y,\, P_z,\, P_w) \delta_{eb}(x,y) \delta_{fc}(x,z) \delta_{gd}(x,w)\,,
    \end{equation}
where, as with the three-point vertex, it is convenient to explicitly extract the open covariant derivatives:
    \begin{equation}
    V_{abcd}(x,\,P_y,\, P_z,\, P_w) = \sum_{m,n,r=0} V^{(\underline{m},\underline{n},\underline{r})}_{abcd}(x) P_y^{\underline{m}} P_z^{\underline{n}} P_w^{\underline{r}}\,.
    \end{equation}
One typically expects these index sums to truncate rather quickly and indeed only $ V^{(0,0,0)}_{abcd}(x) $ is nonzero for renormalizable theories.

With the familiar use of integration by parts, we cast the figure-8 contribution as
    \begin{equation}
    \begin{split}
    G_\mathrm{f8.} &= \mathcal{V}^{(0)}_{IJKL} \mathcal{Q}_{IJ}^{\eminus1} \mathcal{Q}_{KL}^{\eminus1}\\
    &= \sum_{m,n,r} (\eminus 1)^n \! \int_{xyzw}  V^{(\underline{m}, \underline{n}, \underline{r})}_{aefg}(x) \delta_{eb}(x,y) \delta_{fc}(x,z)\delta_{gd}(x,w) \\
    &\qquad\qquad\qquad  \times \big[ Q_{ah}^{\eminus1}(x,\,P_x) P_x^{\underline{m}} \delta_{hb}(x,y)\big] \big[ P_z^{\underleftarrow{n}} Q_{ci}^{\eminus1}(z,\,P_z) P_z^{\underline{r}} \delta_{id}(z,w)\big]\,.
    \end{split}
    \end{equation}
Here, we have a single vertex so the integrand goes to the coincidence limit without additional manipulations. Casting the two deltas in momentum space, one readily obtains
    \begin{equation}
    \begin{split}
    G_\mathrm{f8.} 
    = \sum_{m,n,r} (\eminus 1)^{n} \! \int_{x}\, &\int_{k\ell}  V^{(\underline{m}, \underline{n}, \underline{r})}_{abcd}(x) \big[ Q_{ae}^{\eminus1}(x,\,P_x+k) (P_x+k)^{\underline{m}} U_{eb}(x,y)\big]_{y=x} \\
    &  \times  \big[ (P_x+\ell)^{\underleftarrow{n}} Q_{cf}^{\eminus1}(x,P_x+\ell) (P_x+\ell)^{\underline{r}} U_{fd}(x,z)\big]_{z=x}\,,
    \end{split} 
    \end{equation}
which, after performing the $Q^{\eminus1}$ expansion, can be directly evaluated similarly to the other topologies.

\subsection{Diagrammatic rules for higher-order contributions}
\label{sec:higher_order}

The methods presented above can be generalized to calculate higher-order corrections. To this end, it becomes useful to frame the discussion in terms of vacuum supergraphs, like the ones in Figure~\ref{fig:vacuum_functional}, and present generic rules for their evaluation, similarly to how it is done with Feynman rules and regular Feynman diagrams.

\subsubsection{Effective action}
\label{sec:RulesEffectiveAction}

\begin{figure}[t]
    \centering
    \includegraphics[width=\textwidth]{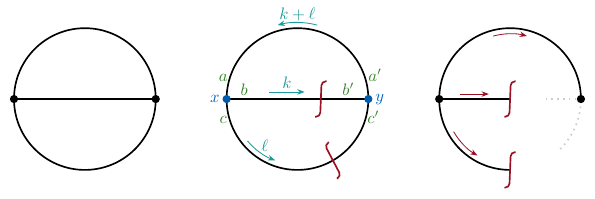}
    \caption{Two-loop sunset supergraph contributing to the quantum effective action (left). The graph graph is cut and labeled for its covariant evaluation (middle) while the $ x $ vertex is chosen as the root in the resulting tree graph, giving direction to all edges (right). }
    \label{fig:SunsetRules}
\end{figure}

As discussed in Section~\ref{sec:Effective_action} for the two-loop case, the quantum effective action is obtained from all possible 1PI vacuum supergraphs. This statement remains valid at all loop orders. The rules to derive the tensorial expressions for $\Gamma[\hat\eta]$ from the corresponding graphs are as follows:
\begin{enumerate}[i.]
    \item Add a prefactor of $-i\hbar^\ell/n_s$, with $\ell$ being the number of loops and $n_s$ the size of the symmetry group permuting vertices and edges of the graph.
    \item Each $n$-point vertex at loop order $\ell'$ contributes with a factor $i\,\widehat{\mathcal{V}}^{(\ell')}_{I_1\dots I_n}$ and each line with a factor $i\,\widehat{\mathcal{Q}}^{\eminus 1}_{IJ}$ with $I$ and $J$ matching the indices of the adjacent vertices. Tree-level vertices consist of 3-point functions and higher, whereas loop-level vertices can have any number of legs, including tadpoles. We remind the reader that the hat indicates that these tensors are functions of the background fields, $ \hat{\eta} $, which are the solutions to the quantum EOMs at the relevant perturbative order.
    \item Finally, the Grassmann signatures, $\zeta_{IJ}$, can be derived by performing the following shifts to the vertices and propagators:
    \begin{align}
    \widehat{\mathcal{V}}^{(\ell)}_{I J}&\to \eta_I\widehat{\mathcal{V}}^{(\ell)}_{IJ}\eta_J\,, &
    \widehat{\mathcal{V}}^{(\ell)}_{I_1\dots I_n}&\to \eta_{I_1}\dots\eta_{I_n}\widehat{\mathcal{V}}^{(\ell)}_{I_1\dots I_n} \; (n\geq3)\,, &
    \widehat{\mathcal{Q}}^{\eminus 1}_{IJ}\to  \widehat{\mathcal{Q}}^{\eminus 1}_{IJ}\,\eta_J\eta_I\,,
    \end{align}
    and moving the $\eta_I$ until two adjacent $\eta_I$ with the same subindex ``annihilate" with each other, while keeping track of the signature.\footnote{The justification for this rule becomes clear after considering the path-integral evaluation in Appendix~\ref{app:Details_Gen_Func}. Indeed, it is pairs of the field multiplets, $\eta$, accompanying the tensors in Eq.~\eqref{eq:expansion_generating-functional} that give rise to the propagators after evaluating the path-integral. The replacements and later annihilation of the $\eta$'s are devised to mimic this result. Special care should be taken not to commute $\eta_I \eta_I$ before annihilation, as doing $ \eta_I \eta_I = \zeta_{II} \eta_I \eta_I$ would introduce an incorrect factor of $\zeta_{II}$.}
    \end{enumerate}    
As an example, for the sunset topology in Figure~\ref{fig:SunsetRules} (left), we have three propagators (6~permutations) and two tree-level three-point vertices (2~permutations). Following our rules above, this means that its tensorial expression, up to a Grassmannian sign reads
\begin{align}
\Gamma[\hat\eta]_\textrm{ss.}\propto\frac{\hbar^2}{12} \widehat{\mathcal{Q}}_{LI}^{\eminus 1} \widehat{\mathcal{Q}}_{MJ}^{\eminus 1} \widehat{\mathcal{Q}}_{NK}^{\eminus 1} \widehat{\mathcal{V}}^{(0)}_{IJK} \widehat{\mathcal{V}}^{(0)}_{LMN}\,.
\end{align}
At this stage, the relative order of the tensors and their indices is arbitrary and we have simply chosen it to match Eq.~\eqref{eq:eff_action_generic_2-loop}. Once the order of the tensors is chosen, we can recover the Grassmannian sign from the last rule above. For the example at hand, we have
\begin{align}
&\widehat{\mathcal{Q}}_{LI}^{\eminus 1}\widehat{\mathcal{Q}}_{MJ}^{\eminus 1}\widehat{\mathcal{Q}}_{NK}^{\eminus 1}\widehat{\mathcal{V}}^{(0)}_{IJK} \widehat{\mathcal{V}}^{(0)}_{LMN}\nonumber\\
&\qquad\to \widehat{\mathcal{Q}}_{LI}^{\eminus 1}\eta_I\eta_L\,\widehat{\mathcal{Q}}_{MJ}^{\eminus 1}\eta_J\eta_M\widehat{\mathcal{Q}}_{NK}^{\eminus 1}\eta_K\eta_N\,\eta_I\eta_J\eta_K\widehat{\mathcal{V}}^{(0)}_{IJK}\,\eta_L\eta_M\eta_N\widehat{\mathcal{V}}^{(0)}_{LMN}\nonumber\\
&\qquad=\zeta_{IN}\zeta_{IM}\zeta_{JN}\,\widehat{\mathcal{Q}}_{LI}^{\eminus 1}\eta_I\eta_I\eta_L\eta_L\widehat{\mathcal{Q}}_{MJ}^{\eminus 1}\eta_J\eta_J\eta_M\eta_M\widehat{\mathcal{Q}}_{NK}^{\eminus 1}\eta_K\eta_K\eta_N\eta_N\widehat{\mathcal{V}}^{(0)}_{IJK}\widehat{\mathcal{V}}^{(0)}_{LMN}\nonumber\\
&\qquad\to\zeta_{IN}\zeta_{IM}\zeta_{JN}\,\widehat{\mathcal{Q}}_{LI}^{\eminus 1}\widehat{\mathcal{Q}}_{MJ}^{\eminus 1}\widehat{\mathcal{Q}}_{NK}^{\eminus 1}\widehat{\mathcal{V}}^{(0)}_{IJK}\widehat{\mathcal{V}}^{(0)}_{LMN}\,,
\end{align}
which reproduces the result in Eq.~\eqref{eq:eff_action_generic_2-loop}. Here, we have used that that $\zeta_{IJ}^2=1$ and that $\eta$ and the tensors $\widehat{\mathcal{V}}^{(\ell)}$ and $\widehat{\mathcal{Q}}^{\eminus 1}$ satisfy the commutation rules (see Appendix~\ref{app:Grassmann})
\begin{align}
\mathcal{T}_{I_1\dots I_n}\eta_J=\zeta_{I_1J}\dots \zeta_{I_nJ}\,\eta_J\mathcal{T}_{I_1\dots I_n}\,,
\end{align}
with $\mathcal T$ being any of the tensors.

\subsubsection{Covariant evaluation of the effective action}

Once the functional contribution of a graph to the effective action has been determined, one can readily evaluate it covariantly following the approach described in Section~\ref{sec:GaugeCov_Evaluation}. As a convenient alternative, we will also establish diagrammatic rules to determine a suitable covariant-evaluation formula directly from an arbitrary $\ell$-loop vacuum supergraph. The reader may want to keep an eye on Figure~\ref{fig:SunsetRules} while reading the rules:
\begin{enumerate}[i.]
    \item \label{item:cuts} Cut $ \ell $ edges in the $ \ell $-loop graph to obtain a connected tree graph. We think of the cut edges as being connected to the original graph in one end (with the cuts forming new vertices in the language of graph theory). Either end is equally good.  Although there are, in general, multiple choices for which edges to cut following this prescription, all choices produce valid covariant expressions. 
    
    \item Choose a spacetime label for each vertex in the graph. Add a loop momentum at each of the cut edges, pointing from the vertex where it is attached to the original graph towards the cut. Next, arrange the loop momenta in the uncut edges so that momentum is conserved at the vertices (of the uncut graph) in the manner of ordinary Feynman diagrams. 
    
    \item \label{item:root_choice} Designate a root vertex in the tree graph resulting from step \ref{item:cuts}. This endows a direction to all edges with the flow direction from the root vertex towards the leaves\footnote{The resulting graph is a directed rooted tree in the language of graph theory.} (which includes the cuts themselves). The direction resulting from this choice \emph{need not} coincide with the sense of direction resulting from the momentum flow. The choice of root vertex is generally not unique, each choice gives a different but equally valid expression. 
    
    \item Next, we need to assign operators for each edge and vertex in the graph. These operators need not commute, even in the absence of Grassmannian signatures, as they generally contain (covariant) derivatives. As the order matters, the different contributions have to be added in the following sequence: uncut edges, vertices, and cut edges. The contributions are as follows:\footnote{As we did in Section~\ref{sec:gauge_inv_evaluation}, we omit here the hat on the operators for notational simplicity. However, an implicit dependence of all the operators on the background field $\hat\eta$ is understood.} 
    \begin{enumerate}[a.]
        \item \textbf{Uncut edges:} Following the flow in the directed rooted tree graph, there is always at most one uncut edge incoming at each of the vertices, while there might be several outgoing from the vertices. For each uncut edge, we add the factor
        \begin{align}
        i\,E_{a_ib_j}^{(\underline{m}_i, \underline{n}_j)}(x, P_x+k)\equiv i\,(\eminus 1)^{m_i}\,(P_x+k)^{\underleftarrow{m}_i} Q_{a_ib_j}^{\eminus 1}(x,P_x+k) (P_x+k)^{\underline{n}_j}\,,
        \end{align}
        where $a_i$ ($b_j$) are the internal indices of the outgoing (incoming) vertex, $x$ is the spacetime label of the incoming vertex, and $k$ is the loop momentum flowing from the outgoing to the incoming vertex. As discussed in Section~\ref{sec:sunset_evaluation}, vertices may contain momentum operators, $P_{x_i}$, which can be moved by IBP to the adjacent propagators.\footnote{
            To illustrate how the momentum operators from the vertices get moved into the propagator consider a propagator connecting two vertices: 
            \begin{multline*}
            \int_{x_i y_j} V^{(\ldots,\underline{m}_i,\ldots)}_{\ldots a'_i\ldots}(x_1) P_{x_i}^{\underline{m}_i} \delta_{a'_i a_i}(x_1,x_i)\cdots \big[Q^{\eminus 1} (x_i, P_{x_i}) \delta(x_i, y_j) \big]_{a_i b_j} V^{(\ldots,\underline{n}_j,\ldots)}_{\ldots b'_j\ldots}(y_1) P_{y_j}^{\underline{n}_j} \delta_{b'_j b_j}(y_1,y_j)\cdots \\
            = V^{(\ldots,\underline{m}_i,\ldots)}_{\ldots a_i\ldots}(x_1) \big[ E^{(\underline{m}_i, \underline{n}_j)}(x_1, P_{x_1}) \delta(x_1, y_1) \big]_{a_i b_j} V^{(\ldots,\underline{n}_j,\ldots)}_{\ldots b_j\ldots}(y_1) \cdots
            \end{multline*}
            For uncut edges, we can use~\eqref{eq:momentum_cord_change} to change $ E $ to act on the $ y_1 $ coordinate, before using IBP and carrying the integration over the delta function. The directionality imposed by the root ensures that at most one uncut edge is incoming at any vertex, which in turn means that the use of IBP from the different uncut edges does not interfere if we start the process on the edges at the leaves of the tree and work our way back to the root. The momenta operators eventually pick up suitable loop momenta, $ P_{x_1} \to P_{x_1} + k $, when acting through one or more Fourier phases from the cut edges. 
        } Reflecting this, the indices $m_i$ and $n_j$ correspond to the indices of the momentum operators (if any) in the outgoing and incoming vertices, respectively.
        
        Note that the differential operators associated with uncut edges all act at different coordinates---due to the ordering of the rooted tree graph---so their relative ordering is irrelevant. We stress that all derivatives in these terms are ``open" in the sense that they act onto anything to their right with the same spacetime dependence.
        
        \item \textbf{Vertices:} Each $n$-point vertex contributes with a factor of
        \begin{align}
        i\,V^{(\underline{m}_2,\dots,\underline{m}_{n})}_{a_1\dots a_n}(x)\,,
        \end{align} 
        having parametrized the functional vertex tensor
        \begin{equation}
        \mathcal{V}_{I_1\ldots I_n}= \sum_{m_2\ldots m_n} V^{(\underline{m}_2, \ldots, \underline{m}_n)}_{a_1a'_2\ldots a_n'}(x_1) \prod_{i=2}^n P_{x_i}^{\underline{m}_i} \delta_{a'_i a_i}(x_1, x_i)\,,
        \end{equation}
        where we suppressed the loop-order index for simplicity.
        The $a_i$ labels are the internal indices associated with the edges connected at the vertex and $m_i$ are the indices associated with the momentum operators at $n-1$ edges in the vertex.\footnote{We remind the reader that, without loss of generality, momentum conservation at the vertex lets us move the momentum operators at one of the vertex edges to all others. In practice, this means that we only need to consider $n-1$ different momentum operators for each $n$-point vertex.} The spacetime coordinate $x$ is, of course, the one carried by the vertex. 
        
        \item \textbf{Cut edges:} Finally, each cut edge contributes with\footnote{
            The propagator of a cut edge picks up momentum operators from the vertices just as the uncut edges. The covariant delta function associated with the edge is $ \delta_{ab}(x,y) = \int_k \,e^{ i(y-x)\cdot k}U_{ab}(x,y)$, with  $ k $ being the loop momenta assigned to that edge. The Fourier phase is responsible for shifting all momentum operators as for the uncut case.}
        \begin{align}
        i\,\left[E^{(\underline{m}_i, \underline{n}_j)}(x,P_x + k) \,U(x,y)\right]_{a_ib_j}\,,
        \end{align}
        where, as before, $a_i$ and $x$ ($b_j$ and $y$) match the internal indices and spacetime label of the outgoing (incoming) vertex, and $k$ is the loop momentum flowing in the edge from the outgoing to the incoming vertex. Similarly, the indices $m_i$ and $n_j$ correspond to the indices of the momentum operators (if any) from the outgoing and incoming vertices, respectively. The closed brackets indicate that all derivatives close at the Wilson line. Thus, the relative ordering between these operators is also irrelevant. 
    \end{enumerate}
    
    \item \label{item:integration} Take the expression in the coincidence limit, where all interaction points are equal, and integrate over the coincident spacetime point and each of the loop momenta. Include also sums over all $ m,n $ indices from the expansion of vertices in powers of momentum operators. 
    
    \item \label{item:prefactors} Include the prefactors of $-i\hbar^\ell /n_s$ and the Grassmannian signatures discussed in the tensorial expression of the effective action to the covariant evaluation formula. Typically, the Grassmannian signatures must be recalculated even if the tensorial expression was determined beforehand, as the covariant evaluation requires a reordering of terms. To this end, one can take $U_{ab}\to\delta_{ab}$, $E^{(\underline{m}, \underline{n})}_{ab} \to Q^{\eminus 1}_{ab} $, and apply the procedure described in rule iii) of Section~\ref{sec:RulesEffectiveAction}, with the DeWitt indices ($I,J,\dots$) now replaced by the internal indices ($a,b,\dots$).
\end{enumerate}

Let us return to the sunset graph as an example to illustrate these rules. As indicated by rule~\ref{item:cuts}, the first step is to label the interaction points. After that, we have to cut two edges of the graph to make a connected tree graph. In Figure~\ref{fig:SunsetRules} (middle), we have chosen to cut the middle and lower edges in the sunset. We emphasize that, while this choice is not unique, the resulting formulas are equally valid. Each cut edge has a unique loop momentum flowing in the edge, the loop momenta in the uncut edges of the graph have to be arranged so that we have momentum conservation at the vertices. Per rule~\ref{item:root_choice}, we designate vertex $ x $ as the root, which endows directionality to both the cut and uncut edges, as shown in Figure~\ref{fig:SunsetRules} (right).
We may then add, in order, the contributions of the uncut edge, the two vertices, and the two cut edges. From the choice of $x$ as the root vertex, the uncut edge is directed from $x$ to $y$. Following rule~\ref{item:integration}, everything is then evaluated at the coincidence limit and integrated over the coincident spacetime coordinate and all loop momenta. We get
\begin{align}
    \frac{\hbar^2}{12}\sum_{m_1,m_2,n_1,n_2}\int_x&\, \int_{k\ell}E^{(0,0)}_{a a'}(y,P_y-k-\ell)V^{(\underline{m}_1,\underline{n}_1)}_{abc}(x)V^{(\underline{m}_2,\underline{n}_2)}_{a' b' c'}(y)\nonumber\\
    &\times [E^{(\underline{m}_1,\underline{m}_2)}(x,P_x+k)U(x,y)]_{bb'}[E^{(\underline{n}_1,\underline{n}_2)}(x,P_x+\ell)U(x,y)]_{cc'}\bigg|_{y=x}\,,
\end{align}
having also included the prefactor from rule~\ref{item:prefactors}. As can be seen, we have arranged the momentum operators from the vertices so that none of them act on the uncut vertex. This result is equivalent to the formula we obtained in Eq.~\eqref{eq:sunset_formula}. Finally, for the Grassmannian signature, we set $U_{ab}\to\delta_{ab}$ and $E^{(\underline{m}, \underline{n})}_{ab} \to Q^{\eminus 1}_{ab} $, to recover
    \begin{align}
    Q_{aa^\prime}^{\eminus 1}V_{abc}V_{a^\prime b^\prime c^\prime}Q_{bb^\prime}^{\eminus 1}Q_{cc^\prime}^{\eminus 1}&\to Q_{aa^\prime}^{\eminus 1}\eta_{a^\prime}\eta_a\,\eta_a\eta_b\eta_cV_{abc}\,\eta_{a^\prime}\eta_{b^\prime}\eta_{c^\prime}V_{a^\prime b^\prime c^\prime}\,Q_{bb^\prime}^{\eminus 1}\eta_{b^\prime}\eta_b\,Q_{cc^\prime}^{\eminus 1}\eta_{c^\prime}\eta_c\nonumber\\
    &\to\zeta_{aa^\prime}\zeta_{bb^\prime}\zeta_{cc^\prime}\zeta_{ab}\zeta_{ac}\zeta_{a^\prime b}\zeta_{a^\prime c}\zeta_{bc^\prime}\zeta_{a^\prime b^\prime}\zeta_{a^\prime c^\prime}\,Q_{aa^\prime}^{\eminus 1}V_{abc}V_{a^\prime b^\prime c^\prime}Q_{bb^\prime}^{\eminus 1}Q_{cc^\prime}^{\eminus 1}\,.
    \end{align}

\begin{figure}[t]
    \centering
    \includegraphics[width=0.7\textwidth]{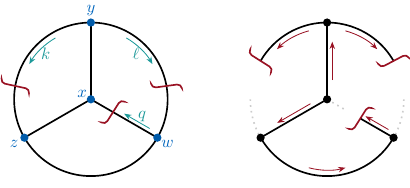}
    \caption{The three-loop tetrahedron supergraph. The left side shows vertex labels, loop momentum assignments, and cuts. The right side shows the direction of the edges in the rooted tree graph, with $ x $ being the chosen root. }
    \label{fig:Tetrahedron_evaluation}
\end{figure}

Let us also briefly illustrate the rules with a more complicated example: the three-loop tetrahedron supergraph in Figure~\ref{fig:Tetrahedron_evaluation} (left). In this case, we have to make three cuts. We have purposely chosen the cuts in a non-symmetric way to highlight that any choice that leads to a tree graph is equally valid and showcase a situation with multiple uncut edges in a chain. As before, we choose the root of the graph at the vertex with spacetime label $x$. The edge direction (flowing towards the leaves) is now a bit more involved, and we illustrate it in Figure~\ref{fig:Tetrahedron_evaluation} (right). We stress, once more, that several different roots are possible and equally valid. Once this is set, we can add the contributions from the three uncut edges, the four vertices, and the three cut edges, yielding
    \begin{align}
    \dfrac{i \hbar^3}{24} &\sum_{m_i,n_i} \int_x \, \int_{k\ell q} E_{a_1b_1}^{(0, 0)}(y, P_y+ k + \ell) \, E_{a_2 c_1}^{(\underline{m}_1, 0)}(z, P_z+q - k - \ell) \, E_{c_2 d_3}^{(\underline{m}_3, \underline{n}_4)}(w, P_w + q - \ell)  \nonumber\\
    &\times V^{(\underline{m}_1, \underline{n}_1)}_{a_1 a_2 a_3}(x)\, V^{(\underline{m}_2, \underline{n}_2)}_{b_1 b_2 b_3}(y)\, V^{(\underline{m}_3, \underline{n}_3)}_{c_1 c_2 c_3}(z) \, V^{(\underline{m}_4, \underline{n}_4)}_{d_1 d_2 d_3}(w) \big[E^{(\underline{m}_2, \underline{n}_3)}(y, P_y + k) U(y,z) \big]_{b_2 c_3} \nonumber \\
    & \times \big[E^{(\underline{n}_2, \underline{m}_4)}(y, P_y + \ell) U(y,w) \big]_{b_3 d_2} \big[E^{(0, \underline{n}_1)}(w, P_w + q) U(w,x) \big]_{d_1 a_3} \bigg|_{y,z,w = x}\,,
    \end{align}
where $0$ in the superscript of $E$ indicates an absence of momentum operators from the vertices.

\section{Renormalization Group and Matching Calculations} 
\label{sec:matching}

The practicality of the functional methods explored here relies on performing an expansion of the dressed propagators. If one were to expand it as a series in the background fields, each term would correspond to a 1PI amplitude. One would then recover the loop integrals associated with the equivalent 1PI Feynman diagrams from diagrammatic approaches and little practical benefit would have been gained. The first functional calculations~\cite{Coleman:1973jx,Jackiw:1974cv,Iliopoulos:1974ur}, on the other hand, relied on an expansion in the derivatives of the field. The leading term of this expansion is the effective potential and the higher-order terms even allow for the derivation of some renormalization constants.
Here, instead, we promote functional methods to evaluate the UV part of the loop integrals, which is what is relevant for RG and matching calculations. Expanding the propagators in this region, the dependence of the integrals on external momenta becomes analytic and manifest gauge invariance can be maintained.   

Functional methods are particularly useful when there exists an operation $ \boldsymbol{R}_\UV $ that expands the dressed propagators in the integrand around the limit where the loop momenta are much larger than both the background fields and the covariant differential operators (which act on those fields and are formally of the size of external momenta in Feynman diagrams), such that the quantity of interest is given by $ \boldsymbol{R}_\UV \widehat{\Gamma} $. In EFT matching, it has been demonstrated that the EFT action can be identified with the hard loop-momentum region of the UV quantum effective action~\cite{Manohar:1997qy,Fuentes-Martin:2016uol,Zhang:2016pja,Fuentes-Martin:2023ljp}. For such calculations, we may identify the operator $ \boldsymbol{R}_\UV = \boldsymbol{R}_\mathrm{hard} $ and apply the functional methods described in the previous sections. Along similar lines, Ref.~\cite{Born:2024mgz} established that the functional formalism can be adapted for counterterm calculations (for $\beta$-functions and anomalous dimensions) by applying the local $ \boldsymbol{R}_\UV= \boldsymbol{R}^\ast $ operation~\cite{Herzog:2017bjx}. We refer to~\cite{Born:2024mgz} for details on this procedure, while we focus on the application to EFT matching in the following. In both RG and matching calculations, one would first cast the functional supergraphs as covariant loop integrals with the techniques described in Section~\ref{sec:gauge_inv_evaluation} before proceeding to apply the relevant $ \boldsymbol{R}_\UV $ operation and evaluating the integrals.

\subsection{EFT matching}

Heavy fields will decouple from a theory at energy scales sufficiently smaller than their masses, i.e., when there is not enough energy available in a process to produce them on-shell. In such cases, it is appropriate to transition to a low-energy EFT without said heavy fields. Any action that describes the same physics as the original UV theory in the low-energy limit is a valid choice; however, it turns out to be useful to impose a stronger condition for the matching calculation---the process of determining an appropriate EFT action---in practice. 
The stronger requirement used in off-shell matching computations is to impose that the EFT should reproduce all low-energy Green's functions of the full theory, even off shell. 

We take the UV fields to be $ \eta_I= (\Phi_\alpha, \phi_i) $, where $ \Phi_\alpha $ are the heavy fields that decouple and $ \phi_i $ the light fields that remain in the EFT. The corresponding sources are $ \mathcal{J}= (\mathcal{J}_\Phi, \, \mathcal{J}_\phi) $.
Given a UV theory $ S_\UV[\Phi, \phi] $, off-shell matching seeks to determine an EFT action $ S_\EFT[\phi] $ such that the relation of the form
    \begin{equation} \label{eq:off-shell_vac_functional_matching}
    W_\EFT[\mathcal{J}_\phi] = W_\UV\big[\mathcal{J}_\Phi = 0,\, \mathcal{J}_\phi \big]\,,
    \end{equation}  
holds for low energies; that is, there should be an equality between all connected Green's functions of the light fields.\footnote{This equality~\eqref{eq:off-shell_vac_functional_matching} is not strict, but understood as a power series in the inverse heavy masses, which is
truncated at the required accuracy.}
No heavy fields are sourced, as there is insufficient energy to produce on-shell heavy particles. 

A more convenient formulation of the matching procedure is obtained by Legendre transforming~\eqref{eq:off-shell_vac_functional_matching} in the light-field sources. The resulting matching condition expressed in terms of the quantum effective actions becomes\footnote{There are some subtleties related to choosing gauge-fixing conditions when using the off-shell matching condition~\cite{Thomsen:2024abg}. Eq.~\eqref{eq:matching_condition_2} is nevertheless valid for suitable choices.} 
    \begin{equation} \label{eq:matching_condition_2}
    \Gamma_\EFT[ \hat{\phi}] = \Gamma_\UV\big[ \widehat{\Phi}[\hat{\phi}],\, \hat{\phi}\big]\,,  \qquad 0 = \dfrac{\delta \Gamma_\sscript{UV}}{\delta \Phi_\alpha} \big[\widehat{\Phi}[\hat{\phi}],\, \hat{\phi}\big]\,,
    \end{equation}
where the heavy fields are solutions to the quantum EOMs in the presence of the light background fields. We can understand condition~\eqref{eq:matching_condition_2} as an equality between all one-light-particle-irreducible\footnote{1LPI means that any tree-level (reducible) propagator is a heavy field.} (1LPI) Green's functions of the UV theory (with exclusively light external states) with the 1PI functions of the EFT. 

At tree level, Eq.~\eqref{eq:matching_condition_2} reproduces the well-known procedure of ``integrating out the heavy fields'' of the UV theory in order to obtain the EFT. It is much less obvious that the matching condition also enables an easy procedure for extracting the one-loop EFT without having to do any loop calculations in the EFT. Using expansion by regions~\cite{Beneke:1997zp,Jantzen:2011nz}, Refs.~\cite{Fuentes-Martin:2016uol,Zhang:2016pja} demonstrated that there is a cancellation between the loop contributions in the EFT and the soft-region loops in the full UV theory at one-loop order. This enables a very direct determination of $ S^{(1)}_\EFT[\phi] $ in terms of the hard region of the UV quantum effective action.\footnote{Variations of this procedure go back much further to, e.g.,~\cite{Dittmaier:1995cr,Dittmaier:1995ee,Manohar:1997qy}, but the more rigorous proofs only appeared later.} 
In a recent paper~\cite{Fuentes-Martin:2023ljp}, we posited that the hard-region matching generalizes  
and presented a master formula for perturbative EFT matching at multi-loop order:
    \begin{equation} \label{eq:gen_matching_formula}
    S_\EFT =  \big(\boldsymbol{R}_\mathrm{hard} \Gamma_\UV \big) \big[ \widehat{\Phi}[\hat{\phi}],\, \hat{\phi}\big] \,,  \qquad 0 = \dfrac{\delta\, \boldsymbol{R}_\mathrm{hard} \Gamma_\sscript{UV} }{\delta \Phi_\alpha} \big[ \widehat{\Phi}[\hat{\phi}] \,,\, \hat{\phi} \big] \,.
    \end{equation}
The $ \boldsymbol{R}_\mathrm{hard} $ operation, when applied to a 1PI graph, expands all propagators in the corresponding loop integral in the limit where the loop momentum is of the order of the heavy masses and all fields and derivatives are small (see App.~\ref{app:regions} for more details). 
Thus, the EFT action is identified with the `hard' part of the UV effective action, which is taken to include all contributions without \emph{any} soft loop momenta and, therefore, includes the tree-level action. The heavy fields are solutions to the EOMs obtained from the `hard' UV effective action, so, in effect, the EFT action is identified with the sum of hard-region UV 1LPI diagrams. This formula was only proven at the two-loop order, so in the next section, we wish to present an all-order proof of its validity. Notably, the general matching formula~\eqref{eq:gen_matching_formula} removes an important theoretical obstacle to performing multi-loop matching with the functional formalism: it means that the UV effective action can be evaluated with the expansion operator $ \boldsymbol{R}_\mathrm{hard} $ in matching calculations, enabling the techniques discussed in Section~\ref{sec:gauge_inv_evaluation}. Obviously, there is no magic here, and practical calculations will still be limited by the availability of master integrals, efficient codes for the evaluations, as well as computational resources.

\subsection{Proof of the matching formula}

The proof sketched out in this section is loosely inspired by the proof for perturbative renormalizability of QFTs given in Chapter~5 of Ref.~\cite{Collins:1984xc}, which demonstrates that the Lagrangian counterterms are sufficient to renormalize all subdivergences of all Feynman diagrams.\footnote{It should not be surprising that techniques from renormalization theory can be adapted for EFT matching given the similarities between the two subjects: a change in the UV behavior of the theory---whether introduction of a UV regulator or removal of heavy fields---is compensated by the introduction of local counterterms/EFT operators.} The main idea behind the proof is quite straightforward: we seek to demonstrate that all non-vanishing graphs with non-hard loops (with at least some soft part) appear one-to-one on both the EFT and UV side of Eq.~\eqref{eq:matching_condition_2}. This leaves just the hard part of the UV effective action to be identified with the EFT action. The main obstacle lies in arguing that the combinatorial factors of both sides agree, thereby facilitating the cancellation.

\subsubsection{Building blocks of the EFT effective action} \label{sec:EFT_building_blocks}
Let us be more deliberate in our formulation of matching condition~\eqref{eq:matching_condition_2}: it is valid only for low-momentum background fields and is understood as a series expansion in heavy masses. More precisely, we may write it as 
    \begin{equation} \label{eq:match_con_2_expanded}
    \Gamma_\EFT[\hat{\phi}] = \boldsymbol{T}_\sscript{EFT} \Gamma_\UV[\hat{\eta}]\,, \qquad 
    0 = \dfrac{\delta \Gamma_\UV}{\delta \Phi}[\hat{\eta}]\,,\qquad 
    \hat{\eta}_I \equiv \big( \boldsymbol{T}_\sscript{EFT} \widehat{\Phi}_\alpha [\hat{\phi}],\, \hat{\phi}_i \big), 
    \end{equation}
where the operator $ \boldsymbol{T}_\sscript{EFT} $ series expands an expression in the limit where all fields, derivatives (external momenta), and light masses are small compared to the heavy masses. Thus, $ \boldsymbol{T}_\sscript{EFT} \widehat{\Phi}[\hat{\phi}] $ is the series expansion of the solution to quantum EOM in the presence of light background fields.  

$ \widehat{\Phi}[\hat{\phi}] $ does not have a definite loop order, since it is a solution to a mixed-order (quantum) EOM. The loop-level contribution to the heavy-field quantum EOM is explicitly accounted for by considering all 1LPI graphs in the UV theory. Matching condition~\eqref{eq:match_con_2_expanded} is, thus, equivalent to
    \begin{equation} \label{eq:match_con_1LPI}
    \Gamma_\EFT[\hat{\phi}]= \boldsymbol{T}_\sscript{EFT} \Gamma_{\sscript{1LPI}}[\tilde{\eta}]\,, \qquad 
    0 = \dfrac{\delta S_\UV^{(0)}}{\delta \Phi_\alpha }[\tilde{\eta}]\, , \qquad \tilde{\eta}_I \equiv \big( \boldsymbol{T}_\sscript{EFT} \overline{\Phi}_\alpha [\hat{\phi}],\, \hat{\phi}_i \big),
    \end{equation}
where $ \overline{\Phi}_\alpha [\hat{\phi}] $ is the solution to the tree-level (classical) EOM of the heavy fields in the presence of the light fields. Meanwhile, $ \Gamma_{\sscript{1LPI}}[\tilde{\eta}] $ denotes the set of all dressed 1LPI diagrams of the UV theory without any tree-level tadpoles of the heavy fields, or, in other words, the set of tree graphs of heavy propagators where all leaves are loop-level 1PI diagrams and other vertices are either tree-level or 1PI diagrams. With the tree-level EOM solution for the heavy fields, the r.h.s. of the first equation in~\eqref{eq:match_con_1LPI} is again the sum over all 1LPI graphs of the UV theory. 

As the heavy fields are no longer independent degrees of freedom in this setup, it will be useful to consider how they depend on the light background fields. Applying a $ \hat{\phi} $ derivative to the heavy-field EOM in Eq.~\eqref{eq:match_con_1LPI} yields
    \begin{equation} \label{eq:heavy_field_light_dev}
    \dfrac{\delta \boldsymbol{T}_\sscript{EFT}  \overline{\Phi}_\alpha[\hat{\phi}]}{\delta \hat{\phi}_i } = 
    \boldsymbol{T}_\sscript{EFT} \dfrac{\delta \overline{\Phi}_\alpha[\hat{\phi}]}{\delta \hat{\phi}_i } =
    - \widetilde{\mathcal{Q}}_{i\beta} \boldsymbol{T}_\sscript{EFT}  \widetilde{\mathcal{Q}}^{\eminus 1}_{\beta \alpha }\,, \qquad 
    \mathcal{Q}_{IJ} \equiv \dfrac{\delta^2 S^{(0)}_\UV}{\delta \eta_I \delta \eta_J}\,.
    \end{equation}
We have used that $ \boldsymbol{T}_\sscript{EFT} $ acts trivially on $ \widetilde{\mathcal{Q}}_{IJ} $ (but not $ \widetilde{\mathcal{Q}}_{IJ}^{\eminus 1} $), as locality of the underlying UV action ensures that the two-point function is polynomial in the derivative and, therefore, already series expanded. 
The well-known tree-level matching formula $ S^{(0)}_\EFT[\hat{\phi}] = S^{(0)}_\UV[\tilde{\eta}] $---the heavy fields being integrated out---follows directly from the general matching formula. 
One may then show~\cite{Zhang:2016pja} that the kinetic operator in the EFT is given by
    \begin{equation} \label{eq:EFT_kinetic_op}
    \mathcal{P}_{ij}[\hat{\phi}] 
    \equiv \dfrac{\delta^2 S_\EFT^{(0)}[\hat{\phi}] }{\delta \hat{\phi}_i \, \delta \hat{\phi}_j } 
    = \widetilde{\mathcal{Q}}_{ij} - \widetilde{\mathcal{Q}}_{i\alpha} \boldsymbol{T}_\sscript{EFT} \widetilde{\mathcal{Q}}^{\eminus 1}_{\alpha \beta}\, \widetilde{\mathcal{Q}}_{\beta j} \,,
    \end{equation}
where, again, the heavy propagator is expanded around its large mass, consistent with $ \widehat{\mathcal{P}}_{ij} $ being a local operator (from the local $ S_\EFT $). It holds that
    \begin{equation}
    \dfrac{\delta \tilde{\eta}_I}{\delta \hat{\phi}_i} = \widetilde{\mathcal{R}}_{Ii}, \qquad \widetilde{\mathcal{R}}_{IJ} = \begin{pmatrix}
    \delta_{\alpha \beta} & -\boldsymbol{T}_\sscript{EFT} \widetilde{\mathcal{Q}}^{\eminus 1}_{\alpha \beta} \widetilde{\mathcal{Q}}_{\beta j} \\ 0 & \delta_{ij}
    \end{pmatrix}\,,
    \end{equation}
as a consequence of~\eqref{eq:heavy_field_light_dev}. The upper triangular matrix $ \widetilde{\mathcal{R}}_{IJ} $ is a matrix that block diagonalizes the dressed kinetic operator $ \mathcal{Q}_{IJ} $. In contrast to the kinetic operator in a free theory, it is not generally diagonal, and one should expect heavy and light propagators to mix in UV supergraphs. The block diagonalized kinetic operator is 
    \begin{equation} \label{eq:def_P-matrix}
    \widehat{\mathcal{P}}_{IJ} \equiv \widetilde{\mathcal{R}}\transpose_{IK} \widetilde{\mathcal{Q}}_{KL} \widetilde{\mathcal{R}}_{LJ} 
    = \begin{pmatrix}
    \widetilde{\mathcal{Q}}_{\alpha \beta} & 0\\
    0 & \widehat{\mathcal{P}}_{ij}
    \end{pmatrix}\,,
    \end{equation}
where the heavy block is the heavy-field kinetic operator, while the light block is the kinetic operator of the EFT, which is given by \eqref{eq:EFT_kinetic_op}.

Let us proceed to examine generic $ \ell $-loop EFT vertices. If the hard-region matching formula~\eqref{eq:gen_matching_formula} is valid up to $ \ell $-loop order, we may equivalently write it as
    \begin{equation} \label{eq:gen_match_formula_1LPI}
     S_{\EFT}^{(\ell)}[\hat{\phi}] = \boldsymbol{T}_\sscript{EFT} \boldsymbol{R}_\mathrm{hard} \Gamma_{\sscript{1LPI}}^{(\ell)}[\tilde{\eta}]\,,
    \end{equation}
in terms of the 1LPI Greens functions. The action of $ \boldsymbol{R}_\mathrm{hard} $ on a 1LPI graph is to expand all loop propagators in the hard region (treating loop momenta the same order as the heavy masses). The subsequent application of $ \boldsymbol{T}_\sscript{EFT}$ expands all the heavy tree-level propagators around the heavy masses (only comparatively small external momenta flow through tree-level propagators). 
The functional $ \ell $-loop vertices in the EFT is then given by
    \begin{equation} \label{eq:general_EFT_vertex_formula}
    \widehat{\mathcal{W}}^{(\ell)}_{i_1\ldots i_n} \equiv \dfrac{\delta^n S_{\EFT}^{(\ell)}}{\delta \phi_{i_1} \cdots \delta \phi_{i_n}} [\hat{\phi}] =  \dfrac{\delta^n \boldsymbol{T}_\sscript{EFT} \boldsymbol{R}_\mathrm{hard} \Gamma_{\sscript{1LPI}}^{(\ell)} [\tilde{\eta}] }{\delta \hat{\phi}_{i_1} \cdots \delta \hat{\phi}_{i_n}} \,.
    \end{equation}
We now inspect the first few EFT vertices to get a better intuition for the implications of~\eqref{eq:general_EFT_vertex_formula}. The one-point vertex is simply
    \begin{equation}  \label{eq:1_EFT_vertex}
    i\, \widehat{\mathcal{W}}^{(\ell)}_{i} = i(\boldsymbol{T}_\sscript{EFT} \boldsymbol{R}_\mathrm{hard} \widetilde{\Gamma}_{\sscript{1LPI},I}^{(\ell)}) \widetilde{\mathcal{R}}_{Ii} = \;
    \ineqgraphics{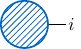}, \qquad  \Gamma_{\sscript{1LPI},I_1\ldots I_n}^{(\ell)} \equiv \dfrac{\delta^n \Gamma[\eta]_{\sscript{1LPI}}}{\delta \eta_{I_1} \cdots \delta \eta_{I_n}}\, .
    \end{equation}
Graphically the hard-region $\ell$-loop 1LPI UV amplitude is represented as a hatched blue blob, while the normal external line, represents the open index associated with a light field. In the graphical interpretation a light leg can be attached to a vertex only through the functional matrix $ \widetilde{\mathcal{R}}_{Ii} $, which is left implied. 

The two-point EFT vertex follows from applying a derivative to~\eqref{eq:1_EFT_vertex}, to which end we observe that  
    \begin{align}
    \dfrac{\delta \widetilde{\mathcal{R}}_{\alpha i}}{\delta \hat{\phi}_j} &= - \boldsymbol{T}_\sscript{EFT} \big( \widetilde{\mathcal{P}}^{\eminus 1}_{\alpha \beta} \widetilde{\mathcal{V}}^{(0)}_{\beta IJ} \big) \widetilde{\mathcal{R}}_{I i} \widetilde{\mathcal{R}}_{Jj}\,,  
    & \dfrac{\delta \widetilde{\mathcal{R}}_{ij}}{\delta \hat{\phi}_k} &= 0\,.
    \end{align}
It follows that the two-point EFT vertex is
    \begin{equation} \label{eq:2_EFT_vertex}
    \begin{split}
    i\,\widehat{\mathcal{W}}^{(\ell)}_{ij} &= i (\boldsymbol{T}_\sscript{EFT} \boldsymbol{R}_\mathrm{hard} \widetilde{\Gamma}_{\sscript{1LPI},IJ}^{(\ell)}) \widetilde{\mathcal{R}}_{Ii} \widetilde{\mathcal{R}}_{Jj} - i  (\boldsymbol{T}_\sscript{EFT} \boldsymbol{R}_\mathrm{hard} \widetilde{\Gamma}_{\sscript{1LPI},\alpha}^{(\ell)}) \boldsymbol{T}_\sscript{EFT} \big( \widetilde{\mathcal{P}}^{\eminus 1}_{\alpha \beta} \widetilde{\mathcal{V}}_{\beta IJ}^{(0)} \big) \widetilde{\mathcal{R}}_{Ii} \widetilde{\mathcal{R}}_{Jj} \\
    &= \ineqgraphics{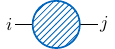} +\; \ineqgraphics{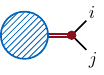},
    \end{split}
    \end{equation}
where a red propagator indicates that it is expanded around the heavy masses as dictated by $ \boldsymbol{T}_\sscript{EFT} $ while a double line is a heavy field propagator (the inverse of the heavy-heavy block of $ \mathcal{Q}_{IJ} $). The small blob signifies a tree-level vertex of the UV theory. 

To fully appreciate how the EFT vertices generalizes, we take one last light-field derivative of \eqref{eq:2_EFT_vertex}.
Observing that  
    \begin{align}
    \dfrac{\delta \boldsymbol{T}_\sscript{EFT} \widetilde{\mathcal{P}}^{\eminus 1}_{\alpha \beta}}{\delta \hat{\phi}_i} &= - \boldsymbol{T}_\sscript{EFT} \big( \widetilde{\mathcal{P}}^{\eminus 1}_{\alpha \gamma} \widetilde{\mathcal{V}}^{(0)}_{\gamma \delta I} \widetilde{\mathcal{P}}^{\eminus 1}_{\delta \beta} \big) \widetilde{\mathcal{R}}_{I i}\,, &
    \dfrac{\delta \widetilde{\mathcal{V}}^{(0)}_{\alpha_1\ldots \alpha_m I_1\ldots I_n}}{\delta \hat{\phi}_j} &= \widetilde{\mathcal{V}}^{(0)}_{\alpha_1\ldots \alpha_m I_1\ldots I_n J} \widetilde{\mathcal{R}}_{J j} \,,
    \end{align}
we find that 	
    \begin{equation} 
    i\, \widehat{\mathcal{W}}^{(\ell)}_{ijk} = 
    \ineqgraphics{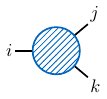} + \sum_{\mathrm{perm.}}\! \ineqgraphics{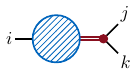} + \sum_{\mathrm{perm.}} \ineqgraphics{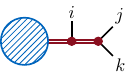} +\; \ineqgraphics{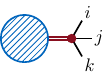}
    \end{equation}
having transitioned to pure graphical notation for convenience. The sums are over the unique permutations of the the external indices. 

A pattern emerges for the EFT vertices, which generalizes to higher-leg vertices: the EFT vertex is the sum of all tree-level graphs produced from a single insertion of a hard-region 1LPI graph from the UV theory, soft-region heavy propagators, and tree-level UV vertices. All the external light indices come with an associated factor $ \widetilde{\mathcal{R}}_{Ii} $. Furthermore, there are no tree-level tadpoles (including from a tree-level 1LPI UV graph) by EOM, nor are there any $ \widetilde{\mathcal{V}}^{(0)}_{\alpha I} \widetilde{\mathcal{R}}_{Ii} =0 $ vertices. In other words, all tree-level vertices have at least three indices. Finally, we emphasize that the symmetry factors associated with $ \widetilde{\Gamma}_{\sscript{1LPI},I_1\ldots I_n}^{(\ell)} $ subgraphs are exactly what one would expect from the corresponding $ n $-point graph. This is a consequence of the vacuum graph $ \widetilde{\Gamma}_{\sscript{1LPI}}^{(\ell)} $ having the appropriate factor and the numerical multiplicity factors associated with the $ \hat{\phi} $ derivative acting at (in)equivalent points in the underlying graph.

\subsubsection{Classifying the UV terms}

To prove the matching master formula~\eqref{eq:gen_matching_formula} (equivalently \eqref{eq:gen_match_formula_1LPI}), our first concern is to determine all the terms that show up in the matching condition. 
The r.h.s. of matching condition~\eqref{eq:match_con_1LPI} consists of 1LPI supergraphs where the external fields carry small momenta. We can write 
    \begin{equation} \label{eq:UV_graph_decomposition}
    i\, \boldsymbol{T}_\sscript{EFT} \Gamma_{\sscript{1LPI}}[\tilde{\eta}] = \sum_{G \in \mathcal{G}_\UV} \!\! G\,,
    \end{equation}
where\footnote{The parenthesized numbers next to the vertices denote the loop order of that vertex.}
    \begin{equation}
    \mathcal{G}_\UV = \Bigg\lbrace
        \ineqgraphics{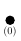},\; 
        \ineqgraphics{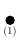},\;
        \ineqgraphics{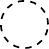},\;
        \ineqgraphics{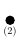},\;
        \ineqgraphics{GUV_CT.pdf},\;
        \ineqgraphics{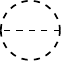},\;
        \ineqgraphics{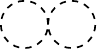},\;
        \ineqgraphics{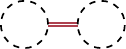},\;
        \ineqgraphics{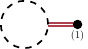},\;
        \ldots
    \Bigg\rbrace
    \end{equation}
is the set of all 1LPI dressed vacuum supergraphs without tree-level tadpoles of the heavy fields and with all tree-level propagators expanded around their heavy masses. All dependence on external fields is through the combination $ \tilde{\eta}$. In the graphical notation black lines have no region associated with their (loop) momentum and dashed lines indicates summation over all UV degrees of freedom, both heavy and light.  
That the internal reducible (tree-level) heavy propagators carry soft momentum can be viewed as a consequence of momentum conservation. 

Our main idea is to decompose each UV graph first by integration region and then splitting its propagators into heavy- or EFT-types. Consider the sunset graph as an illustrative example of what this looks like: 
    \begin{equation}
    G_\mathrm{ss.} = \dfrac{i}{12} \widetilde{\mathcal{V}}^{(0)}_{IJK} \widetilde{\mathcal{Q}}^{\eminus 1}_{IL} \widetilde{\mathcal{Q}}^{\eminus 1}_{JM} \widetilde{\mathcal{Q}}^{\eminus 1}_{KN} \widetilde{\mathcal{V}}^{(0)}_{LMN} 
    = \ineqgraphics{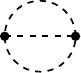}\,.
    \end{equation}
Expansion by regions (applied to the momentum space representation of the graph) implies that 
    \begin{equation} \label{eq:sunset_mixed_prop_split}
    \ineqgraphics{sunset.pdf} \; =  \; 
    \ineqgraphics{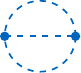} \; + \;
    \ineqgraphics{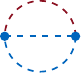} \; + \; 
    \ineqgraphics{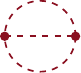}\,,
    \end{equation}
where as per usual blue propagators are expanded in the hard region while red propagators are expanded with small loop momenta compared to the heavy masses through the action of $ \boldsymbol{T}_\sscript{EFT} $.\footnote{Observe also that the symmetry factor for the mixed-region graph is $ 1/ 4$ as opposed to $ 1/12 $ for the soft- and hard-region graphs.} In the usual language of expansion by regions, the red propagators are expanded in the soft region, and conservation of hard momentum ensures that there is no graph with a single hard (blue) propagator. Next, we will want to decompose any soft propagator into an EFT and a heavy type. This is done by inserting identities in the form $ \widetilde{\mathcal{R}}_{IJ} \widetilde{\mathcal{R}}^{\eminus 1}_{JK} $. For instance for the mixed region sunset graph, we have 
    \begin{multline}
    G_\mathrm{ss.} \supset \dfrac{i}{4} \boldsymbol{R}_{\mathrm{hard}} \big( \widetilde{\mathcal{V}}^{(0)}_{IJK} \widetilde{\mathcal{Q}}^{\eminus 1}_{JM} \widetilde{\mathcal{Q}}^{\eminus 1}_{KN} \widetilde{\mathcal{V}}^{(0)}_{LMN} \big) \boldsymbol{T}_\sscript{EFT} \big( \widetilde{\mathcal{Q}}^{\eminus 1}_{IL} \big) \\
    = \dfrac{i}{4} \boldsymbol{R}_{\mathrm{hard}} \big( \widetilde{\mathcal{V}}^{(0)}_{IJK} \widetilde{\mathcal{Q}}^{\eminus 1}_{JM} \widetilde{\mathcal{Q}}^{\eminus 1}_{KN} \widetilde{\mathcal{V}}^{(0)}_{LMN} \big) \boldsymbol{T}_\sscript{EFT} \big(  \widetilde{\mathcal{R}}_{Ii} \widetilde{\mathcal{P}}^{\eminus 1}_{i\ell } \widetilde{\mathcal{R}}_{L\ell} + \widetilde{\mathcal{R}}_{I\alpha} \widetilde{\mathcal{P}}^{\eminus 1}_{\alpha \beta } \widetilde{\mathcal{R}}_{L\beta} \big),
    \end{multline}
which in graphical notation reads  
    \begin{equation}
    \ineqgraphics{sunset_mixed_regions.pdf}\; = \; 
    \ineqgraphics{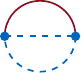} \; + \; 
    \ineqgraphics{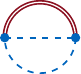}\,.
    \end{equation}
Generally, hard loop integrals are polynomial in any soft loop momenta passing through them. The same holds for soft-expanded heavy propagators. As a result, the second term with the soft hard-type propagator will result in a scaleless integral and, thus, vanish under loop-integration in dimensional regularization.\footnote{See also App.~\ref{app:regions} for additional details.} Only the contribution with the EFT-type propagator remains. 

We return now to the case of a generic graph $ G\in \mathcal{G}_\UV $ in the UV effective action~\eqref{eq:UV_graph_decomposition}. Expansion by regions lets us write such a graph as
    \begin{equation} \label{eq:sup_graphs_by_region}
    G= \sum_{\substack{\gamma \subseteq G\\ \mathrm{loops} } } \boldsymbol{T}_\sscript{EFT}^{(G\setminus\gamma)} \circ \boldsymbol{R}_\mathrm{hard}^{(\gamma) }(G)\, .
    \end{equation}
The sum is over the set of all sub-loops $ \gamma $ of $ G $ and $ \boldsymbol{R}_\mathrm{hard}^{(\gamma)} $ expands all corresponding propagators with the loop momenta of the order of the heavy masses while $ \boldsymbol{T}_\sscript{EFT}^{(G\setminus \gamma)}$ takes the combination of loop momenta in the remaining propagators to be soft and then expands them (and also acts on the tree-level propagators which were already expanded). The subgraphs $ \gamma $ are generically a union of disjoint 1PI subgraphs $ \gamma = \cup_i \gamma_i $ and the hard region operator satisfies $ \boldsymbol{R}_\mathrm{hard}^{(\gamma)} = \circ_i \boldsymbol{R}_\mathrm{hard}^{(\gamma_i)} $. We have constructed a detailed proof for the expansion by region formula underlying~\eqref{eq:sup_graphs_by_region} for momentum space Feynman integrals to arbitrary loop order. For a proof of this expansion we
refer the reader to Appendix~\ref{app:regions}.

\begin{figure}
    \centering
    \includegraphics[]{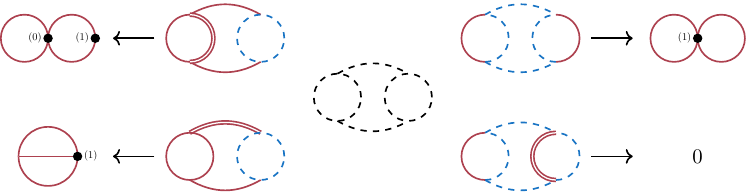}
    \caption{Non-exhaustive examples of some soft and heavy regions in a 3-loop supergraph $ G\in \mathcal{G}_\UV $ and combinations of heavy and EFT-like propagators in said graphs. For each combination of regions and propagator choices, the EFT counterpart graph is shown, obtained by shrinking hard loops and heavy propagators to points.
    Double lines denote heavy propagators, single lines the EFT-like propagators, and dashed lines full UV propagators. Black lines are taken to be the full region of the loop, while red lines carry soft and blue lines hard momentum.  
    \label{fig:sup_graph_decomposition}}
\end{figure}

Our next consideration is that of decomposing the soft-region propagators of the UV graphs into a heavy and an EFT type, the latter equaling the tree-level propagators of the EFT. While the off-diagonal parts of the UV kinetic operator, mix light and heavy propagators, inserting $ \widetilde{\mathcal{R}} \widetilde{\mathcal{R}}^{\eminus 1} $ between every vertex and soft propagator in the diagrams block-diagonalizes the propagators as per~\eqref{eq:def_P-matrix}.  We may then separate the heavy and light blocks of the resulting propagator into different terms, one with an EFT-type and one with a heavy-type propagator. Repeating this process for all soft propagators, we can think of summing explicitly over the various choices for heavy and EFT-like propagators in the expression for the UV supergraph~\eqref{eq:sup_graphs_by_region}. This sum goes over all possible sets of propagators $ \pi \subseteq G \setminus \gamma $ not in the hard loops (for convenience we sum over all kinds of propagators in the hard loops): 
    \begin{equation} \label{eq:sup_graph_decomposition_init}
    G= \sum_{\substack{\gamma \subseteq G\\ \mathrm{loops} } } \sum_{\substack{\pi \subseteq G \setminus \gamma \\ \mathrm{red.} \subseteq \pi} }     \boldsymbol{T}_\sscript{EFT}^{(G\setminus \gamma)} \circ \boldsymbol{P}_\EFT^{(G\setminus \gamma \setminus \pi)} \circ \boldsymbol{P}_\mathrm{heavy}^{(\pi)} \circ \boldsymbol{R}_\mathrm{hard}^{(\gamma) }(G)\,, 	
    \end{equation} 
where $ \boldsymbol{P}_\mathrm{heavy}^{(\pi)} $ picks out the heavy part of the propagators in $ \pi $, while $ \boldsymbol{P}_\EFT^{(G\setminus \gamma \setminus \pi)} $ picks the EFT-like part of the remaining propagators. \emph{All} reducible propagators in the UV supergraphs are always taken to be heavy; in fact, they were heavy already in the original $ G \in \mathcal{G}_\UV $. Some additional examples of this decomposition are shown in Figure~\ref{fig:sup_graph_decomposition}.

We now observe that some of the terms in the decomposition~\eqref{eq:sup_graph_decomposition_init} vanish trivially; namely, the terms where one or more loop becomes scaleless. This happens whenever a soft loop exclusively contains heavy propagators, which can be identified after shrinking the hard loops to points and considering the remaining soft loops one by one. We saw an example of this for the second term in~\eqref{eq:sunset_mixed_prop_split}. Another example of a vanishing term is given in the lower right of Figure~\ref{fig:sup_graph_decomposition}. This observation allows us to rewrite the decomposition of $ G $ as
    \begin{equation} \label{eq:sup_graph_decomposition}
    G= \boldsymbol{L}^{(G)}(G) + \sum_{\substack{\gamma \subsetneq G \\ \mathrm{red.} \subseteq \gamma} } \boldsymbol{T}_\EFT^{(G\setminus \gamma)}  \circ 
    \boldsymbol{P}^{(G\setminus\gamma)}_\EFT \circ \boldsymbol{L}^{(\gamma)}(G)	\,.
    \end{equation} 
The sum over $ \gamma $ is now over \emph{all} subgraphs of $ G $ subject to the constraint that they contain the reducible lines of $ G $. For future convenience, we have explicitly extracted the term $ \gamma = G $ from the sum. The operator $ \boldsymbol{L}^{(\gamma)} $  sets all loops in $ \gamma $ to the hard region, takes the heavy part of the remaining tree-level propagators in $ \gamma $, and sets these tree-level propagators to be soft (thus, ensuring that the subgraph is local).\footnote{In terms of the previously defined operators, $ \boldsymbol{L}^{(\gamma)} $ is defined as follows: Let $ \gamma_\mathrm{loop}\subseteq \gamma $ be the maximal union of disconnected loops (1PI subgraphs) in the sense that any other such $ \gamma_\mathrm{loop}' $ with $ \gamma_\mathrm{loop} \subseteq \gamma_\mathrm{loop}' \subseteq \gamma  \implies \gamma_\mathrm{loop}' = \gamma_\mathrm{loop} $. Then $ \boldsymbol{L}^{(\gamma)} = \boldsymbol{T}_\sscript{EFT}^{(\gamma \setminus \gamma_\mathrm{loop})} \circ \boldsymbol{P}^{(\gamma \setminus \gamma_\mathrm{loop})}_\mathrm{heavy} \circ \boldsymbol{R}^{(\gamma_\mathrm{loop})}_\mathrm{hard} $.}
Observe that if $ \gamma = \cup_i \gamma_i $, where $ \gamma_i $ are disjoint connected elements, then $ \boldsymbol{L}^{(\gamma)} = \circ_i \boldsymbol{L}^{(\gamma_i)} $. In obtaining~\eqref{eq:sup_graph_decomposition} from~\eqref{eq:sup_graph_decomposition_init}, we have merged the sum over loops ($ \gamma $) and remaining edges ($ \pi $) into a single sum over subgraphs (new $ \gamma $). This fails to account for the terms where the edges $ \pi $ form a loop (possibly with the hard loops), but these are exactly the terms that contain scaleless integrals and therefore vanish identically.

All told, for the UV side of the matching equation, we obtain 
    \begin{equation} \label{eq:matching_eq_UV_side}
    i \, \boldsymbol{T}_\sscript{EFT} \widetilde{\Gamma}_\UV^{\sscript{1LPI}} = \sum_{G \in \mathcal{G}_\UV} \boldsymbol{L}^{(G)}(G) + \sum_{G \in \mathcal{G}_\UV} \sum_{\substack{\gamma \subsetneq G \\ \mathrm{red.} \subseteq \gamma} } \boldsymbol{T}_\sscript{EFT}^{(G\setminus \gamma)} \circ \boldsymbol{P}^{(G\setminus\gamma)}_\EFT \circ \boldsymbol{L}^{(\gamma)}(G)\,.
    \end{equation}
The result of acting with $ \boldsymbol{L}^{(\gamma)} $ on a subgraph is readily seen to produce the structure of the EFT vertices discussed in Section~\ref{sec:EFT_building_blocks}: the loops are in the hard-region, while all tree-level propagators are heavy and soft-region. 

\subsubsection{Classifying the EFT terms}

We now turn to the EFT side of matching condition~\eqref{eq:match_con_1LPI}. The eventual goal is to prove the matching formula~\eqref{eq:gen_matching_formula}/\eqref{eq:gen_match_formula_1LPI} by induction in the loop order. Clearly, it holds at tree-level and at one-loop order~\cite{Fuentes-Martin:2016uol,Zhang:2016pja}, which provides the induction start. For the induction step to order $ \ell+1 $, we use the inductive assumption 
    \begin{equation} \label{eq:induction_assumption}
    i \,S_\EFT^{(k)} = \sum_{G \in \mathcal{G}_\UV^{(k)}} \boldsymbol{L}^{(G)} (G), \qquad \mathrm{for} \quad k \leq \ell\,, 
    \end{equation}
where $ \mathcal{G}_\UV^{(k)} $ is the set of dressed 1LPI vacuum supergraphs at $ k $-loop order. We aim to use this inductive assumption to relate the vertices of EFT supergraphs to UV graphs at order $ \ell + 1$.

The EFT quantum effective action is given by 
    \begin{equation} \label{eq:Gamma_EFT_diag}
    i\, \Gamma_\EFT = i\,S_\EFT + \sum_{g  \in \widetilde{\mathcal{G}}_\EFT} \boldsymbol{T}_\sscript{EFT} (g)\,,
    \end{equation}
where the sum runs over the set of all \emph{genuine} loop 1PI EFT vacuum graphs:
    \begin{equation} \label{eq:G_EFT}
    \boldsymbol{T}_\sscript{EFT} \widetilde{\mathcal{G}}_\EFT = \Bigg\lbrace
    \ineqgraphics{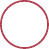},\;
    \ineqgraphics{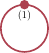},\; \ldots,\;
    \ineqgraphics{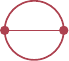},\; \ldots,\; 
    \ineqgraphics{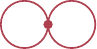}, \;
    \ldots
    \Bigg\rbrace\,.
    \end{equation}
The reader might wonder at the appearance of $ \boldsymbol{T}_\sscript{EFT} $ in~\eqref{eq:Gamma_EFT_diag}. The expansion operator has merely been made explicit  based on the observation that it acts trivially on EFT propagators,
    \begin{equation}
    \widetilde{\mathcal{P}}_{ij}^{\eminus 1} = \boldsymbol{T}_\sscript{EFT} \widetilde{\mathcal{P}}_{ij}^{\eminus 1}, 
    \end{equation}
which follows from no heavy scales appearing in the EFT propagator. Alternatively, one might explicitly invoke expansion by region on the EFT graphs: with no heavy masses in the propagators, any hard region loop inevitably becomes scaleless and vanish.  

The EFT vertices stem from the application of functional derivatives w.r.t. $ \hat{\phi} $ to the EFT action $ S_\EFT[\hat{\phi}] $. From the discussion in Section~\ref{sec:EFT_building_blocks}, it follows that each EFT vertex corresponds to a sum of connected $ n $-point supergraph from the UV theory. All loops in said graphs are in the hard region and summed over all propagator types while the tree-level, reducible propagators are heavy type (expanded in soft momenta). The symmetry factors of the resulting $ n $-point graphs agree with what we would get from ordinary Feynman rules (the derivatives insertion may lift some of the symmetry of the vacuum graph). Observe that the structure of the EFT vertices is identical to what is produced by applying the $ \boldsymbol{L}^{(\gamma)} $-operator to the UV supergraphs. This applies to all EFT vertices up to $ \ell $-loop order as a consequence of the induction hypothesis~\eqref{eq:induction_assumption}. 
We can discriminate between distinct contributions to each $ n $-point $ k $-loop EFT vertex (red blobs in~\eqref{eq:G_EFT}) from each $ G \in \mathcal{G}_\UV^{(k)} $ as per~\eqref{eq:induction_assumption}; supergraphs with all unique vertex combinations are included in $ \widetilde{\mathcal{G}}_\EFT $. 
In other words for each topology shown in~\eqref{eq:G_EFT} there is a supergraph for each combination of vertices drawn from the appropriate $ \mathcal{G}_\UV^{(k)} $.

It follows that any EFT vertices showing up in EFT vacuum graphs at loop order $ \ell+1 $ (counting also the loop-order of the vertices, which are at most order $ \ell $) corresponds to a subregion $ \gamma_i $ in a corresponding UV vacuum graph. 
As the ordinary graphs include symmetry factors, we employ the notation 
    \begin{equation}
    G = \dfrac{1}{N(G)} \overline{G}\,,
    \end{equation}
where $ N(G) $ is the number of symmetry permutations of the graphs (size of its symmetry group) and $ \overline{G} $ the graph without a symmetry factor. 
We then have 
    \begin{equation}
    \forall g \in \widetilde{\mathcal{G}}^{(\ell+1)}_\EFT \; \exists G\in \mathcal{G}_\UV^{(\ell+1)}, \, \gamma \subseteq G: \quad 
    \boldsymbol{T}_\sscript{EFT}^{(g)} (\overline{g}) = \boldsymbol{T}_\sscript{EFT}^{(G\setminus \gamma)} \circ \boldsymbol{P}^{(G\setminus\gamma)}_\EFT \circ \boldsymbol{L}^{(\gamma)} (\overline{G})\,,
    \end{equation} 
not counting the symmetry factors of the graphs (as indicated by the bars). Conversely, it is obvious that there is an equivalent EFT vertex for any subgraph of the UV theory:
    \begin{equation}
    \forall G\in \mathcal{G}_\UV^{(\ell+1)}, \, \gamma \subseteq G \; \exists g \in \widetilde{\mathcal{G}}^{(\ell+1)}_\EFT: \quad 
    \boldsymbol{T}_\sscript{EFT}^{(G\setminus \gamma)} \circ \boldsymbol{P}^{(G\setminus\gamma)}_\EFT \circ \boldsymbol{L}^{(\gamma)} (\overline{G}) = \boldsymbol{T}_\sscript{EFT}^{(g)} (\overline{g}) \,.
    \end{equation} 
In both cases, $ \gamma $ is restricted to contain all reducible propagators in $ G $.
Using this one-to-one correspondence between the EFT and UV vacuum graphs (discounting multiplicity), we may cast the $ (\ell+1) $-loop EFT quantum effective action as 
    \begin{equation} \label{eq:matching_eq_EFT_side}
    i\, \Gamma_\EFT^{(\ell+1)} = i\, S_\EFT^{(\ell+1)} + \sum_{G \in \mathcal{G}_\UV^{(\ell+1)}} \sum_{\substack{\gamma \subsetneq G \\ \mathrm{unique}}} \dfrac{1}{N(G,\gamma)} \boldsymbol{T}_\sscript{EFT}^{(G\setminus \gamma)} \circ \boldsymbol{P}^{(G\setminus\gamma)}_\EFT \circ \boldsymbol{L}^{(\gamma)}(\overline{G})\,,
    \end{equation}
for some appropriate symmetry factors $ N(G,\gamma) $.
In contrast to the second sum on the UV side, Eq.~\eqref{eq:matching_eq_UV_side}, of the matching equation, the sum over $ \gamma $ runs over only unique subgraphs; each equivalent choice of $ \gamma $ corresponds to the same EFT graph and only one appears in the summation over EFT vacuum graphs.
To complete our argument, we need to demonstrate that the second sum in~\eqref{eq:matching_eq_UV_side} and the sum in  Eq.~\eqref{eq:matching_eq_EFT_side} coincide when including symmetry factors: now for the combinatorics.

\subsubsection{Accounting for the combinatorics}

Let us determine the combinatorial factor associated to a graph $ g \in \widetilde{\mathcal{G}}^{(\ell+1)}_\EFT $ with a corresponding UV equivalent $ G $ and subgraph $ \gamma = \cup_i \gamma_i $. The size of the symmetry group of the EFT graph is $ N(g) = N(G\setminus \!\gamma) $, the symmetry group associated with $ G $ after factoring out the subgroup associated with $ \gamma $; that is, after having replaced the disjoint components of $ \gamma $ with appropriate vertices. As we have seen, the EFT action is identified with the hard region of UV graphs, and the vertices obtained from them carry the symmetry factor of the associated $ n $-point graph. In this case, the sizes of the symmetry group of the graphs constituting the EFT vertices are $ N(\gamma_i) $. Thus, the symmetry factor appearing in Eq.~\eqref{eq:matching_eq_EFT_side} is 
    \begin{equation}
    N(G, \gamma) = N(G\setminus \!\gamma) \prod_i N(\gamma_i) = N(G\setminus \!\gamma)  N(\gamma)\,. 
    \end{equation} 

Next, we elucidate the symmetry factor for the UV side of matching equation~\eqref{eq:matching_eq_UV_side}. Let us also collapse the sum over subgraphs to include only the unique terms. For the $ (\ell+1) $-loop contribution, we obtain 
    \begin{equation} \label{eq:matching_eq_UV_side_final}
    i\, \boldsymbol{T}_\sscript{EFT} \Gamma_\UV^{\sscript{1LPI}\, (\ell+1)} = \sum_{G \in \mathcal{G}^{(\ell+1)}_\UV} \boldsymbol{L}^{(G)}(G) + \sum_{G \in \mathcal{G}^{(\ell+1)}_\UV} \sum_{\substack{\gamma \subsetneq G \\ \mathrm{unique}}} \dfrac{K(G, \gamma)}{N(G)} \boldsymbol{T}_\sscript{EFT}^{(G\setminus \gamma)} \circ \boldsymbol{P}^{(G\setminus\gamma)}_\EFT \circ \boldsymbol{L}^{(\gamma)}(\overline{G})\,,
    \end{equation}
where 
    \begin{equation}
    K(G, \gamma) = \big| \{ \gamma'\subsetneq G:\; G \setminus \!\gamma' = G\setminus \!\gamma \} \big|\,,
    \end{equation}
stands for the number of subgraphs equivalent to $ \gamma $. Now the argument is that the symmetry factors on the UV and EFT sides agree, namely that 
    \begin{equation} \label{eq:combinatorial_factors_equality}
    N(G) = K(G,\gamma) N(G\setminus\!\gamma) N(\gamma)\,.
    \end{equation}
This can be shown by framing the question in terms of the symmetry groups. 

We let $ H(G) $ denote the symmetry group associated with the graph $ G $ and observe that $ N(G) = | H(G) | $ for all graphs. The two subgroups $ H(G\setminus \! \gamma) $ and $ H(\gamma) $ commute ($ H(G\setminus \! \gamma) \, H(\gamma) = H(\gamma) H(G\setminus \! \gamma) $). It follows that $ H_{G,\gamma} \equiv H(G\setminus \! \gamma) \, H(\gamma) \subseteq H(G) $ is a subgroup of the symmetry group of the full graph. The norm of $ H_{G,\gamma} $ is 
    \begin{equation}
    |H_{G,\gamma}| = \dfrac{\big| H(G\setminus \! \gamma)\big| \, \big|H(\gamma) \big|}{\big| H(G\setminus \! \gamma) \cap H(\gamma) \big|} = \big| H(G\setminus \! \gamma)\big| \, \big| H(\gamma) \big|\,,
    \end{equation}
since there is no overlap between $ \gamma $ and $ G\setminus \gamma $. Lagrange's index theorem states that 
    \begin{equation}
    \big|H(G) : H_{G,\gamma} \big| = \dfrac{\big| H(G) \big|}{\big| H_{G,\gamma} \big|}\,,	 
    \end{equation}
where the index $ \big|H(G) :  H_{G,\gamma}\big| $ denotes the number of cosets of $ H_{G,\gamma} $ in $ H(G) $. These cosets, on the other hand, are in one-to-one correspondence with the equivalent ways to choose $ \gamma $ in $ G $, meaning that $ K(G,\gamma) = \big|H(G)\, : \, H_{G,\gamma}\big| $. This proves Eq.~\eqref{eq:combinatorial_factors_equality}.

With equality of the combinatorial factors, we use the matching condition to equate the quantum effective actions~\eqref{eq:matching_eq_EFT_side} and~\eqref{eq:matching_eq_UV_side_final}. The result is 
    \begin{equation}
    i\, S_\EFT^{(\ell+1)} = \sum_{G \in \mathcal{G}^{(\ell+1)}_\UV} \! \boldsymbol{L}^{(G)}(G)\,,
    \end{equation}
which concludes the inductive step for the EFT action at order $ \ell + 1 $. This, in turn, proves matching formula~\eqref{eq:gen_matching_formula} to all orders in perturbation theory.

\section{Matching QED to the Euler--Heisenberg Theory}
\label{sec:EH}

Having presented the formalism of functional multi-loop matching and RG calculations, we turn to a concrete example of two-loop matching in a theory that simultaneously includes fermionic degrees of freedom and a gauge symmetry. We take QED as our UV theory and decouple the electron in the infrared. The resulting EFT is commonly referred to as the Euler--Heisenberg theory~\cite{Heisenberg:1936nmg}, which describes self-interacting photons through higher-dimensional effective operators.
The QED Lagrangian is given by 
\begin{equation}\label{EH:QED_Lagrangian}
    \cL_{\sscript{QED}}=\bar\psi(i\gamma^\mu D_\mu -m)\psi -\frac{1}{4}F^{\mu\nu}F_{\mu\nu}-\frac{1}{2\xi}(\partial_\mu A^\mu)^2 + \mathcal{L}_\mathrm{ct.}\,,
\end{equation}
where $m$ is the fermion mass, $D_\mu=\partial_\mu + ieA_\mu$ is the usual covariant derivative with coupling constant $e$ (and charge $ \eminus 1$), and $ \xi $ is the gauge-fixing parameter.  We write the counterterm Lagrangian as\footnote{An alternative way of writing the counterterm Lagrangian shifts the Dirac spinor according to $\psi\to(1+\delta_\psi)^{\eminus 1/2}\psi$. Thus, the wavefunction counterterm is absorbed into a shift of the mass counterterm and a counterterm for the fermion source, the latter being irrelevant in the absence of external fermion fields.}
\begin{equation}\label{EH:CountertermLagrangian}
	\cL_{\mathrm{ct.}}=-\frac{\delta_F}{4}F^{\mu\nu}F_{\mu\nu}+\delta_\psi \bar\psi i\gamma^\mu D_\mu \psi -\delta_m\bar\psi\psi\,, 
\end{equation}
where, as usual, the counterterms $\delta_i$ are given in terms of a loop expansion, that is, $\delta_i=\sum_{\ell=1}^\infty \delta_i^{(\ell)}$. The one-loop counterterms and the two-loop photon-wavefunction counterterm, relevant for two-loop matching, read
\begin{align}
\label{EH:Counterterms}
\begin{aligned}
	\delta_F^{(1)}&=-\frac{e^2}{12\pi^2}\frac{1}{\epsilon}\,,&
    \delta_F^{(2)}&=-\frac{e^4}{128\pi^4}\frac{1}{\epsilon}\,,&
	\delta_\psi^{(1)}&=-\frac{e^2}{16\pi^2}\frac{1}{\epsilon}\,,&
	\delta_m^{(1)}&=-\frac{me^2}{4\pi^2}\frac{1}{\epsilon}\,.
\end{aligned}
\end{align}
These have been computed functionally using the methods described in~\cite{Born:2024mgz} and crosschecked with \texttt{RGBeta}~\cite{Thomsen:2021ncy}.

The Euler--Heisenberg theory is valid for processes at energy scales much smaller than the fermion mass $m$. Once the fermion $\psi$ is integrated out, the EFT incorporates operators constructed solely from the field-strength tensor $F^{\mu\nu}$.
Since the starting UV theory is CP-even, EFT operators containing odd powers of $\widetilde F^{\mu\nu}= \tfrac{1}{2} \epsilon^{\mu\nu\rho\sigma}F_{\rho\sigma}$  are absent. On top of this, it holds that $F\ud{\mu}{\nu}F\ud{\nu}{\rho}F\ud{\rho}{\mu}=0$, and the leading-order gauge-field EOM takes the form $D_\mu F^{\mu\nu}=0$. As a consequence, the leading effective operators in the Euler--Heisenberg theory have dimension eight. 

Before proceeding with the matching at one- and two-loop orders, we point out that the EOM for the heavy fermion reduces to the Dirac equation, with the trivial solution $\psi=0$. This implies that the tree-level matching is simply given by 
\begin{equation}
	\cL_{\sscript{EFT}}^{(0)}=-\frac{1}{4}F^{\mu\nu}F_{\mu\nu}\,.
\end{equation}
The photon self-interactions in the Euler--Heisenberg theory are, thus, purely a loop effect.

\subsection{One-loop matching}
\label{secEH: one-loop matching}

We collect the degrees of freedom in the UV theory in a multiplet of the form
\begin{align}
    \eta=(A^\mu\;\; \psi\;\; \psi^c)\,,
\end{align}
where $\psi^c=C\bar\psi^\intercal$ is the charge-conjugated Dirac spinor. The non-vanishing components of the fluctuation operator dictated by the QED Lagrangian~\eqref{EH:QED_Lagrangian} are
\begin{equation}
\label{eq:fluc_ops}
	Q_{A_\mu A_\nu}=-g^{\mu\nu}P_x^2\,,\qquad 
	Q_{\psi\psi^c}=-Q_{\psi^c\psi}=C(\slashed P_x-m)\,.
\end{equation}
As the theory is invariant under charge conjugation, the fermionic part of the fluctuation operator is degenerate. 

With these results, we have that only the one-loop QED counterterms and log-type trace contribute to the one-loop matching. Following Eq.~\eqref{eq: log-trace}, we have
\begin{align}
\begin{aligned}
	\int_x\cL_{\sscript{EFT}}^{(1)}=\int_x\cL_{\mathrm{ct.}}^{(1)}-\frac{i}{2}\mathrm{STr} \log \cQ 
	&=\int_x\cL_{\mathrm{ct.}}^{(1)}-\frac{i}{2}\Tr\log\lzs -g^{\mu\nu}P_x^2 \, \delta_{AA}(x,y)\dzs\\
	&\quad-\frac{i}{2}\Tr\log\lzs (  \slashed P_x-m )( \slashed P_x+m)\delta_{\psi \psi^c}(x,y) \dzs\,,
\end{aligned}
\end{align}
where the first trace contributes only to the path-integral normalization and, as such, can be neglected. The second trace is computed using Eq.~\eqref{eq:logtrace_fermions} yielding
\begin{align}\label{EH:OneLoopMatching}
\begin{aligned}
    \cL^{(0)}_{\sscript{EFT}}+\cL^{(1)}_{\sscript{EFT}}&=-\frac{1}{4}\left(1+\frac{e^2}{12\pi^2}\log\frac{\mu^2}{m^2}\right)F^{\mu\nu}F_{\mu\nu} \\
    &\quad+ \frac{e^4}{16\pi^2}\frac{1}{m^4}\left[\frac{7}{90}(F^{\mu\nu}F_{\nu\sigma})(F^{\rho\sigma}F_{\mu\rho})  -\frac{1}{36}(F^{\mu\nu}F_{\mu\nu})^2\right] \,.
\end{aligned}
\end{align}
This result has been verified using \texttt{Matchete}~\cite{Fuentes-Martin:2022jrf}. Note that the Euler-Heisenberg Lagrangian does not renormalize at dimension 8. Indeed, any one-loop contribution in the EFT involves at least two dimension-8 vertices and, therefore, divergent contributions can only start appearing at dimension 12. For the same reason, the removal of $\cO(\epsilon)$ terms from the one-loop matching, which contributes at two loops, only enters at higher-order in the power counting. This is the reason why we have omitted these terms in Eq.~\eqref{EH:OneLoopMatching}.

\subsection{Two-loop matching}

Proceeding to the two-loop computation, it is evident from the UV Lagrangian that the only contributing topologies are the one-loop counterterm insertion and the sunset, as there are no four-point interactions for the figure-8 topology. In what follows, we evaluate the contributing topologies separately.

\subsubsection{One-loop counterterm insertion}

From the Eq.~\eqref{EH:CountertermLagrangian}, it follows that the non-vanishing terms of the one-loop two-point vertex are
\begin{equation}
	V^{(1)}_{A_\mu A_\nu}=-g^{\mu\nu}\delta_F^{(1)} P_x^2\,,
	\qquad
	V^{(1)}_{\psi\psi^c}=V^{(1)}_{\psi^c\psi}=C\Big(\delta_\psi^{(1)} \slashed P_x-\delta_m^{(1)}\Big)\,.
\end{equation}
The one-loop counterterm topology then becomes 
\begin{equation}\label{EH:OneLoopB}
	G_{\mathrm{ct.}}=\cQ^{\eminus1}_{IJ}\mathcal{V}^{(1)}_{JI}
	=\cQ^{\eminus1}_{A_\mu A_\nu} \mathcal{V}^{(1)}_{A_\nu A_\mu}+2\times \cQ^{\eminus1}_{\psi^c\psi}\mathcal{V}^{(1)}_{\psi\psi^c}\,,
\end{equation}
where the factor 2 accounts for the two degrees of freedom of the fermion, i.e., $\psi$ and $\psi^c$, which contribute the same. Once again, the first term in Eq.~\eqref{EH:OneLoopB} only contributes to the path-integral normalization and can be neglected. The remaining contribution yields 
    \begin{align}
    \begin{aligned}
    G_{\mathrm{ct.}}=2\sum_{s=0}^\infty\int_{x\,}&\int_{k\,}\,  \frac{(-1)^s}{(k^2-m^2)^{s+1}} \\ 
    &\times \tr \! \Big\{ (\slashed P_x+\slashed k+m) (\slashed P_x^2+2k\cdot P_x)^s \big[ \delta_\psi (\slashed P_x+\slashed k)-\delta_m \big] \Big\} U_{\psi^c\psi}(x,y)\Big|_{y=x}\,.
    \end{aligned}
    \end{align} 
The truncation of the EFT expansion at dimension 8 implies that only terms up to $s=7$ contribute.\footnote{The term with $s=8$ is proportional to a fully-symmetric product of derivatives acting on the PDP, which vanishes in the coincidence limit, see Eq.~\eqref{eq:sym_sum_U}.} The sum of all terms at dimension 4 and 8 yields
    \begin{align}
    \cL_{\sscript{EFT}}^{(2)}\supset
    &-\frac{2e^4}{(16\pi^2)^2} \lzm\frac{1}{\epsilon}+\log\frac{\mu^2}{m^2}\dzm F^{\mu\nu}F_{\mu\nu}
    -\frac{1}{(16\pi^2)^2}\frac{e^6}{3m^4}\lzm\frac{1}{\epsilon}+\log\frac{\mu^2}{m^2}+\frac{3}{2}\dzm (F^{\mu\nu}F_{\mu\nu})^2
    \nonumber \\
    &+\frac{1}{(16\pi^2)^2}\frac{14}{15}\frac{e^6}{m^4} \lzm \frac{1}{\epsilon}+\log\frac{\mu^2}{m^2}+\frac{3}{2} \dzm (F^{\mu\nu}F_{\nu\sigma})(F^{\rho\sigma}F_{\mu\rho})\,,
    \end{align}
in the on-shell EFT basis, that is, after removing terms that vanish by the application of the gauge-field EOM.

\subsubsection{Sunset topology}

The only non-vanishing three-point vertices are
\begin{equation}
	V_{A_\mu \psi\psi^c}=-V_{A_\mu \psi^c\psi}=e\,C\gamma^\mu\,.
\end{equation}
Taking into account the degeneracy, the contribution from the sunset topology is given by
\begin{equation}
    \begin{alignedat}{2}
    G_{\mathrm{ss.}}&=\bigg( \mathcal{V}_{A_\mu \psi^c\psi}\cQ^{\eminus1}_{A_\mu A_\nu}\cQ^{\eminus1}_{\psi^c\psi}\cQ^{\eminus1}_{\psi\psi^c}\mathcal{V}_{A_\nu \psi\psi^c}
    +\mathcal{V}_{ \psi^c A_\mu\psi}\cQ^{\eminus1}_{\psi^c\psi}\cQ^{\eminus1}_{A_\mu A_\nu}\cQ^{\eminus1}_{\psi\psi^c}\mathcal{V}_{ \psi A_\nu\psi^c}\\
    &\phantom{...}+\mathcal{V}_{ \psi^c \psi A_\mu}\cQ^{\eminus1}_{\psi^c\psi}\cQ^{\eminus1}_{\psi\psi^c}\cQ^{\eminus1}_{A_\mu A_\nu}\mathcal{V}_{ \psi \psi^c A_\nu}\bigg) + (\psi^c\leftrightarrow\psi)
    \\ &=6\times \mathcal{V}_{ \psi^c \psi A_\mu}\cQ^{\eminus1}_{\psi^c\psi}\cQ^{\eminus1}_{\psi\psi^c}\cQ^{\eminus1}_{A_\mu A_\nu} \mathcal{V}_{ \psi \psi^c A_\nu}\,.
    \label{sunset0}
    \end{alignedat}
\end{equation}
Notably, the presence of charge-conjugation matrices $C$ will effectively reverse the sign for the loop momentum of the $Q^{\eminus 1}_{\psi\psi^c}$ propagator. This property can be confirmed diagrammatically by applying the standard Feynman rules, as one of the loop momenta naturally assumes an opposite orientation relative to its associated fermionic line. Thus, employing the formula for the sunset topology~\eqref{eq:sunset_formula} along with the QED-specific expressions for $V_{IJK}$ and $Q_{IJ}$, we have
\begin{align}\label{eq:expr-sunset-EH}
\begin{aligned}
    G_{\mathrm{ss.}} &=6\sum_{n,m}(-1)^{n+m+1}\int_x\ \int_{k,\ell}\ \frac{1}{(k+\ell)^2}\\
    &\times\tr \! \Bigg[ \gamma^\mu \frac{\slashed{P}_x+ \slashed{k}+ m}{k^2-m^2}\left(\frac{\slashed{P}_x^2+2k\cdot P_x}{k^2-m^2}\right)^{\!\! n}  \gamma_\mu \frac{\slashed{P}_x-\slashed{\ell}+m}{\ell^2-m^2}\left(\frac{\slashed{P}_x^2-2\ell\cdot P_x}{\ell^2-m^2}\right)^{\!\!m} \Bigg] U_{\psi^c\psi}(x,y)\bigl|_{y=x}\,. 
\end{aligned}
\end{align}
Evaluating the $n$ and $m$ sums, truncating the expansion at dimension 8, and evaluating the two-loop integrals using the usual IBP relations~\cite{Chetyrkin:1997fm, Martin:2016bgz}, we find that the sunset contribution to the two-loop EFT Lagrangian is 
\begin{equation}
	\begin{alignedat}{2}
		\cL_{\sscript{EFT}}^{(2)}\supset&\frac{e^4}{(16\pi^2)^2}\frac{3}{2}\lzs\frac{1}{\epsilon}+2\log\frac{\mu^2}{m^2}-\frac{13}{18}\dzs  F^{\mu\nu}F_{\mu\nu}
		\\
		&-\frac{1}{(16\pi^2)^2}\frac{14}{15}\frac{e^6}{m^4}\lzs \frac{1}{\epsilon}+2\log\frac{\mu^2}{m^2}+\frac{887}{756} \dzs  (F^{\mu\nu}F_{\nu\sigma})(F^{\rho\sigma}F_{\mu\rho})
		\\
		&+\frac{1}{(16\pi^2)^2}\frac{1}{3}\frac{e^6}{m^4}\lzs \frac{1}{\epsilon}+2\log\frac{\mu^2}{m^2}+\frac{113}{108} \dzs  (F^{\mu\nu}F_{\mu\nu})^2\,.
	\end{alignedat}
\end{equation}

\subsection{Full matching result}

As for the one-loop matching, we find that, as expected, the divergent terms in the expressions above are identically canceled against the two-loop UV counterterms. The full matching Lagrangian up to two-loop order is
\begin{equation}
\label{eq:2-loop EFT full}
    \begin{alignedat}{2}
        \cL_{\sscript{EFT}}=&
        -\frac{1}{4} \! \lzs 1+ \frac{e^2}{16\pi^2} \frac{4}{3} \log\frac{\mu^2}{m^2} +\frac{e^4}{(16\pi^2)^2} \lzm\frac{13}{3} -4\log\frac{\mu^2}{m^2}\dzm \dzs F^{\mu\nu}F_{\mu\nu}
        \\&
        +\lzs \frac{1}{16\pi^2} \frac{7}{90} \frac{e^4}{m^4} +\frac{1}{(16\pi^2)^2} \frac{e^6}{m^4}\lzm \frac{247}{810}-\frac{14}{15} \log\frac{\mu^2}{m^2} \dzm \dzs (F^{\mu\nu}F_{\nu\sigma})(F^{\rho\sigma}F_{\mu\rho})
        \\&
        -\lzs \frac{1}{16\pi^2} \frac{1}{36}\frac{e^4}{m^4} +\frac{1}{(16\pi^2)^2} \frac{e^6}{m^4}\lzm\frac{49}{324}- \frac{1}{3} \log\frac{\mu^2}{m^2}  \dzm \dzs(F^{\mu\nu}F_{\mu\nu})^2\,.
    \end{alignedat}
\end{equation}
We have verified that our two-loop result for the dimension-4 terms is in agreement with the one presented in Ref.~\cite{Grozin:2012ec}. Furthermore, we have checked that the Wilson coefficients in Eq.~\eqref{eq:2-loop EFT full} are scale-invariant also at two-loop order, indicating the absence of running in the EFT, which as we argued is expected at dimension 8
\begin{equation}
   \frac{d c_i^{(2)}}{dt}= \beta_e^{(1)}\frac{\partial c_i^{(1)}}{\partial e}+\beta_m^{(1)}\frac{\partial c_i^{(1)}}{\partial m}+\frac{\partial c_i^{(2)}}{\partial t}=0\,,
\end{equation}
where the $\beta$ functions of the QED parameters can be obtained with, e.g., \texttt{RGBeta}~\cite{Thomsen:2021ncy}. 
Our determination of the finite parts for the two-loop dimension-8 contributions cannot be directly compared to previous results in the literature. To our knowledge, they have been determined for the first time in \msbar here.\footnote{Other works~\cite{Fliegner:1997ra,Kors:1998ew,Dunne:2001pp,Huet:2010nt,Huet:2011kd,Dunne:2012vv,Gies:2016yaa} have determined higher-order corrections for the Euler-Heisenberg Lagrangian; however, the approaches or approximations used in those works make it difficult to compare our results with the ones provided in those references.}

\section{Summary and Conclusion}
\label{sec:conclusion}

In this work, we extended modern functional methods to multi-loop RG and EFT matching calculations in arbitrary weakly-coupled gauge theories with mixed bosonic and fermionic degrees of freedom. 
Tensorial expressions for the effective action with all Grassmannian signs provide the starting point for the formalism, and we give general rules for their construction at any loop order. 
It is possible to cast the two-loop tensor contractions in terms of an expansion in fields and derivatives in position space while retaining manifest gauge invariance. This allows for direct and systematic calculations of EFT matching conditions and RG equations. The provided formulas are regulated in the manner of ordinary loop integrals, which in the hard-momenta limit become vacuum integrals, results for which are known analytically up to three-loop order~\cite{Martin:2016bgz}. The covariant formulas at higher-loop order can be obtained from vacuum supergraphs through the application of a set of simple rules.

We have further demonstrated the validity of a master formula for off-shell perturbative EFT matching at multi-loop orders that relates the EFT action to the hard part of the UV-theory effective action. The proof is based on the decomposition of the UV-theory loop integrals using expansion by regions and the identification of the corresponding contributions in the EFT. This result provides a rigorous foundation for the use of functional methods in EFT matching calculations while being similarly relevant for other approaches based on amplitude matching. 

The viability of the functional methods presented here has been demonstrated with the two-loop matching calculation for the Euler-Heisenberg Lagrangian, which describes the effective dynamics of photons at low energies after integrating out the electron in QED. In particular, we provide the complete matching conditions for the dimension-8 Wilson coefficients and verify the expected scale invariance of the result. Our methods for covariant evaluation of the functional supergraphs have also recently been applied to the calculation of the two-loop RG equations for a bosonic version of the SMEFT~\cite{Born:2024mgz}, bringing these methods an important step toward realistic applications.

The functional formalism presented in this work provides a powerful set of tools for precise calculations in EFTs. It offers several advantages over traditional diagrammatic methods, such as manifest covariance and a very systematic prescription, making it particularly suitable for automation. These advantages become increasingly pronounced at higher-loop orders, where the complexity of the calculations grows significantly. The automation of this formalism will enable efficient higher-order matching and RG calculations for generic UV theories, facilitating the precise exploration of a broader range of theories. The methods explored here can, thus, play an important role in advancing our understanding of physics in the SM and beyond in the upcoming precision era heralded by future colliders experiments.

\subsection*{Acknowledgments}
We extend our thanks to Toni Pich for comments on the manuscript and Renato Fonseca for useful discussions.
The work of JFM and AMS is supported by the grant PID2022-139466NB-C21 funded by MICIU/AEI/10.13039/501100011033 and FEDER/UE, and by the Junta de Andaluc\'ia grants P21\_00199 and FQM101. The work of JFM is further supported by the grants IJC2020-043549-I and EUR2024.153549 funded by MCIN/AEI/10.13039/501100011033 and the European Union NextGenerationEU/PRTR. The work of AMS is further supported by the Junta de Andaluc\'ia grant AST22\_6.5.
The work of AP has received funding from the Swiss National Science Foundation (SNF) through the Eccellenza Professorial Fellowship ``Flavor Physics at the High Energy Frontier'' project number 186866. 
The work of AET is funded by the Swiss National Science Foundation (SNSF) through the Ambizione grant ``Matching and Running: Improved Precision in the Hunt for New Physics,'' project number 209042.

\renewcommand{\thesection}{\Alph{section}}
\appendix 

\section{Superfields and Path Integrals} \label{app:superfields}

The perturbative expansion of path integrals is a textbook topic in QFT, both for bosonic and Grassmannian fields. It is less commonly discussed in theories with fields of mixed spin statistics. To obtain the required generalization of our formalism to realistic theories, such as the SM and its extensions, we sketch the derivation of the path integrals in this appendix. We begin by reviewing some concepts from supervector spaces. The reader may refer to~\cite{Witten:2012bg,DeWitt:2012mdz,Wegner:2016ahw} for more mathematical details and rigor.

\subsection{Working with supervectors}
\label{app:Grassmann}

We consider the supervector space $ \mathbb{R}_c^m \times \mathbb{R}_a^n $ with elements $ z_i = (x_a| \theta_\alpha) $, where $ x_a $ are ordinary numbers and $ \theta_\alpha $ are Grassmannian. From the commutative properties of the coordinates $ z_i $, one can easily verify that 
    \begin{equation}
    z_i z_j = \zeta_{ij} z_j z_i\,, \qquad \zeta = \begin{pmatrix}
    1 & 1 \\ 1 & -1 
    \end{pmatrix} \implies \zeta_{ij} ^2 = 1\,,  
    \end{equation}
where the indices of $ \zeta_{ij} $ do not count when determining if Einstein summation is implied.\footnote{Summation is still performed if an index of $ \zeta $ is repeated twice on objects other than $ \zeta $.}
This relation can be generalized to supertensors. Taking two tensors $M_{i_1 \ldots i_m} $ and $ N_{j_1 \ldots j_n} $ to commute as the product of their indices (meaning that $ z_{i_1} \cdots z_{i_m} M_{i_1 \ldots i_m} $ and $ z_{j_1} \cdots z_{j_n} N_{j_1 \ldots j_n} $ are both Grassmann even), the tensors commute as 
    \begin{equation}\label{eq:super_tensor_commutation}
    M_{i_1 \ldots i_m} N_{j_1 \ldots j_n} = N_{j_1 \ldots j_n} M_{i_1 \ldots i_m} \prod_{a =1}^m \prod_{b = 1}^n \zeta_{i_a j_b}\,.
    \end{equation}

The integrand in the path integral is a Grassmann--even function of the superspace coordinates. The quadratic monomial, which appears exponentiated in Gaussian integrals, is of particular interest: $ f(z) = \exp\!\big[ - \tfrac{1}{2} z_i Q_{ij} z_j \big] $. The symmetries of the supermatrix $ Q_{ij} $ are dictated by the scalar product $ z_i Q_{ij} z_j $:
    \begin{equation} \label{eq:GrassmanQ_symmetry}
    z_j Q_{ji} z_i = z_i Q_{ij} z_j = \zeta_{jj} z_j z_i Q_{ij}= \zeta_{ii} \zeta_{ij} \zeta_{jj} z_j Q_{ij} z_i \implies Q_{ij} = \zeta_{ii} \zeta_{ij} \zeta_{jj} Q_{ji}\,.
    \end{equation}
In block form, the supermatrix reads
    \begin{equation} \label{eq:mixed_stat_Q}
    Q= \begin{pmatrix}
    A & B \\ -B\transpose & C 
    \end{pmatrix}, \qquad A= A\transpose\,, \quad C= - C\transpose\,,
    \end{equation}
where $ B $ is Grassmannian and $A,C$ matrices of ordinary numbers. 
The inverse of the supermatrix $ Q $ can be determined by solving for the blocks of $ Q^{\eminus 1} $, subject to $ Q^{\eminus 1} Q= Q\, Q^{\eminus 1} = \mathds{1} $. One finds 
    \begin{equation}
    Q^{\eminus 1} = \begin{pmatrix} a & - a B C^{\eminus 1} \\ C^{\eminus 1} B\transpose a & c \end{pmatrix}, \qquad \begin{cases}
    a &=  \big(A + B C^{\eminus 1} B\transpose \big)^{\eminus 1} \\ c &=  \big(C + B\transpose A^{\eminus 1} B \big)^{\eminus 1} 
    \end{cases}\,.
    \end{equation}
We observe that the symmetries $ a= a\transpose $ and $ c = - c\transpose $ are those of the original blocks of $ Q $. However, the symmetries of $ Q^{\eminus 1} $ are \emph{not} identical to those of $ Q $. Indeed, it holds that 
    \begin{equation} \label{eq:Qinv_symmetry}
    Q^{\eminus 1}_{ij} = \zeta_{ij} Q^{\eminus 1}_{ji}\,.
    \end{equation}

Derivatives w.r.t. supervectors generally do not commute.\footnote{In this paper, it is understood that derivatives always act from the left. Consequently, the derivative first acts on the leftmost elements of products.} Indeed, they inherit the commutative properties of the individual coordinates, so 
    \begin{equation}
        \dfrac{\partial^2 }{\partial z_i \partial z_j} \equiv \dfrac{\partial }{\partial z_i} \dfrac{\partial }{\partial z_j}
        = \zeta_{ij} \dfrac{\partial }{\partial z_j} \dfrac{\partial }{\partial z_i}\,.
    \end{equation}
The integration measure over the superspace is chosen to be  
    \begin{equation}
    \int \,[\dd z] \equiv \! \int \dd x^m \!\! \int \dd \theta_1 \cdots \dd \theta_n\,.
    \end{equation} 
The Gaussian integration for even $n$ then evaluates as 
    \begin{equation}\label{eq:def_berenzinian}
    \int \,[\dd z] \, e^{\eminus\tfrac{1}{2} z_i Q_{ij} z_j} = (2\pi)^{m/2} (\mathrm{Ber}\,Q)^{\eminus 1/2}\,.
    \end{equation}
It is easy to see that the integral vanishes when there is an odd number, $ n $, of Grassmannian coordinates. For QFT applications, this does not matter, as fermionic degrees of freedom always come with their corresponding conjugates. The kinetic terms formed between two conjugate fermions then form part of the anti-symmetric sub-block $ C $ of  
Eq.~\eqref{eq:mixed_stat_Q}. The Berezinian (or superdeterminant) is defined from the supertrace as $\mathrm{Ber}\,W=e^{\mathrm{str}\ln W}$, with $\mathrm{str}\, W=\zeta_{ii}\, W_{ii}$, and it presents similar properties to those of the regular determinant but in superspace. The most notable difference is the property 
    \begin{align}\label{eq:Ber_prop}
    \mathrm{Ber}\,
    \begin{pmatrix}
    A & B\\ 
    0 & D
    \end{pmatrix}
    =\mathrm{Ber}\,
    \begin{pmatrix}
    A & 0\\ 
    C & D
    \end{pmatrix}
    =\det A\;\det{}^{\eminus 1} D\,,
    \end{align}
which follows from the different signs in the supertrace. The Berezinian satisfies the usual multiplication property $ \mathrm{Ber}(V W) = \mathrm{Ber}V \,\mathrm{Ber} W$ and is related to the regular determinant by means of the following expression
    \begin{equation}
    \mathrm{Ber} \! \begin{pmatrix}
    A &B \\ C & D
    \end{pmatrix} = \det(A- B D^{\eminus 1} C) \det{}^{\eminus 1} D = \det A \det{}^{\eminus 1} (D - C A^{\eminus 1} B)\,,
    \end{equation}
which assumes that either $D$ or $A$ are invertible. Indeed, in this case, we can decompose our original supermatrix as
\begin{align}
\begin{pmatrix}
A & B\\
C & D\\
\end{pmatrix}
=
\begin{pmatrix}
I & B\\
0 & D\\
\end{pmatrix}
\begin{pmatrix}
A-BD^{-1}C & 0\\
D^{-1}C & I\\
\end{pmatrix}
=
\begin{pmatrix}
A & 0\\
C & I\\
\end{pmatrix}
\begin{pmatrix}
I & A^{-1}B\\
0 & D-C A^{\eminus 1} B\\
\end{pmatrix}
\,.
\end{align}
yielding the expressions above once we use the property in Eq.~\eqref{eq:Ber_prop}. Under a coordinate transformation $ x= x(y, \xi) $ and $ \theta = \theta(y, \xi) $, the integral transforms as 
    \begin{equation}
    \int \dd x^m \int \dd \theta^n f(x, \theta) = \int \dd y^m \int \dd \xi^n \,\mathrm{Ber} \!\begin{pmatrix}
    \partial x/ \partial y & \partial x/ \partial \xi \\ \partial \theta/ \partial y & \partial \theta / \partial \xi  
    \end{pmatrix} f\big(x(y,\xi),\, \theta(y,\xi)\big)\,.
    \end{equation}
With this, we can now proceed to evaluate QFT path integrals, assuming that the properties of the supervector space generalize to spaces of infinite dimensions.

\subsection{Derivation of the generating functional}
\label{app:Details_Gen_Func}

In QFT, the fields $ \eta = (\phi, \chi) $ (bosons and fermions) make up the infinite-dimensional superspace vectors subject to the commutation relations~\eqref{eq:field_commutation}. 
We begin by shifting the integration variable $\eta\to \bar\eta+\hbar^{1/2}\eta$ in the path integral of Eq.~\eqref{eq:vac_func_def} and expanding the action as shown in Eq.~\eqref{eq:action_expansion}. The saddlepoint approximation of the path integral then gives
\begin{align}\label{eq:expansion_generating-functional}
    e^{iW[\mathcal{J}]/\hbar} = 
    e^{i\bar{S}/\hbar} \!\! \int \, [\mathcal{D}\eta]  \, e^{\frac{i}{2} \eta_I \overline{\cQ}_{IJ}\, \eta_J}  
    \bigg[ 1 &+\frac{i\hbar}{2} \eta_I \overline{\mathcal{V}}^{(1)}_{IJ}\eta_J
    +\frac{i\hbar}{24}\eta_I\eta_J\eta_K\eta_L \overline{\mathcal{V}}^{(0)}_{IJKL}\\
    &-\frac{\hbar}{72}
     \eta_L\eta_M\eta_N      \eta_I\eta_J\eta_K    \overline{\mathcal{V}}^{(0)}_{IJK} \overline{\mathcal{V}}^{(0)}_{LMN}\notag\\
    &-\frac{\hbar}{2}\zeta_{II} \overline{\mathcal{V}}^{(1)}_I\eta_I\eta_J\overline{\mathcal{V}}^{(1)}_J
    -\frac{\hbar}{6}\zeta_{II}\overline{\mathcal{V}}^{(1)}_I \eta_I\eta_J\eta_K\eta_L \overline{\mathcal{V}}^{(0)}_{JKL} + \cO(\hbar^2)\bigg]\,, \nonumber
\end{align}
where, in our convention, the bar denotes evaluation at $\bar{\eta}$, the solution to the classical equations of motion.

To evaluate the integrals in Eq.~\eqref{eq:expansion_generating-functional}, we begin with the Gaussian integral:\footnote{This is, in fact, the limit of an analytic continuation of Eq.~\eqref{eq:def_berenzinian} with a quadratic dampening term. There is an irrelevant normalization in the extension to infinite-dimensional space, which is absorbed into the path-integral measure.}
\begin{align}
    \int \, [\mathcal{D}  \eta ] \,
    e^{\frac{i}{2} \eta_I\overline{\cQ}_{IJ}\, \eta_J} = (\ber \overline\cQ)^{\eminus 1/2}\,.
\end{align}
The remaining integrals are best evaluated by introducing an auxiliary source to the Gaussian integral\footnote{The linear term in the exponential is eliminated with the variable change $\eta_I\to \eta_I-\mathcal{K}_J\overline\cQ^{\eminus 1}_{JI}$.} 
\begin{align}
    I[\mathcal{K}]=\int \, [\mathcal{D}  \eta]\, e^{\frac{i}{2}\eta_I\overline{\cQ}_{IJ}\, \eta_J+i\mathcal{K}_I\eta_I}
    = e^{\eminus  i v[\mathcal{K}]} (\ber\overline\cQ)^{\eminus 1/2}, \qquad v[\mathcal{K}]
    = \frac{1}{2} \zeta_{JJ}\mathcal{K}_I \overline{\cQ}^{\eminus 1}_{IJ} \mathcal{K}_J\,.
\end{align}
The integrals of field monomials multiplying the Gaussian are then identified as derivatives of $I[\mathcal{K}]$:
\begin{align}
    (-i)^n\frac{\partial}{\partial \mathcal{K}_{I_1}} \cdots \frac{\partial}{\partial \mathcal{K}_{I_n}} I[\mathcal{K}] \bigg|_{\mathcal{K}=0}
    = \int \, [\mathcal{D} \eta]\, \eta_{I_1} \cdots \eta_{I_n} e^{\frac{i}{2}\eta_I\overline{\cQ}_{IJ}\, \eta_J}\,.
\end{align}
We have
\begin{align}
    (\ber \overline\cQ)^{1/2}& \int\, [\mathcal{D} \eta]\, e^{\frac{i}{2} \eta_I \overline{\cQ}_{IJ}\, \eta_J} \eta_I \overline{\mathcal{V}}^{(1)}_{IJ}  \eta_J
    = -\zeta_{JJ}(\ber \overline\cQ)^{1/2} \frac{\partial}{\partial \mathcal{K}_{I}} \frac{\partial}{\partial \mathcal{K}_{J}} e^{-iv[\mathcal{K}]} \bigg|_{\mathcal{K}=0} \overline{\mathcal{V}}^{(1)}_{JI} 
    = i\, \zeta_{II} \overline{\cQ}^{\eminus1}_{IJ} \overline{\mathcal{V}}^{(1)}_{JI}, \\[5pt]
    (\ber \overline\cQ)^{1/2}& \int\, [\mathcal{D} \eta] \, e^{-\frac{i}{2} \eta_I \overline{\cQ}_{IJ}\, \eta_J} 
    \eta_I \eta_J\eta_K \eta_L \overline{\mathcal{V}}^{(1)}_{IJKL}
    =\, -3 \overline\cQ^{\eminus1}_{IJ} \overline\cQ^{\eminus 1}_{KL} \overline{\mathcal{V}}^{(1)}_{IJKL}\,,\\[5pt]
    (\ber \overline\cQ)^{1/2}& \int\, [\mathcal{D} \eta] \, e^{-\frac{i}{2} \eta_I \overline{\cQ}_{IJ}\, \eta_J} \eta_L \eta_M \eta_N \eta_I \eta_J \eta_K  \overline{\mathcal{V}}^{(0)}_{IJK}\overline{\mathcal{V}}^{(0)}_{LMN}\\[5pt]
    &\phantom{...}= - 9 i\,  \overline\cQ^{\eminus 1}_{LM} \overline\cQ^{\eminus 1}_{NI} \overline\cQ^{\eminus 1}_{JK} \overline{\mathcal{V}}^{(0)}_{IJK}\overline{\mathcal{V}}^{(0)}_{LMN}
    - 6 i\, \zeta_{IN}\zeta_{IM}\zeta_{JN} \overline\cQ^{\eminus 1}_{LI} \overline\cQ^{\eminus 1}_{MJ} \overline\cQ^{\eminus 1}_{NK}\overline{\mathcal{V}}^{(0)}_{IJK}  \overline{\mathcal{V}}^{(0)}_{LMN}\,.\nonumber
\end{align}
These integrals along with Eq.~\eqref{eq:expansion_generating-functional} lead to Eq.~\eqref{eq:vac_func_2-loop} after taking a logarithm of both sides of the equation. 

\subsection{Derivation of the effective action}
\label{app:Details_Eff_Action}

We detail here the derivation of the effective action from the Legendre transform of the generating functional. To this end, we make the dependence on $ \hat{\eta} $ explicit in the definition of $ \Gamma[\hat{\eta}] $. This can be achieved perturbatively, since the background field and the classical fields agree at tree level. 
We take the derivative of the tree-level EOM~\eqref{eq:bkg_field_EOM} w.r.t. the source to obtain
    \begin{equation} \label{eq:bkg_field_derivative}
    \dfrac{\delta \overline{\eta}_J}{\delta \mathcal{J}_I} = - \overline{\mathcal{Q}}^{\eminus1}_{IJ}\,.
    \end{equation}
Before proceeding, we will need the derivative of the supertrace of a logarithm. This must be treated with care, and we find 
\begin{equation}
    \begin{split}
    \frac{\delta}{\delta\cJ_I}\mathrm{STr} \log \overline{\cQ}&=\zeta_{JJ} \frac{\delta}{\delta\cJ_I} [\log \overline{\cQ}]_{JJ}
    =-\zeta_{JJ} \frac{\delta}{\delta\cJ_I} \bigg[\sum_{n=1}^\infty\frac{1}{n}(\cI-\overline{\cQ})^n_{JJ}\bigg]\\
    &= \zeta_{JJ} \frac{\delta\overline\cQ_{JK}}{\delta\cJ_I} \sum_{n=1}^\infty(\cI-\overline{\cQ})^{n-1}_{KJ}
    = \zeta_{JJ} \frac{\delta\overline\cQ_{JK}}{\delta\cJ_I} \overline{\cQ}^{\eminus 1}_{KJ}\,,
    \end{split}
\end{equation}
where the second-to-last step effectively relies on the cyclicity of the supertrace (which can be verified with commutation rule~\eqref{eq:super_tensor_commutation}). Next, we determine $\hat\eta$ in terms of $ \overline{\eta} $ by applying a source derivative to the vacuum functional~\eqref{eq:vac_func_2-loop} and using Eq.~\eqref{eq:bkg_field_derivative}:
\begin{align} 
    \hat{\eta}_I \equiv \dfrac{\delta W[\mathcal{J}]}{\delta \mathcal{J}_I} &= \overline{\eta}_I 
    + \hbar\,\frac{\delta\bar\eta_J}{\delta\cJ_I} \left(\frac{\delta\overline{\mathcal{S}}^{(1)}}{\delta\bar\eta_J}+\dfrac{i}{2} \zeta_{KK}  \frac{\delta\overline\cQ_{KL}}{\delta\bar\eta_J} \overline{\cQ}^{\eminus 1}_{LK} \right)  + \ord{\hbar^2}\notag\\
    &=\overline{\eta}_I 
    - \hbar\,\overline{\cQ}^{\eminus 1}_{IJ}  \left(\overline{\mathcal{V}}^{(1)}_J+\dfrac{i}{2}   \overline{\cQ}^{\eminus 1}_{KL}\overline{\mathcal{V}}^{(0)}_{JKL} \right)  + \ord{\hbar^2}\,.
    \end{align}
Inverting this relation, we have 
    \begin{equation}
    \overline{\eta}_I = \hat{\eta}_I 
    + \hbar\, \widehat{\mathcal{Q}}_{IJ}^{\eminus1} \left(\widehat{\mathcal{V}}^{(1)}_J+\dfrac{i}{2}   \widehat{\mathcal{Q}}_{KL}^{\eminus1} \widehat{\mathcal{V}}^{(0)}_{JKL} \right)  + \ord{\hbar^2}\,.
    \end{equation}
Finally, inserting the vacuum functional~\eqref{eq:vac_func_2-loop} into the definition of the effective action~\eqref{eq:eff_action_def}, and evaluating all terms around $ \hat{\eta} $ yields Eq.~\eqref{eq:eff_action_generic_2-loop}.

\section{The Parallel Displacement Propagator} 
\label{app:pdp}

Parallel displacement propagators (PDPs) have been used elsewhere in the literature and many important properties and proofs thereof can be found in, e.g., Refs.~\cite{Barvinsky:1985an,Kuzenko:2003eb}. Since the formalism and notation deviate significantly from ours at points, we consider it worthwhile to review some properties and proofs in this appendix. Additionally, we have derived some formulas for repeated derivatives acting on PDPs, which can be used in the functional evaluation of the quantum effective action.

\subsection{Basic properties}

With the boundary condition that $ U(x, x) = \mathds{1} $, the PDP may be defined as the unique solution to the equation
    \begin{equation} \label{eq:parallel_disp_prop}
    (x-y)_\mu D_x^\mu U_{ab}(x,y) = 0 \implies U(x,y) = \mathscr{P} \exp \!\left[i \! \int_{y}^{x} \dd x^\mu A_\mu  \right]\,.
    \end{equation}
We easily see that its inverse is given by
    \begin{equation}
    U^{\eminus 1}(x,y) = U^\dagger(x,y) = U(y,x)\,,
    \end{equation}
and can verify that also 
    \begin{equation} \label{eq:parallel_disp_prop_alt}
    (x-y)_\mu D_y^\mu U_{ab}(x,y) = 0\,.
    \end{equation}
The PDP has the property that
    \begin{equation} \label{eq:sym_devs_on_U}
    (x-y)_{\underline{n}} D_x^{\underline{n}}\, U_{ab}(x,y) = (x-y)_{\underline{n}} D_y^{\underline{n}}\, U_{ab}(x,y) = 0\,, \qquad \forall n\geq 1\,.
    \end{equation}
This can be proven with induction starting from Eqs.~\eqref{eq:parallel_disp_prop} and~\eqref{eq:parallel_disp_prop_alt} for $ n =1 $. The induction step follows by multiplying $ (x-y)_\nu D_x^\nu $ to the known identity for $ n $, and applying $ (x-y)_\nu D_x^\nu (x-y)_\mu = (x-y)_\mu $. In the coincidence limit of the PDP, Eq.~\eqref{eq:sym_devs_on_U} produces the corollary that\footnote{The symmetrization of indices denoted with parentheses contain the customary normalization factor, e.g., 
    \[ 
        D_x^{(\mu_1}\cdots D_x^{\mu_n)} = 
        \dfrac{1}{n!} \sum_\sigma D_x^{\mu_{\sigma(1)}} \cdots D_x^{\mu_{\sigma(n)}}\,.
    \]
}
    \begin{equation}\label{eq:sym_sum_U}
    \begin{split}
    0&= \dfrac{1}{n!} D_x^{\mu_1}\cdots D_x^{\mu_n} \big[(x-y)_{\underline{n}} D_x^{\underline{n}}\, U_{ab}(x,y)\big] \Big|_{y=x}\,, \qquad \forall n\geq 1\,,\\
    &= D_x^{(\mu_1}\cdots D_x^{\mu_n)} U_{ab}(x,y) \Big|_{y=x}\,.
    \end{split}
    \end{equation}  
It follows from this corollary that\footnote{We let $ D_\mu= \partial_\mu - iA_\mu $, so $ [D_\mu, D_\nu] = -i G_{\mu\nu} $. }
    \begin{align} \label{eq:sym_sum_identity}
    0 &= (n+1) D_x^{(\mu_1}\cdots D_x^{\mu_n } D_x^{\nu)} U_{ab}(x,y) \Big|_{y=x} \\
    &= \bigg[ D_x^{(\mu_1}\cdots D_x^{\mu_n)} D_x^{\nu} U_{ab}(x,y)  +  \sum_{k=1}^{n} \dfrac{1}{n!} \sum_\sigma D_x^{\mu_{\sigma(1)}} \cdots D_x^{\mu_{\sigma(k-1)}} D_x^{\nu} D_x^{\mu_{\sigma(k)}} \cdots D_x^{\mu_{\sigma(n)}} U_{ab}(x,y) \bigg]_{y=x} \nonumber \\
    &= \Big[(n+1) D_x^{(\mu_1}\cdots D_x^{\mu_n)} D_x^{\nu} U_{ab}(x,y)  + i n D_x^{(\mu_1}\cdots D_x^{\mu_{n-1}} G \ud{\mu_n) \nu}{ac}(x) U_{cb}(x,y) \Big]_{y=x}\,, \nonumber
    \end{align}
where the sum is over all permutations $ \sigma $ of the first $ n $ integers.
The last equality follows from another application of identity~\eqref{eq:sym_sum_U}, which implies that 
    \begin{equation}
    \sum_\sigma D_x^{\mu_{\sigma(1)}} \cdots D_x^{\mu_{\sigma(k-1)}} G \ud{\mu_{\sigma(k)} \nu}{ac}(x) D_x^{\mu_{\sigma(k)}} \cdots D_x^{\mu_{\sigma(n)}} U_{cb}(x,y) \Big|_{y=x} = 0\,,\quad  \forall k<n\,.
    \end{equation}
This kills all commutators in Eq.~\eqref{eq:sym_sum_identity} but for the last as we move $ D^\nu_x $ to the right. Rearranging the term in the last line of Eq.~\eqref{eq:sym_sum_identity}, it follows that 
    \begin{equation} \label{eq:partially_sym_devs_on_U}
    D_x^{(\mu_1}\cdots D_x^{\mu_n)} D_x^{\nu} U_{ab}(x,y) \Big|_{y=x} = - i \dfrac{n}{n+1} D_x^{(\mu_1}\cdots D_x^{\mu_{n-1}} G \ud{\mu_n) \nu\,}{ab}(x)\,.
    \end{equation}
This is an important starting point towards taking generic derivatives of the PDP. Before getting to that, we first need to discuss how to do Taylor expansions in a manifestly-covariant way by using the PDP.

\subsection{Covariant Taylor expansion}

For a gauge-invariant function $ f(x) $, we can perform the usual Taylor expansion to write	
    \begin{equation}
    f(x) = \sum_{n=0}^{\infty} \dfrac{1}{n!} (x-y)_{\underline{n}} \partial_y^{\underline{n}} f(y)\,.
    \end{equation} 
Here the r.h.s. transforms as the l.h.s., namely as a singlet at position $ x $.
The ordinary Taylor expansion of some gauge-covariant function (in a non-trivial representation) does not maintain the transformation properties under background-gauge variations and must be amended accordingly. 

For a field $ f_a(x) $ in some representation of the gauge group, the combination $ U\ud{a}{b}(y,x) f^b(x) $ is invariant w.r.t. to gauge transformations in $ x $. Thus, 
    \begin{equation}
    U_{ab}(y,x) f_b(x) = \sum_{n=0}^{\infty} \dfrac{1}{n!} (x-z)_{\underline{n}} \partial_z^{\underline{n}}  [U_{ab}(y,z) f_b(z)]\,.
    \end{equation}
Setting $ y=x $, we find 
    \begin{equation} \label{eq:cov_Taylor_exp}
    \begin{split}
    f_a(x) &= \sum_{n=0}^{\infty} \dfrac{1}{n!} (x-z)_{\underline{n}} \partial_z^{\underline{n}}  [U_{ab}(x,z) f_b(z)]= U_{ab}(x,z) \sum_{n=0}^{\infty} \dfrac{1}{n!} (x-z)_{\underline{n}} D_z^{\underline{n}} f_b(z)\,,
    \end{split}
    \end{equation}
which follows from the identity~\eqref{eq:sym_devs_on_U}. Here, we have utilized that the PDP transforms covariantly such that the chain rule for covariant derivatives is applicable (the partial derivative is a covariant derivative when acting on a product forming a singlet). The r.h.s. of Eq.~\eqref{eq:cov_Taylor_exp} is a singlet at $ z $ and transforms like $ f_a(x) $ at $ x $. We may think of this as the covariant Taylor expansion of $ f_a $. 

One can easily adapt the covariant Taylor expansion~\eqref{eq:cov_Taylor_exp} to the case where the function transforms in a product representation, as it simply holds that, e.g., $ U_{ab|cd}(x, y) = U_{ac}(x, y) U_{bd}(x, y)$ for two-index representations. 
It will be especially useful to consider the case of a function with two open indices, e.g., a field strength tensor transforming in the adjoint representation. We may write
    \begin{equation} \label{eq:cov_taylor_adjoint}
    \begin{split}
    U_{ab}(x,y) f_{bc}(y) U_{cd}(y,x) &= U_{ab}(x,y) U_{bc}(y,z) \sum_{n=0}^{\infty} \dfrac{1}{n!} (y-z)_{\underline{n}} D_z^{\underline{n}} \big[f(z) U(z,x) \big]_{cd}\bigg|_{z=x}\\
    &=\sum_{n=0}^{\infty} \dfrac{1}{n!} (y-x)_{\underline{n}} D_x^{\underline{n}} f_{ad}(x)\,,
    \end{split}
    \end{equation}
where the second equality follows from Eq.~\eqref{eq:sym_sum_U}.

The covariant Taylor expansion allows us to demonstrate properties of covariant derivatives acting on the covariant delta function. Under integration, it holds that
    \begin{equation} \label{eq:diff_op_on_cov_delta}
    \begin{split}
    \int_y D_x^\mu \delta_{ab}(x,y) f_b(y) 
    &= \int_y D_x^\mu \bigg[ \delta(x-y) \sum_{m=0}^{\infty}	\dfrac{1}{m!} (y-x)_{\underline{m}} D_x^{\underline{m}} f_a(x) \bigg] \\
    &= \int_y \delta(x-y) \sum_{m=0}^{\infty}	\dfrac{1}{m!} (y-x)_{\underline{m}} D_x^{\mu} D_x^{\underline{m}} f_a(x)
    =  D_x^\mu f_{a}(x)\,,
    \end{split} 
    \end{equation}
where the first equality follows from the covariant Taylor expansion~\eqref{eq:cov_Taylor_exp}. The second step is due to $ \int_y\; \partial_x [(x-y)^n \delta(x-y) f(x)] = \int_y \; (x-y)^n \delta (x-y) \partial_x f(x) $.\footnote{This in turn follows from 
    \[
    \int_y \partial_x [(x-y)^n \delta(x-y)] = \int_y [\partial_x (x-y)^n \delta(x-y) + (x-y)^n \partial_x \delta(x-y)]
    = \int_y  \delta(x-y) (\partial_x +\partial_y)(x-y)^n  
    = 0\,.  \nonumber
    \]
}
On the other hand, 
    \begin{equation}
    D_x^\mu f_{a}(x) = \int_y D^\mu_x\delta_{ab}(x,y) f_{b}(y) =
    - \int_y D_y^\mu \delta_{ab}(x,y)f_{b}(y)\,,
    \end{equation}
which then proves the behavior~\eqref{eq:dev_on_cov_delta} (under integration). We note that this is a property associated with using the geodesic Wilson line.

\subsection{Derivatives of the Wilson line} \label{app:PDP_devs}

We can now determine the action of a covariant derivative on the parallel displacement propagator~\cite{Kuzenko:2003eb}. We use the covariant Taylor expansion on the derivative of the propagator itself to obtain
    \begin{equation} 
    D_x^{\nu} U_{ab}(x,y) = U_{ac}(x,z) \sum_{n=0}^{\infty} \dfrac{1}{n!} (x-z)_{\underline{n}} D_z^{\underline{n}} D_z^{\nu} U_{cb}(z,y) \Big|_{z=y}\,.
    \end{equation} 
Using Eq.~\eqref{eq:partially_sym_devs_on_U} for the highly-symmetrized derivatives acting on $ U $ in the coincidence limit, we find 
    \begin{equation} \label{eq:D_on_U}
    D_x^{\nu} U_{ab}(x,y) = - i \, U_{ac}(x,y) \sum_{n=1}^{\infty} \dfrac{n}{(n+1)!} (x-y)_{\underline{n}} D_y^{\mu_1}\cdots D_y^{\mu_{n-1}} G^{\mu_n \nu}_{cb}(y)\,.
    \end{equation}  
Derivatives w.r.t. the second coordinate is performed with the simple observation 
    \begin{equation}
    0 = D_y^\mu \big(U_{ab} (x,y) U_{bc}(y,x) \big)  \implies  D_y^\mu U_{ab} (x,y) = - U_{ac} (x,y) D_y^\mu U_{cd} (y,x) U_{db} (x,y)\,.
    \end{equation}
In conjunction with Eq.~\eqref{eq:D_on_U}, we find that
    \begin{equation} \label{eq:Dy_on_U}
    0 =  D_y^\nu U_{ab} (x,y) =  i \sum_{n=1}^{\infty} \dfrac{n}{(n+1)!} (y-x)_{\underline{n}} \big[D_x^{\mu_1} \cdots D_x^{\mu_{n-1}} G^{\mu_n \nu}_{ac}(x) \big] U_{cb} (x,y)\,.
    \end{equation}

In some situations, we might want to extract the PDP on the other side of the field strength tensor in Eq.~\eqref{eq:D_on_U}. Starting from~\eqref{eq:D_on_U} and applying the Taylor expansion~\eqref{eq:cov_taylor_adjoint} to $ D_x^{\underline{n-1}} G^{\mu_n \nu} $yields 
    \begin{equation}
    D^\nu_x U(x,y) = -i \sum_{n=1}^{\infty} \dfrac{n}{(n+1)!} (x-y)_{\underline{n}} \sum_{m=0}^{\infty} \dfrac{1}{m!} (y-x)_{\underline{m}} \big[D_x^{\underline{m}} D_x^{\underline{n-1}} G^{\mu_n \nu}(x) \big] U(x,y)\,.
    \end{equation} 
To manipulate the double sum, we consider 
    \begin{align}
    \sum_{n=1}^{\infty} \sum_{m=0}^{\infty} \dfrac{n}{(n+1)!}\dfrac{(\eminus 1)^{m}}{m!} s^{m+n}
    &= \sum_{n=1}^{\infty} \sum_{k=0}^{n-1} \dfrac{n-k}{(n-k+1)!} \dfrac{(\eminus 1)^{k}}{k!} s^{n} \\
    &= \sum_{n=1}^{\infty} \dfrac{s^n}{(n+1)!} \sum_{k=0}^{n-1} (\eminus1)^k (n-k) \binom{n+1}{k} 
    = - \sum_{n=1}^{\infty}  \dfrac{(\eminus 1)^n s^n}{(n+1)!}\,. \nonumber
    \end{align}
It then follows that 
    \begin{equation} \label{eq:D_on_U_right}
    D^\nu_x U(x,y) = i \sum_{n=1}^{\infty} \dfrac{(-1)^n}{(n+1)!} (x-y)_{\underline{n}} \big[D_x^{\underline{n-1}} G^{\mu_n \nu}(x) \big] U(x,y)\,,
    \end{equation} 
which is a useful alternative to~\eqref{eq:D_on_U}. The formulas presented here suffice to compute all derivatives acting on PDPs in the covariant evaluation formulas presented in Section~\ref{sec:GaugeCov_Evaluation}. It is also possible to directly present the results of repeated applications as shown in Note~\ref{note:coincidence_limit_formula}.

\begin{note}[note:coincidence_limit_formula]{Coincidence-limit master formula} 
Consider two ordered sets of Lorentz indices $ \boldsymbol{\mu} = \{\mu_i\}_{i=1}^m $ and $ \boldsymbol{\nu} = \{\nu_i\}_{i=1}^n $ for integers $ m,n$. We introduce the notation 
    \begin{equation}
    D^{\boldsymbol{\mu}} = D^{\mu_1} \cdots D^{\mu_m}\,, \qquad G^{\boldsymbol{\mu}}= D^{(\mu_1} \cdots D^{\mu_{m-2}} G^{\mu_{m-1}) \mu_m}\,,
    \end{equation}
the last of which is non-zero only if there are at least two elements in the list.
	
We may now formulate a compact formula for mixed derivatives of a PDP in the coincidence limit:\footnote{
    The argument is a combinatorial one that accounts for all ways of applying repeated derivatives to the PDP such that the resulting Taylor series has a zeroth order term in $ (x-y) $. The last elements of $ a_i $ sets are the indices of the $ x $-derivatives that act on the Wilson line using Eq.~\eqref{eq:D_on_U}. The remaining elements in $ a_i $ are $ x $- and $ y $-derivatives acting on the powers of $ (x-y) $ in the corresponding application of~\eqref{eq:D_on_U}, resulting in a zeroth order term. Next, the derivatives $ D^{\boldsymbol{\nu} \setminus\! (A\cup B)} $ stem from $ y $-derivatives acting on the $ G(y) $ created by $ x $-derivatives of the PDPs. Finally, the product over $ B $ stems from the remaining $ y $-derivatives acting on the Wilson line per~\eqref{eq:Dy_on_U} with additional derivatives acting on the powers of
    $ (y-x) $.
}
    \begin{equation}\label{eq:pdp_at_coincidence}
    D_x^{\boldsymbol{\mu}} D_y^{\boldsymbol{\nu}} U(x,y) \Big|_{y=x} = \sum_{A,B} i^{|A| + |B|} \prod_{b\in B}\!  \left( \dfrac{|b| -1}{|b|} G^b \right)\! D^{\boldsymbol{\nu} \setminus\!  (A\cup B)} \! \left( \prod_{a\in A} (\eminus 1)^{1+ |\boldsymbol{\nu} \cap a|} \dfrac{|a| -1 }{|a|} G^a\right)\,,
    \end{equation}
where the non-commuting products over the ordered sets $ A,B $ are arranged with the first element being the left-most in the product, etc. 
The ordered sets $ A= \{a_i\} $ and $ B= \{b_i\} $ of disjoint sets of Lorentz indices are chosen such that $ \bigcup_i a_i \cup \bigcup_i b_i \subseteq \boldsymbol{\mu} \cup \boldsymbol{\nu} $ and $  \boldsymbol{\mu} \subseteq \bigcup_i a_i $. Meanwhile, the sets of Lorentz indices satisfy $ |a_i|, |b_i| \geq 2 $ and are ordered according to
    \begin{align}
    a_i &= \{ \nu_{j_1},\, \nu_{j_2},\ldots, \mu_{k_1},\, \mu_{k_2},\ldots \}\,, \quad \mathrm{where} \; j_{1} < j_2 <\ldots \quad \mathrm{and} \; k_1<k_2< \ldots\,, \nonumber \\
    b_i &= \{ \nu_{j_1},\, \nu_{j_2},\ldots \}\,. 
    \end{align}
The sets further satisfy constraints on their last element, the index on the field-strength tensor that is not symmetrized over:
    \begin{equation} \label{eq:coincidence_limit_last_element}
    \begin{split}
    a_{i,\eminus 1} &= \mu_j\,, \quad j = \max \big\{k:\; \mu_k \notin \cup_{\ell > i} a_\ell \big\}\,,\\
    b_{i, \eminus1} &= \nu_j\,, \quad j = \max \big\{k:\; \nu_k \notin (\cup_{\ell < i} b_\ell) \cup (\cup_{a\in A } a) \big\}\,,
    \end{split}
    \end{equation}
this also implies that $ a_i \cap \boldsymbol{\mu} \neq \emptyset $. In other words, the last element of $ a_{|A|} $ is $ \mu_m $; the last element of $ a_{|A|-1} $ is the last element of $ \boldsymbol{\mu} $ not in $ a_{|A|} $; and so on. Meanwhile, the last element of $ b_1 $ is the last element of $ \boldsymbol{\nu} $ not used in any $ a_i $ nor in $ \boldsymbol{\nu} \setminus\!  (A\cup B) $; that is, it is the last element of $ \boldsymbol{\nu} $ in any of the $ b_i $ sets.
	
Evaluating the derivatives of the PDP in the coincidence limit, thus, comes down to constructing all sets $ A,B $ satisfying the constraints given here. One approach could be to determine all partitions $ A $ of $ \boldsymbol{\mu} \cup I $ and $ B$ of $\boldsymbol{\nu} \setminus I $ for all $ I\subseteq \boldsymbol{\nu} $ with at least two elements in each set. Afterward, the partitions are sorted and ordered according to their last elements so as to satisfy Eq.~\eqref{eq:coincidence_limit_last_element}.
\end{note}

\section{Alternative Evaluation of the Sunset Diagram} \label{app:alt_sunset}
In this appendix, we derive an alternative formula for the sunset contribution $ G_\mathrm{ss.} $ to the effective action, which might prove more useful in some circumstances. Our starting point is~\eqref{eq:sunset_formula_ibp}.
In this case, we introduce momentum integration in place of all remaining deltas, giving
    \begin{align}
    G_\mathrm{ss.} = \!\! \sum_{m,n,m',n'} \!\!(\eminus 1)^{m+n} &\int_{xx'} \int_{k\ell q} e^{i (k+\ell+q) \cdot (x'-x)} V^{(\underline{m}, \underline{n})}_{abc}(x) V^{(\underline{m}'\,, \underline{n}')}_{a'b'c'}(x') \big[Q_{ad}^{\eminus 1}(x,\,P_x+q) U_{da'}(x,x') \big] \nonumber \\
    &\quad \times \big[(P_x+k)^{\underleftarrow{m}} Q_{be}^{\eminus 1}(x,\,P_x +k) (P_x+k)^{\underline{m}'} U_{eb'}(x,x') \big]  \nonumber \\
    &\quad \times \big[(P_x+\ell)^{\underleftarrow{n}} Q_{cf}^{\eminus 1}(x,\, P_x+ \ell ) (P_x+\ell)^{\underline{n}'} U_{fc'}(x,x')  \big]\,.
    \end{align} 
We would ideally like to carry out the $ x' $ integration to convert the exponential into a delta function in momentum space; however, this is not yet possible seeing as the integrand depends on fields evaluated at $x' $ through the $ V^{(\underline{m}', \underline{n}')}_{a'b'c'}(x') $ function and through the PDPs, which are not in the coincidence limit. The former is addressed by moving the vertex back to $ x $ with the covariant Taylor expansion~\eqref{eq:cov_Taylor_exp}, from which\footnote{The PDP of a product representation simply factors. For instance, $$ U_{a'b'c'|abc}(x',x) = U_{a'a}(x',x) U_{b'b}(x',x) U_{c'c}(x',x)\,.$$ }
    \begin{equation}
    V^{(\underline{m}', \underline{n}')}_{a'b'c'}(x') = U_{a'b'c'|def}(x',x) \sum_{r=0}^{\infty} \dfrac{(\eminus i)^r}{r!} (x'-x)_{\underline{r}} P_x^{\underline{r}} V^{(\underline{m}', \underline{n}')}_{def}(x)\,.
    \end{equation}
Thus, the sunset superdiagram is given by
    \begin{align}
    G_\mathrm{ss.} = \!\! \sum_{m,n,m',n'} \!\!(\eminus 1)^{m+n} & \sum_{r=0}^{\infty} \dfrac{(\eminus i)^r}{r!}\int_{xx'} \int_{k\ell q} e^{i (k+\ell+q) \cdot (x'-x)} (x'-x)_{\underline{r}} V^{(\underline{m}, \underline{n})}_{abc}(x) P_x^{\underline{r}} V^{(\underline{m}', \underline{n}')}_{a'b'c'}(x) \nonumber\\
    &\times \big[Q_{ad}^{\eminus 1}(x,\,P_x+q) U_{dd'}(x,x') \big] U_{d'a'}(x',x) \nonumber\\
    & \times \big[(P_x+k)^{\underleftarrow{m}} Q_{be}^{\eminus 1}(x,\,P_x +k) (P_x+k)^{\underline{m}'} U_{ee'}(x,x') \big] U_{e'b'}(x',x) \nonumber\\
    &\times \big[(P_x+\ell)^{\underleftarrow{n}} Q_{cf}^{\eminus 1}(x,\, P_x+ \ell ) (P_x+\ell)^{\underline{n}'} U_{ff'}(x,x') \big] U_{f'c'}(x',x)\,.
    \end{align}
At this stage, all field dependence on $ x' $ is isolated to the PDPs. We have arranged them such that they occur only through the combinations $ [P_x^{\underline{s}} U(x,y)] U(y,x) $, which is just $ \mathds{1} $ for $ s=0 $. When $ s > 0 $, repeated application of Eq.~\eqref{eq:D_on_U_right} will give powers of field-strength tensors and their derivatives evaluated at $ x $ multiplied by $ (x-x')_\mu $ to some power, while the PDPs can be taken to cancel in the end. Hence, we can parameterize the covariant propagators as 
    \begin{equation} \label{eq:Z_prop}
    \sum^{\infty}_{s=0} (x-x')_{\underline{s}} Z^{(\underline{m}, \underline{m}', \underline{s}) }_{bb'}(x, k) \equiv \big[(P_x+k)^{\underleftarrow{m}} Q_{be}^{\eminus 1}(x,\,P_x +k) (P_x+k)^{\underline{m}'} U_{ee'}(x,x') \big] U_{e'b'}(x',x)\,. 
    \end{equation} 
All background fields in the covariant functions $ Z(x,k) $ are evaluated at $ x $ and there are no remaining open derivatives. We proceed to write the sunset sum as 
    \begin{align}
    G_\mathrm{ss.} = \!\! \sum_{m,n,m',n'} &\!\!(\eminus 1)^{m+n} \!\! \! \sum_{r,s,t,u=0}^{\infty}   \dfrac{i^r }{r!}\int_{xx'} \, \int_{k\ell q} e^{i (k+\ell+q) \cdot (x'-x)} (x-x')_{\underline{r}+ \underline{s}+ \underline{t} + \underline{u}} \nonumber\\
    &\times V^{(\underline{m}, \underline{n})}_{abc}(x) P_x^{\underline{r}} V^{(\underline{m}', \underline{n}')}_{a'b'c'}(x)  Z^{(0,0,\underline{s})}_{aa'}(x,q)\,
    Z^{(\underline{m}, \underline{m}', \underline{t})}_{bb'}(x,k) \,
    Z^{(\underline{n}, \underline{n}', \underline{u})}_{cc'}(x,\ell)\,.
    \end{align}
The difference between the spacetime coordinates is traded for momentum derivatives through the relation $ x^\mu e^{iq \cdot x} = -i \partial^\mu_q e^{iq \cdot x} $. With a final use of integration by parts, those derivatives are moved to the corresponding propagator. We arrive at the formula 
    \begin{align}
    G_\mathrm{ss.} = \!\! \sum_{m,n,m',n'} \!\!(\eminus1)^{m+n} & \!\! \sum_{r,s,t,u=0}^{\infty} \dfrac{(\eminus i)^{s+t+u}}{r!} \int_{x} \, \int_{k\ell q } \delta(q+k+\ell) V^{(\underline{m}, \underline{n})}_{abc}(x) P_x^{\underline{r}} V^{(\underline{m}', \underline{n}')}_{a'b'c'}(x) \nonumber\\
    &\times \big[\partial_q^{{\underline{r}+ \underline{s}+ \underline{t} + \underline{u}}} Z^{(0,0,\underline{s})}_{aa'}(x,q) \big]\,
    Z^{(\underline{m}, \underline{m}', \underline{t})}_{bb'}(x,k) \,
    Z^{(\underline{n}, \underline{n}', \underline{u})}_{cc'}(x,\ell)\,.
    \end{align}
After performing the relevant $q $ derivatives and integrating over the momentum-space delta function, this looks like an ordinary two-loop Feynman integral. 
In practice, we can expect this rather intimidating expression to be more manageable than might be initially feared, due to the $ m^{(\prime)}, n^{(\prime)} $ sums having only a few non-trivial terms. Each power in $ r,s,t,u$ increases the canonical order of the operators in the integrand, so we may truncate their expansion by EFT order when doing matching or renormalization calculations.

\section{Expansion by Regions at Arbitrary Loop Order}
\label{app:regions}
Expansion by regions~\cite{Beneke:1997zp,Jantzen:2011nz} refers to techniques for writing a loop integral in dimensional regularization as the sum of integrals in which the original integrand is written as (truncated) series expansions according to different assumptions of the loop momenta, i.e., with different regions for the loop momenta. Determining the relevant regions can be very complicated for physical amplitudes, but it is relatively straightforward for the heavy-mass expansion, which is relevant to matching calculations. At one-loop order, it is well known that the relevant regions are one where the loop momentum is soft (small compared to the heavy masses) and one where it is hard (of the order of the heavy masses). 
Our aim in this appendix is to establish the generalization of this result to arbitrary loop orders, thereby justifying Eq.~\eqref{eq:sup_graphs_by_region}.   

Let us begin by setting up some notation: any momentum-space $\ell$-loop integral can be cast in the form
    \begin{equation} \label{eq:gen_loop_integral}
    I= \int_{k_1 \ldots k_n} \,\, \prod_{\sigma \in \Sigma} \delta_\sigma \prod_{i=1}^n D_{i}(k_i), \qquad \delta_\sigma = \delta \! \left( \sum_{i\in \sigma} k_i\right)\,,
    \end{equation}
where $ D_i $ are the (chain of) propagator(s) of the $ i $'th loop momenta and $ \sigma \in \Sigma$ are the sets of momenta appearing in the momentum-conserving delta functions ($ \ell = n - |\Sigma |$) from the vertices. Any factors of loop momenta $ k_i $ stemming from the interactions are also absorbed into the numerator of the corresponding $ D_i $. The delta functions are kept explicit rather than used to perform some of the momentum integrals in order to simplify the momenta decomposition and thus our derivations below. The situation relevant for EFT matching is that there is a heavy mass scale $ \Lambda $, setting the scale of one or more heavy masses. All external momenta (and light masses) are small compared to $ \Lambda $. When speaking about the size of the external momenta, it is assumed that we work in Euclidean space, such that the triangle equality is preserved. Without this restriction, two soft momenta could add to become hard. We assume that we may always go to Euclidean momenta by performing a Wick rotation of the integral. Working off shell for matching (or counterterm) calculations, we can restrict ourselves to consider only background fields with energy component $ q_0 = 0 $. That way, any linear combination of external momenta will have a non-positive Minkowski norm, and there will be no physical thresholds in the loop integral, which might have introduced non-analyticities and prevented the Wick rotation.

We consider momentum regions referred to as e.g., $ r = [\mathrm{hshhs\ldots}] $, to denote the region of Euclidean space\footnote{We use `$ \gtrsim $' as the negation of `$ \ll $.' One can imagine choosing some concrete scale between the soft scale and $ \Lambda $ and using that as the cutoff between the soft and the hard momenta. The boundary between the two regions drops out from the expansion by region formula in the end and is, thus, irrelevant anyway. }
    \begin{equation}
    r = \left\{  (k_1,\, \ldots, k_n) \in \mathds{R}^{n\cdot d}: 
    \begin{dcases}
    k_i \gtrsim \Lambda & \text{if} \; r^{(i)}  = \mathrm{h} \\
    k_i \ll \Lambda & \text{if} \; r^{(i)}  = \mathrm{s} 
    \end{dcases}  
    \right\}\,, 
    \end{equation}
where $ r^{(i)} $ denotes the $ i $'th letter in $ r $ or, equivalently, whether $ k_i $ is soft or hard in that region. 
Between them, the $ 2^n $ non-overlapping regions $ R = \big\{[\mathrm{ss\ldots}] ,\, \ldots, \,[\mathrm{hh\ldots}] \big\} $ make up the entire space $ \mathds{R}^{n\cdot d} = \cup_{r\in R}\,  r $. For additional notation, $ I_r $ refers to the original integral but with the integration region restricted to $ (k_i) \in r $, and $ \boldsymbol{T}_r $ is the expansion operator, which acts on the integrand by expanding it around the appropriate region of loop momenta. The expansion is such that the integral converges absolutely in the appropriate region, meaning that
    \begin{equation} \label{eq:converging_expansion}
    I_r = \boldsymbol{T}_r I_r\,.
    \end{equation}
The well-known one-loop expansion by regions (integrals with $n =1$ and $ \Sigma = \emptyset$) reads\footnote{We have omitted square brackets around the region letters in the subscripts, as there is little risk of mistaking its meaning.} 
    \begin{equation} \label{eq:one-loop_exp_by_regions}
    I = \boldsymbol{T}_\mathrm{h} I + \boldsymbol{T}_\mathrm{s} I\,.
    \end{equation}
The expansions on the r.h.s. appear as a power series in $ 1/\Lambda $ and the equality is strict only if all terms are included; however, the result is systematically improvable by including more terms. Incidentally, the power series in $ 1/\Lambda $ coincides with the EFT expansion. 
    
\begin{note}[note:one-loop_formula]{One-loop expansion by regions}
    To prove the one-loop expansion by regions formula~\eqref{eq:one-loop_exp_by_regions}, we note that the integration space  $ \mathds{R}^d = [\mathrm{h}] \cup [\mathrm{s}] $ is made up of the hard and the soft region, and so $ I = I_\mathrm{h} + I_\mathrm{s} $. It follows from~\eqref{eq:converging_expansion} that
        \begin{equation}
        I = \boldsymbol{T}_\mathrm{h} I_\mathrm{h} + \boldsymbol{T}_\mathrm{s} I_\mathrm{s} = \boldsymbol{T}_\mathrm{h} (I - \boldsymbol{T}_\mathrm{s} I_\mathrm{s}) + \boldsymbol{T}_\mathrm{s} (I- \boldsymbol{T}_\mathrm{h}  I_\mathrm{h})\,.
        \end{equation}
    The two expansion operators commute in the one-loop case (see Section~\ref{app:integrand_expansion}), and so the expansion can be organized as
        \begin{equation}
        I = \boldsymbol{T}_\mathrm{h} I + \boldsymbol{T}_\mathrm{s} I - \boldsymbol{T}_\mathrm{h} \boldsymbol{T}_\mathrm{s} I\,.
        \end{equation}
    The final term, with the double expansion, vanishes because the double expansion of the integrand results in a scaleless integral. 
\end{note}

\subsection{Expansions of the integrand} \label{app:integrand_expansion}
The propagator chains $ D_i(k_i) $ in the loop integral~\eqref{eq:gen_loop_integral} are composed of a product of ordinary propagators\footnote{Anything appearing in the numerator (from e.g. fermion propagators or vertex Feynman rules) is a polynomial in the loop momentum and does not change under expansion, nor does it give scale to scaleless integrals.}
    \begin{equation}
    \Delta(k+q,\, \mu^2) = \dfrac{1}{(k+q)^2 - \mu^2}\,,
    \end{equation}
for some external momentum $ q $, with $ q^2 \ll \Lambda $. When expanding the integrand with the application of an operator $ \boldsymbol{T}_r $, the individual propagators simply expand according to the region of their loop momentum. 
We discriminate between two cases, depending on the mass in the propagator. In the event that the mass is light, $ m\ll \Lambda $, the expansion of the propagator is 
    \begin{align}
    \boldsymbol{T}_\mathrm{s} \Delta(k+q,\, m^2) &= \Delta(k+q,\, m^2)\,, \\
    \boldsymbol{T}_\mathrm{h} \Delta(k+q,\, m^2) &= \sum_{n=0}^{\infty} \dfrac{ \big(m^2 - 2\, k\cdot q - q^2 \big)^n}{ (k^2 )^{n+1} }\,,\\
    \boldsymbol{T}_\mathrm{h} \boldsymbol{T}_\mathrm{s} \Delta(k+q,\, m^2) &= \boldsymbol{T}_\mathrm{s} \boldsymbol{T}_\mathrm{h} \Delta(k+q,\, m^2) = \boldsymbol{T}_\mathrm{h} \Delta(k+q,\, m^2)\,.
    \end{align} 
When the mass is heavy, $ M\gtrsim \Lambda $, instead, the regions are 
    \begin{align}
    \boldsymbol{T}_\mathrm{s} \Delta(k+q,\, M^2) &= \sum_{n=0}^{\infty} \dfrac{ -\big( k+q \big)^{\!2n} }{ \big(M^2 \big)^{\!n+1} }\,,\\
    \boldsymbol{T}_\mathrm{h} \Delta(k+q,\, M^2) &= \sum_{n=0}^{\infty} \dfrac{ \big( -2\,k\cdot q -q^2 \big)^{\!n} }{ \big(k^2 -M^2 \big)^{\!n+1} }\,,\\
    \boldsymbol{T}_\mathrm{h} \boldsymbol{T}_\mathrm{s} \Delta(k+q,\, M^2) &= \boldsymbol{T}_\mathrm{s} \boldsymbol{T}_\mathrm{h} \Delta(k+q,\, M^2) = \boldsymbol{T}_\mathrm{s} \Delta(k+q,\, M^2)\,.
    \end{align}
In both cases, we observe that the double-expansion w.r.t. the loop momenta results in a scaleless function, meaning that each term in the series is a simple power $ k^2 $ (after doing tensor decomposition of the numerator of the full integrand). 

The expansion of the delta functions is more subtle than that of the propagators. A delta function depends on multiple loop momenta in a non-factorizable manner. Consider the typical situation where some momenta are taken soft and some hard in the expansion:
    \begin{equation}
    \boldsymbol{T}_\mathrm{h\ldots s\ldots} \delta(k_1 + \ldots +\ell_1 + \ldots) = \sum_{n=0}^{\infty} \frac{1}{n!}\dfrac{\partial^n \delta(k_1 + \ldots)}{\partial k_1^{\mu_1} \cdots \partial k_1^{\mu_n}} (\ell_1 + \ldots)^{\mu_1} \cdots (\ell_1+\ldots)^{\mu_n}\,. 
    \end{equation}
When integrating over any of the loop momenta still in the delta function, the derivatives can be removed as per usual with integration by parts. Crucially, such integration by parts (when acting on scaleless functions), does not reintroduce a scale when acting on other delta functions or already expanded propagators. As such, all the soft momenta have essentially been factored out of the delta function. There is an exception to this rule when all momenta are in the same region; that is, 
    \begin{equation}
    \boldsymbol{T}_\mathrm{hh\ldots} \delta(k_1 + \ldots) = \boldsymbol{T}_\mathrm{ss\ldots} \delta(k_1 + \ldots) = \delta(k_1 + \ldots)\,.
    \end{equation}

When applying a second expansion to a delta function, the result is a delta function expanded over all momenta that were soft in either of the two expansions and the two expansions will commute. The notable exception occurs when none of the momenta are hard in both expansions; in this case, the two expansions will generally \emph{not commute}. In fact, we have
    \begin{multline}
    \boldsymbol{T}_\mathrm{s\ldots s\ldots h\ldots } \boldsymbol{T}_\mathrm{s\ldots h\ldots s\ldots } \delta(k_1 + \ldots + \ell_1 + \ldots + q_1 + \ldots) \\
    = \sum_{n=0}^{\infty} \dfrac{\partial^n \delta(\ell_1 + \ldots)}{\partial \ell_1^{\mu_1} \cdots \partial \ell_1^{\mu_n}} (k_1 + \ldots + q_1+\ldots)^{\mu_1} \cdots (k_1 + \ldots + q_1+\ldots)^{\mu_n}\,,
    \end{multline}
whereas
    \begin{multline}
    \boldsymbol{T}_\mathrm{s\ldots h\ldots s\ldots } \boldsymbol{T}_\mathrm{s\ldots s\ldots h\ldots } \delta(k_1 + \ldots + \ell_1 + \ldots + q_1 + \ldots) \\
    = \sum_{n=0}^{\infty} \dfrac{\partial^n \delta(q_1 + \ldots)}{\partial q_1^{\mu_1} \cdots \partial q_1^{\mu_n}} (k_1 + \ldots + \ell_1+\ldots)^{\mu_1} \cdots (k_1 + \ldots + \ell_1+ \ldots)^{\mu_n}\,.
    \end{multline}
This non-commutation of the expansions is part of what makes it challenging to establish a useful expansion by regions in the full integral.

Establishing expansion by regions in dimensional regularization relies heavily on scaleless integrals vanishing. In general, we have
    \begin{equation}
    \int_k \,(k^2)^{n} = 0, \qquad \text{for}\quad  n\neq -\tfrac{d}{2}\,.  
    \end{equation}
For our application, $ n $ is a result of propagators, vertex rules, and region expansions, and will always be an integer, meaning that such integrals vanish. By scaling arguments, one can also establish that
    \begin{equation}
    \int_k \,\delta(k) (k^2)^{n} = \begin{dcases}
    1 & \text{for} \quad n=0 \\
    0 & \text{for} \quad n\neq 0
    \end{dcases} \,.
    \end{equation}

\subsection{Expansion by regions for a generic loop integral}
We now aim to show how the one-loop expansion by region formula~\eqref{eq:one-loop_exp_by_regions} for the heavy-mass expansion generalizes to arbitrary loop integrals of the form~\eqref{eq:gen_loop_integral}. Delta functions involving exactly two momenta (from vertices with two edges) can be integrated out directly, collapsing the two neighboring propagators into a propagator chain of a single loop momentum. In this view, all one-loop integrals are of the form of a single propagator chain (with one loop momentum) and no delta functions.  

In general, we will discriminate between propagator (chains) entering and exiting the same vertex and those that reach between two different vertices. We can, without loss of generality, take the momenta entering and exiting the same vertex to be $ \{k_i\}_{m < i \leq n} $ for some $ m\leq n $. These momenta do not enter in any of the momentum-conserving delta functions ($ \forall \sigma \in \Sigma, m<i \leq n: i \notin \sigma $). Accordingly, we may factorize the loop integral~\eqref{eq:gen_loop_integral} as  
    \begin{equation} \label{eq:factorizing_loop_momenta}
    I= J \prod_{j=m+1}^n \int_{k_j} \! D_j(k_j), \qquad 
    J= \int_{k_1 \ldots k_m} \,\, \prod_{\sigma \in \Sigma} \delta_\sigma \prod_{i=1}^m D_{i}(k_i)\,.
    \end{equation}
Independently of the other integrals, the one-loop expansion by regions formula~\eqref{eq:one-loop_exp_by_regions} can be applied to each of the integrals over $ \{k_i\}_{m < i \leq n} $, giving
    \begin{equation}
    \int_{k_i} \! D_i(k_i) = \boldsymbol{T}_\mathrm{h} \! \int_{k_i} \! D_i(k_i) + \boldsymbol{T}_\mathrm{s} \! \int_{k_i} \! D_i(k_i), \qquad m< i \leq n\,.
    \end{equation}
For the momenta $ \{k_i\}_{m < i \leq n} $, the region expansion is nothing but a direct product of the expansion of each momentum individually. 
This leaves us to focus on the integral $ J $ over the remaining momenta $ \{k_i\}_{i \leq m} $. By assumption, all of these propagator chains in $ J $ go between multiple different vertices, meaning that there are at least two distinct vertices in $ J $. 

\begin{figure}
    \centering
    \includegraphics{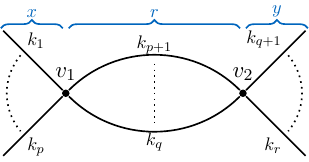}
    \caption{Region decomposition of the momenta entering the two adjacent vertices $ v_{1,2} $. The regions $ x,y, r $ cover disjoint sets of momenta.
    \label{fig:exp_by_regions}}
\end{figure}

Consider two adjacent vertices, $ v_1 $ and $ v_2 $, in the graph underlying the loop integral $ J $. We may assume, without loss of generality, that the incoming momenta to $ v_1 $ is $ \{k_i\}_{i\leq q} $ and $ \{k_i\}_{p< i \leq r} $ to $ v_2 $ ($ 0\leq p<q \leq r\leq m $) as depicted on Figure~\ref{fig:exp_by_regions}. The two vertices, thus, share momenta $ \{k_i\}_{p<i \leq q} $, and no other delta functions will involve these. The regions can be written as a direct product of the regions from the spaces $ \mathds{R}^{p \cdot d} \times \mathds{R}^{(q-p) \cdot d} \times \mathds{R}^{(r-q) \cdot d} \times \mathds{R}^{(m-r) \cdot d} $ and will be referred to as, e.g., $ x|r|y|a $, where $ a $ covers all momenta $ \{ k_i \}_{r< i \leq n} $, which are not involved in the two vertices. Then, the integral can be written as 
    \begin{equation} \label{eq:full_int_decomposition}
    J = \sum_{x,y} \sum_{r\in R} \sum_a J_{x|r|y|a} = \sum_{x,y} \sum_{r\in R} \sum_a  \boldsymbol{T}_{x|r|y|a} J_{x|r|y|a}\,,
    \end{equation}
writing explicitly $ R $ as the set of all regions for the momenta $ \{k_i\}_{p<i \leq q} $. 

We will now keep $ x,y $ (and $ a $) fixed, while ignoring $ a $ for the following argument, as this region is nothing but a spectator. 
Among the regions in $ R $, we may distinguish the regions $ \mathbf{h} = [\mathrm{hh\ldots}], \mathbf{s} = [\mathrm{ss\ldots}] $ where all momenta are either hard or all of them soft, respectively, and the regions $ R_\mathrm{nc.} = R\setminus\{\mathbf{h},\, \mathbf{s}\} $. 
The regions $ \mathbf{h}, \mathbf{s} $ are said to be commuting in the sense that 
    \begin{equation}
    \boldsymbol{T}_{x|r|y} \boldsymbol{T}_{x|\mathbf{h},\mathbf{s}|y} I = \boldsymbol{T}_{x|\mathbf{h},\mathbf{s}|y} \boldsymbol{T}_{x|r|y} I\,, \qquad r \in R\,,
    \end{equation}
on the integrand. By contrast, if $ r \in R_\mathrm{nc.} $, there is always another region for which the expansions do not commute. $ R_\mathrm{nc.} $ are said to be the non-commuting regions. The full integration region for the momenta $ \{k_i\}_{p<i \leq q} $ is denoted with $ F= \cup_{r\in R}\, r $.

We now wish to consider the sum over regions $ r\in R $ given the regions $ x,y $. This sum can be extended to span all of space in the following manner: 
    \allowdisplaybreaks[0] 
    \begin{subequations} \label{eq:Jr_sum}
    \begin{align} 
    \sum_{r\in R} \boldsymbol{T}_{x|r|y} J_{X|r|Y} = \,& \phantom{+}
    \big(\boldsymbol{T}_{x|\mathbf{s}|y} + \boldsymbol{T}_{x|\mathbf{h}|y} - \boldsymbol{T}_{x|\mathbf{s}|y} \boldsymbol{T}_{x|\mathbf{h}|y} \big) J_{X|F|Y} 
    \label{subeq:Jr_1}\\ 
    &+ \sum_{r \in R_{\mathrm{nc.}} } \big(1 - \boldsymbol{T}_{x|\mathbf{s}|y}\big) \big(1 - \boldsymbol{T}_{x|\mathbf{h} |y} \big) \boldsymbol{T}_{x|r|y} J_{X|F|Y}
    \label{subeq:Jr_2}\\ 
    &+ \sum_{\substack{r,r' \in R_{\mathrm{nc.}} \\r \neq r'}} \big(1 - \boldsymbol{T}_{x|\mathbf{s}|y}\big) \big(1 - \boldsymbol{T}_{x|\mathbf{h} |y} \big) \boldsymbol{T}_{x|r'|y} \boldsymbol{T}_{x|r|y} J_{X|r'|Y}\,.
    \label{subeq:Jr_3}
    \end{align}
    \end{subequations}
    \allowdisplaybreaks
This expansion is basically a special case of the results presented in~\cite{Jantzen:2011nz}, but we also give a concrete proof in Note~\ref{note:decomposition_formula} to keep the discussion self-contained.
It does not commit to the integration regions $ X,Y $ associated with $ x,y $ to be the same as the expansion regions. The following arguments work independently of these regions, and we are particularly interested in the cases $ X= x $ and $ X=F $ (similarly for $ y $). It will become convenient to introduce the notation $ r \cap r' = \braces{\mathrm{s}_{i_1}, \mathrm{h}_{i_2}, \ldots} $ to denote all the individual regions for the momenta that are identical between $ r $ and $ r' $. For instance, $ [\mathrm{hssh}] \cap [\mathrm{sshh}] = \braces{\mathrm{s}_2, \mathrm{h}_4} $. The aim is now to demonstrate a large degree of cancellation among the terms in~\eqref{eq:Jr_sum}. We will need to discriminate between two cases for the $ x,y $ regions: 
    \begin{enumerate}[i)]
    \item \label{case:hard} $ \exists i,j:\, \mathrm{h}_i \in x, \mathrm{h}_j \in y $; that is, at least one of the momenta in each of $ x $ and $ y $ are hard. Thus, any doubly-expanded momentum in $ \{k_i\}_{p<i\leq q} $ gets factored out of the vertex delta function(s). 
    \item \label{case:soft} At least one of $ x,y $ is purely soft; that is, without loss of generality $ y=\mathbf{s} $. This also includes the case $ p=0 $ and/or $ r-q=0 $, where there are no momenta and the region is empty. In this event, two non-commuting regions $ r,r'\in R_\mathrm{nc.} $ will not commute when acting on a delta associated with the second vertex when they do not share a hard momentum between them ($ r \cap r' \cap \mathbf{h} = \emptyset $). 
    \end{enumerate}

We begin with the first terms~\eqref{subeq:Jr_1}. In case~\ref{case:hard}, the doubly-expanded propagators are scaleless, and the associated momenta always gets factored out of the vertex delta function(s). With integration over the full space, it holds that
    \begin{equation}
    \boldsymbol{T}_{x|\mathbf{s}|y} \boldsymbol{T}_{x|\mathbf{h}|y} J_{X|F|Y} = 0\,.
    \end{equation}
The same holds in case~\ref{case:soft} when there is more than one propagator between the two vertices ($ q-p > 1 $). In this event, the momenta $ \{k_i\}_{p<i \leq q} $ do not factor out of the delta function of the second vertex, but with multiple scaleless propagators the integral over $ F $ still vanishes. 
The notable exception occurs in case~\ref{case:soft} when there is only a single propagator between the two vertices ($ q-p=1 $). The integral over the $ k_q $ propagator is the same regardless of whether it is doubly-expanded (denoted $ \mathrm{d} $) or expanded in the hard region:
    \begin{equation} \label{eq:int_single_hard}
    \int_{k_{q}} \delta_{\mathrm{h}|\mathbf{s}} (k_q+ \ldots + k_r) D_{q,\mathrm{h}}(k_q) = \int_{k_{q}} \delta_{\mathrm{h}|\mathbf{s}} (k_q+ \ldots + k_r) D_{q,\mathrm{d}}(k_q)\,.
    \end{equation}
It follows that 
    \begin{equation}
    \boldsymbol{T}_{x|\mathbf{s}|y} \boldsymbol{T}_{x|\mathbf{h}|y} J_{X|F|Y} = \boldsymbol{T}_{x|\mathbf{h}|y} J_{X|F|Y}\,,
    \end{equation}
and the two terms cancel in~\eqref{subeq:Jr_1}. This may be viewed as an instance of conservation of momentum in the second vertex: there cannot be a single hard momentum entering a vertex with nowhere to go.

Proceeding to~\eqref{subeq:Jr_2}, observe that the sum is non-trivial only if $ |R_\mathrm{nc.}| = 2^{q-p} -2 > 0 $, i.e., when there are at least two propagators shared between the vertices.
Furthermore, any $ r \in R_\mathrm{nc.} $ contains at least one soft and one hard momentum. In case~\ref{case:hard}, a double expansion of $ r $ with $ \mathbf{s},\mathbf{h} $ always results in at least one scaleless propagator (with no overlapping delta function), and it holds that
    \begin{equation}
    \boldsymbol{T}_{x|\mathbf{s}|y} \boldsymbol{T}_{x|\mathbf{h}|y} \boldsymbol{T}_{x|r|y} J_{X|F|Y} = \boldsymbol{T}_{x|\mathbf{s}|y}  \boldsymbol{T}_{x|r|y} J_{X|F|Y} = \boldsymbol{T}_{x|\mathbf{h}|y} \boldsymbol{T}_{x|r|y} J_{X|F|Y} = 0\,, \qquad r\in R_\mathrm{nc.}\,.
    \end{equation}
In case~\ref{case:soft}, the loop momenta $ k_i $ with $ \mathrm{s}_i \in r $ still factors out of the vertex delta functions. There is always one such, and the double-expansion with the hard region
    \begin{equation}
    \boldsymbol{T}_{x|\mathbf{h}|y} \boldsymbol{T}_{x|r|y} J_{X|F|Y} = 0, \qquad r\in R_\mathrm{nc.},
    \end{equation}
remains scaleless.
By contrast, doubly-expanded propagators with $ h_i \in r $ do not factor out of the delta function of $ v_2 $.\footnote{If also $ x=\mathbf{s} $, we might worry that both delta functions would contain the hard momenta of $ r $; however, then the two expanded delta functions would be identical (giving a meaningless $ \delta^2 $ function). Instead of performing the expansion directly on each delta function, one should rewrite
    \[
    \delta(k_1 + \ldots + k_q)\, \delta(k_{q+1} + \ldots + k_r) = \delta(k_{1} + \ldots + k_p - k_{q+1} - \ldots - k_r) \delta(k_{q+1} + \ldots + k_r)\,,
    \] 
before the expansion.}
We observe that
    \begin{equation}
    \boldsymbol{T}_{x|\mathbf{s}|y}  \boldsymbol{T}_{x|r|y} J_{X|F|Y}= \begin{dcases}
    \boldsymbol{T}_{x|r|y} J_{X|F|Y} & \text{for} \quad |\mathbf{h} \cap r| = 1\\
    0 & \text{for} \quad |\mathbf{h} \cap r| > 1
    \end{dcases}\,.
    \end{equation}
In the first case, where there is exactly one hard momentum in $ r $, the double expansion with the soft region acts trivially by~\eqref{eq:int_single_hard}. By contrast, when multiple lines get doubly expanded, there are multiple scaleless propagators with a single delta function. We conclude that	
    \begin{equation}
    \sum_{r\in R_{\mathrm{nc.}}} \big(1 - \boldsymbol{T}_{x|\mathbf{s}|y}\big) \big(1 - \boldsymbol{T}_{x|\mathbf{h} |y} \big) \boldsymbol{T}_{x|r|y} J_{X|F|Y} = \sum_{r\in \overline{R}_{\mathrm{nc.}}} \boldsymbol{T}_{x|r|y} J_{X|F|Y}\,,
    \end{equation}
where $ \overline{R}_{\mathrm{nc.}} = \overline{R} \setminus \{\mathbf{h},\mathbf{s} \} $ are the momentum-conserving non-commuting regions. 

This brings us to the overlap integrals~\eqref{subeq:Jr_3}, where one can distinguish between two cases 
\begin{enumerate}[a.]
    \item Case $ r,r' \in R_\mathrm{nc.} $ and $ r \neq r' $ such that $ r \cap r' \cap \mathbf{h} = \emptyset $ or $ r \cap r' \cap \mathbf{s} = \emptyset $. In this event, the propagators associated with momenta $ \{k_i\}_{p<i\leq q} $ have either all been expanded in the soft or all in the hard region with the expansion $ \boldsymbol{T}_{x|r'|y} \boldsymbol{T}_{x|r|y} $. An additional full soft (hard) expansion, $ \boldsymbol{T}_{x|\mathbf{s} (\mathbf{h})|y}  $, acts trivially on both the propagators and the delta function(s):
        \begin{equation}
        \boldsymbol{T}_{x|\mathbf{s} (\mathbf{h})|y} \boldsymbol{T}_{x|r'|y} \boldsymbol{T}_{x|r|y} J_{X|r'|Y} =  \boldsymbol{T}_{x|r'|y} \boldsymbol{T}_{x|r|y} J_{X|r'|Y}\,.
        \end{equation}
    It follows that the term vanishes in the sum \eqref{subeq:Jr_3}.
    
    \item Case $ r,r' \in R_\mathrm{nc.}$ and $ r \neq r' $ such that $ \exists p<i,j\leq q: \mathrm{h}_i, \mathrm{s}_j \in r \cap r' $. Take, without loss of generality, $ p+2\leq s < q $ such that 
        \begin{equation}
        r^{\prime (i)} = \begin{dcases}
        r^{(i)} & \mathrm{for}\; p < i \leq s \\
        !r^{(i)} & \mathrm{for}\; s< i\leq q
        \end{dcases} \,,
        \end{equation} 
    where again $ r^{(i)} $ denotes the region of the $ i $'th momentum of $ r $ and the notation $ !\mathrm{h} = \mathrm{s} $, $ !\mathrm{s} = \mathrm{h} $ implies the complementary region of that momenta. Observe that there are $ 2^{q-s} $ ordered pairs of regions $ r_i \neq r_i' $ with the same overlap as $ r,r' $,\footnote{This counts $ r,r' $ among them.} namely 
        \begin{equation}
        \begin{split}
        (r_1, r_1') &= \big( \big[{r}^{(p+1)} \ldots r^{(s)} \mathrm{s} \ldots \mathrm{s}\big],\,  \big[{r}^{(p+1)}\ldots r^{(s)} \mathrm{h} \ldots \mathrm{h}\big] \big)\,,\\ 
        \vdots \quad & = \qquad \qquad \vdots \\
        (r_{2^{q-s}}, r_{2^{q-s}}') &= \big( \big[{r}^{(p+1)} \ldots r^{(s)} \mathrm{h} \ldots \mathrm{h}\big],\, \big[{r}^{(p+1)} \ldots r^{(s)} \mathrm{s} \ldots \mathrm{s} \big] \big)\,.
        \end{split}
        \end{equation}
    Our assumption of both a hard and a soft momentum present in $ r^{(p< i\leq s)} $ ensures that  $ \forall i\leq 2^{q-s} :\, r_i^{(\prime)} \notin \{\mathbf{h}, \mathbf{s}\} $.
    
    When acting with the associated expansions on the integrand, it is crucial that there is a hard momentum among $ r^{(p< i\leq s)} $; this ensures that the repeated expansions on the vertex delta function commute and expand out all the momenta $ \{k_i\}_{s<i \leq q} $. Observe, therefore, that 
        \begin{equation}
        \boldsymbol{T}_{x|r'_i|y} \boldsymbol{T}_{x|r_i|y} = 
        \boldsymbol{T}_{x|r_i|y} \boldsymbol{T}_{x|r_i'|y} = 
        \boldsymbol{T}_{x|r'|y} \boldsymbol{T}_{x|r|y} \,, 
        \qquad \forall i\leq 2^{q-s}\,,
        \end{equation}
    when acting on the integrand. The full sum over all the $ 2^{q-s} $ overlaps now neatly combines into a single integration region with $ \{k_i\}_{s<i\leq q} $ integrated in all of space:
        \begin{equation} \label{eq:sum_patially_overlapping_regions}
        \sum_{i=1}^{2^{q-s}} \boldsymbol{T}_{x|r'_i|y} \boldsymbol{T}_{x|r_i|y} J_{X|r'_i|Y} = \boldsymbol{T}_{x|r'|y} \boldsymbol{T}_{x|r|y} J_{X|\mathbf{h}_{p+1}\mathbf{s}_{p+2} \ldots F_{s+1}\ldots F_{q}|Y} = 0\,.
        \end{equation}
    The associated propagators are doubly expanded and the momenta factored out of both vertex delta functions, implying the integrals are scaleless. 	
    The sum~\eqref{eq:sum_patially_overlapping_regions} accounts for all regions sharing the overlap and, thus, exhausts all regions of this type without double counting. Additional expansions with $ \boldsymbol{T}_{x|\mathbf{s}(\mathbf{h})|y} $ act trivially on the already scaleless propagators and will not cause the sum to be non-zero. 
\end{enumerate}
We conclude that~\eqref{subeq:Jr_3} vanishes regardless of $ x,y $. 

With all the proceeding arguments for cancellation among the terms in~\eqref{eq:Jr_sum}, we find that it reduces to
    \begin{equation}
    \sum_{r\in R} \boldsymbol{T}_{x|r|y} J_{X|r|Y} = \sum_{r\in \overline{R}} \boldsymbol{T}_{x|r|y} J_{X|F|Y}\,,
    \end{equation}
where $ \overline{R} $ are the momentum conserving regions (with implicit dependence on $ x,y $). The implications are that the full integral decomposition~\eqref{eq:full_int_decomposition} (reintroducing any spectator momenta) reads 
    \begin{equation} 
    J = \sum_a\sum_{x,y} \sum_{r\in R}   \boldsymbol{T}_{x|r|y|a} J_{x|r|y|a} = \sum_a\sum_{x,y} \sum_{r\in \overline{R} } \boldsymbol{T}_{x|r|y|a} J_{x|F|y|a}\,.
    \end{equation}

At this stage, we can move to consider the momenta propagating between two other adjacent vertices (of which one---but not both---could be $ v_{1,2} $). From this perspective the momenta $ \{k_i\}_{p<i\leq q} $ could be either spectators or part of the new regions $ x,y $. Similarly, the momenta that are part of $ x,y,a $ get reshuffled. Crucially, there can be no overlap between the momenta between the two new vertices and the momenta between the two old ones. Since none of our arguments relied on the integration regions for $ X,Y,a $ it does not matter that the momenta $ \{k_i\}_{p<i\leq q} $ are now integrated over all of space. The arguments can, thus, be repeated until all momenta have been extended to all of space. We conclude that
    \begin{equation} 
    J = \sum_{r\in \overline{R}_J} \boldsymbol{T}_r J\,,
    \end{equation}
where $ \overline{R}_J $ are the momentum-conserving regions of the $ J $ integral. In terms of the full $ I $ integral~\eqref{eq:factorizing_loop_momenta}, the region expansion of the factorized loop momenta can also be said to be momentum conserving, as they are not involved in any momentum-conserving delta functions. It follows then that 
    \begin{equation} \label{eq:exp_by_region_formula}
    I = \sum_{r\in \overline{R}_I} \boldsymbol{T}_r I\,,
    \end{equation}
is valid for the generic loop integral.

The heavy-mass expansion by regions~\eqref{eq:exp_by_region_formula} is the key result needed to facilitate perturbative EFT matching when decoupling heavy fields. In the notation of Section~\ref{sec:matching}, the hard-region expansion operator is identified with 
    \begin{equation}
    \boldsymbol{R}_\mathrm{hard} = \boldsymbol{T}_\mathbf{h}\,,
    \end{equation}
when acting on any loop integral $ I $, where $ \mathbf{h} \in R_I $ is the region where all loop momenta are hard. The decomposition of any graph/loop integral of the UV theory~\eqref{eq:sup_graphs_by_region}, is nothing but an application of~\eqref{eq:exp_by_region_formula}. The momentum conservation requirement of the regions $ \overline{R}_I $ implies that the hard momenta of the region must always form sub-loops $ \gamma $ inside the graph $ G $; in other words, the graph formed by the hard propagators must be a disjoint union of 1PI graphs. Any tree-level pieces would terminate in momentum-violating vertices with only one incoming hard momentum.

\begin{note}[note:decomposition_formula]{Decomposition formula} 
Here we prove~\eqref{eq:Jr_sum}, using the techniques of~\cite{Jantzen:2011nz}. We will presently ignore all momenta other than $ \{k_i\}_{p<i\leq q} $ which are present in the regions $ r\in R = R_\mathrm{nc.} \cup \{\mathbf{h}, \mathbf{s}\} $. All other momenta are simply harmless spectators when belonging to fixed regions. Expansion operators $ \boldsymbol{T}_r $ produce an absolutely convergent series when the integration region is restricted to $ r $. Thus, we have e.g. 
    \begin{equation}
    J_r = \boldsymbol{T}_r J_r\,, \qquad \forall r \in R\,.
    \end{equation}
The basic approach behind the region expansion is to extend the integral from a specific region $ r\in R $ to all of space. An example of this might look like 
    \begin{equation}
    J_r = \boldsymbol{T}_r J_r = \boldsymbol{T}_r J - \sum_{ \substack{r' \in R \\ r'\neq r}} \boldsymbol{T}_r J_{r'} = \boldsymbol{T}_r J - \sum_{ \substack{r' \in R \\ r'\neq r}} \boldsymbol{T}_{r'} \boldsymbol{T}_r J_{r'}\,, \qquad \forall r \in R\,.
    \end{equation}
Seeing as the expansion in the two regions $ r, r' $ may not commute, the ordering of the expansions matter; $ \boldsymbol{T}_r $ is performed before $ \boldsymbol{T}_{r'} $ in the sum.

Now we are ready to decompose the integral in the regions:
    \begin{equation}
    J = \sum_{r\in R} \boldsymbol{T}_r J_r = 
    \sum_{r\in R} \boldsymbol{T}_r J - \boldsymbol{T}_\mathbf{h} \boldsymbol{T}_\mathbf{s} J_{\mathbf{h} \cup \mathbf{s}} 
    - \sum_{r \in R_\mathrm{nc.}} \big( \boldsymbol{T}_{\mathbf{h}} \boldsymbol{T}_{r} J_{\mathbf{h} \cup r} + \boldsymbol{T}_{\mathbf{s}} \boldsymbol{T}_{r} J_{\mathbf{s} \cup r} \big) 
    - \sum_{ \substack{r, r' \in R_\mathrm{nc.} \\ r\neq r'}} \boldsymbol{T}_{r'} \boldsymbol{T}_r J_{r'}\,,
    \end{equation}
as $ \boldsymbol{T}_{\mathbf{h},\mathbf{s}} $ commute with all expansions. We proceed to extend the integration region of all integrals except for those that are expanded with two non-commuting regions. Thus, 
    \begin{multline}
    J = \sum_{r \in R} \boldsymbol{T}_r J - \boldsymbol{T}_\mathbf{h} \boldsymbol{T}_\mathbf{s} J 
    + \sum_{r \in R_\mathrm{nc.}} \big( \boldsymbol{T}_\mathbf{h} \boldsymbol{T}_\mathbf{s} \boldsymbol{T}_r J_r -\boldsymbol{T}_{\mathbf{h}} \boldsymbol{T}_{r} J - \boldsymbol{T}_{\mathbf{s}} \boldsymbol{T}_{r} J + \boldsymbol{T}_\mathbf{h} \boldsymbol{T}_\mathbf{s} \boldsymbol{T}_r J_{\mathbf{h} \cup \mathbf{s}} \big) \\
    - \sum_{_{ \substack{r, r' \in R_\mathrm{nc.} \\ r\neq r'}} } \big( \boldsymbol{T}_{r'} \boldsymbol{T}_r J_{r'} - \boldsymbol{T}_{\mathbf{h}} \boldsymbol{T}_{r'} \boldsymbol{T}_r J_{r'} - \boldsymbol{T}_{\mathbf{s}} \boldsymbol{T}_{r'} \boldsymbol{T}_r J_{r'} \big)\,.
    \end{multline}
With a final extension of the regions in the single sum, one gets 
    \begin{multline}
    J = \sum_{r\in R} \boldsymbol{T}_r J - \boldsymbol{T}_\mathbf{h} \boldsymbol{T}_\mathbf{s} J 
    + \sum_{r\in R_\mathrm{nc.}} \big( \boldsymbol{T}_\mathbf{h} \boldsymbol{T}_\mathbf{s} \boldsymbol{T}_r J -\boldsymbol{T}_{\mathbf{h}} \boldsymbol{T}_{r} J - \boldsymbol{T}_{\mathbf{s}} \boldsymbol{T}_{r} J \big) \\
    - \sum_{_{ \substack{r, r' \in R_\mathrm{nc.} \\ r\neq r'}} } \big( \boldsymbol{T}_{r'} \boldsymbol{T}_r J_{r'} - \boldsymbol{T}_{\mathbf{h}} \boldsymbol{T}_{r'} \boldsymbol{T}_r J_{r'} - \boldsymbol{T}_{\mathbf{s}} \boldsymbol{T}_{r'} \boldsymbol{T}_r J_{r'} + \boldsymbol{T}_{\mathbf{h}} \boldsymbol{T}_{\mathbf{s}} \boldsymbol{T}_{r'} \boldsymbol{T}_r J_{r'}  \big)\,.
    \end{multline}
Applying some convenient rearrangements, this may be collected as 
    \begin{equation}
    J =  \big(\boldsymbol{T}_\mathbf{s} + \boldsymbol{T}_\mathbf{h} - \boldsymbol{T}_\mathbf{s} \boldsymbol{T}_\mathbf{h} \big) J 
    + \sum_{r \in R_\mathrm{nc.}} \big(1- \boldsymbol{T}_\mathbf{s} \big) \big(1- \boldsymbol{T}_\mathbf{h} \big) \boldsymbol{T}_{r} J  
    - \sum_{_{ \substack{r, r' \in R_\mathrm{nc.} \\ r\neq r'}} } \big(1- \boldsymbol{T}_\mathbf{s} \big) \big(1- \boldsymbol{T}_\mathbf{h} \big) \boldsymbol{T}_{r'} \boldsymbol{T}_r J_{r'} \,,
    \end{equation}
which in turn justifies~\eqref{eq:Jr_sum}.
\end{note}

\sectionlike{References}
\vspace{-10pt}
\bibliography{References}

\end{document}

%% file: PreRamble.tex
\usepackage[english]{babel}
\usepackage{microtype,xspace}
\usepackage[utf8]{inputenc}	

\usepackage{amsfonts,amsmath,amssymb,bm,mathrsfs,mathtools,dsfont} 	
\allowdisplaybreaks 	
\usepackage[table]{xcolor}
\usepackage{hyperref}
\usepackage{slashed}
\usepackage{multicol}

\usepackage{siunitx}
\usepackage{geometry}
\usepackage[sort&compress,numbers,merge]{natbib}
\usepackage{chapterbib}
\addto\captionsenglish{\renewcommand*{\bibname}{References}}
\makeatletter
\renewcommand{\@memb@bchap}{ 
\bibmark \prebibhook
}
\makeatother

\usepackage{graphicx}
\graphicspath{{./Figures/}}
\usepackage{caption,subcaption}
\usepackage{multirow}

\usepackage{tabularx}
\newcolumntype{Y}{>{\centering\arraybackslash}X}

\usepackage[shortlabels]{enumitem}
\setlist{itemsep=.1em,topsep=.5em}
\SetEnumerateShortLabel{i}{\textit{\roman*}}
\SetEnumerateShortLabel{a}{\textit{\alph*}}

\definecolor{red}{rgb}{0.6,.0706,.1373}
\definecolor{blue}{rgb}{0,0.396,0.741}
\definecolor{green}{rgb}{0.25,0.6,0.2}
\definecolor{teal}{rgb}{0.11,0.6,0.6}
\definecolor{orange}{rgb}{.8, .4806, 0.173}
\definecolor{yellow}{rgb}{.8, .7, 0.05}
\colorlet{blueref}{blue!80!black}
\colorlet{bluelink}{blue!90!black}
\hypersetup{
	colorlinks, 
	bookmarksopen, 
	bookmarksnumbered,
	citecolor=blueref, 		
	linkcolor=bluelink,	
	urlcolor=bluelink,			
	pdfauthor={Fuentes-Martín, Palavrić, and Thomsen},    	
}

\addto\captionsenglish{}

\renewcommand{\contentsname}{Contents}
\renewcommand{\printtoctitle}[1]{}

\settocdepth{subsection}
\newcommand{\toc}{ {
	\hypersetup{linkcolor = black} 
	\vspace*{-.06\textheight}	
	\tableofcontents*
	\thispagestyle{empty} 
} }

\makeatletter
\newcommand*\ifthispageodd{%
  \checkoddpage
  \ifoddpage
    \expandafter\@firstoftwo
  \else
    \expandafter\@secondoftwo
  \fi
}
\makeatother

\usepackage[noindentafter]{titlesec} 
\counterwithout{section}{chapter} 
\counterwithout{figure}{chapter} 
\counterwithout{table}{chapter} 
\numberwithin{equation}{section} 
\setsecnumdepth{subsubsection}

\DeclareMathVersion{sans}
\SetSymbolFont{operators}{sans}{OT1}{cmbr}{m}{n}
\SetSymbolFont{letters}{sans}{OML}{cmbrm}{m}{it}
\SetSymbolFont{symbols}{sans}{OMS}{cmbrs}{m}{n}
\SetMathAlphabet{\mathit}{sans}{OT1}{cmbr}{m}{sl}
\SetMathAlphabet{\mathbf}{sans}{OT1}{cmbr}{bx}{n}
\SetMathAlphabet{\mathtt}{sans}{OT1}{cmtl}{m}{n}
\SetSymbolFont{largesymbols}{sans}{OMX}{iwona}{m}{n}

\DeclareMathVersion{boldsans}
\SetSymbolFont{operators}{boldsans}{OT1}{cmbr}{b}{n}
\SetSymbolFont{letters}{boldsans}{OML}{cmbrm}{b}{it}
\SetSymbolFont{symbols}{boldsans}{OMS}{cmbrs}{b}{n}
\SetMathAlphabet{\mathit}{boldsans}{OT1}{cmbr}{b}{sl}
\SetMathAlphabet{\mathbf}{boldsans}{OT1}{cmbr}{bx}{n}
\SetMathAlphabet{\mathtt}{boldsans}{OT1}{cmtl}{b}{n}
\SetSymbolFont{largesymbols}{boldsans}{OMX}{iwona}{bx}{n}

\titleformat{\section}{\centering \needspace{4\baselineskip}\Large \bfseries \sffamily \mathversion{boldsans} \color{blue!80!black} }{\thesection}{15pt}{}{}
\titlespacing{\section}{0pt}{15pt}{5pt}
\titleformat{\subsection}{\needspace{3\baselineskip} \large \sffamily \mathversion{sans} \color{blue!70!black} }{\thesubsection}{10pt}{}{}
\titlespacing{\subsection}{0pt}{10pt}{5pt}
\titleformat{\subsubsection}{\normalsize \sffamily \itshape \mathversion{sans} \color{blue!70!black} }{\thesubsubsection}{10pt}{}{}
\titlespacing{\subsubsection}{0pt}{10pt}{0pt}

\newcommand{\sectionlike}[1]{\phantomsection \addcontentsline{toc}{section}{#1} \setcounter{subsection}{0} \sectionmark{#1}
		\begin{center}
		\needspace{3\baselineskip}
		\Large \bfseries \sffamily \mathversion{boldsans} \color{blue!80!black} #1  
		\end{center}
	\vspace{-5pt} 
}

\ifnum\fullheadfoot=1
	\makepagestyle{standardstyle}
	\makeoddhead{standardstyle}{\sffamily \mathversion{subsectionmath} \rightmark}{}{\sffamily \bfseries{\thepage}}
	\makeevenhead{standardstyle}{\sffamily \bfseries{\thepage}}{}{\sffamily \mathversion{subsectionmath} \leftmark}
	\makeheadrule{standardstyle}{\textwidth}{\normalrulethickness}
	\makepsmarks{standardstyle}{
		\nouppercaseheads
		\createmark{section}{both}{shownumber}{\ }{\hspace{1.2em}  }
		\createmark{subsection}{right}{shownumber}{}{\hspace{1.2em} }
	}
	
	\copypagestyle{appstyle}{standardstyle}
	\makeevenhead{appstyle}{\sffamily \bfseries{\thepage}}{}{ \sffamily \mathversion{subsectionmath} \leftmark}
	\makepsmarks{appstyle}{
		\nouppercaseheads
		\createmark{section}{both}{nonumber}{\ }{\ \ }
		\createmark{subsection}{right}{shownumber}{}{\ \ }
	}
\else 
	\makepagestyle{standardstyle}
	\makeoddfoot{standardstyle}{}{\sffamily -- {\bfseries\thepage} --}{} 
	\makeevenfoot{standardstyle}{}{\sffamily -- {\bfseries\thepage} --}{}
	\copypagestyle{appstyle}{standardstyle}
\fi

\createplainmark{toc}{both}{\contentsname}
\createplainmark{bib}{both}{\bibname}

\makeatletter
\let\MyIntOrig\int
\def\MyIntSpace{\hspace{-.35em}} 
\def\int{\MyInt}
\def\MyInt{\MyIntOrig\MyIntSkipMaybe}
\def\MyIntSkipMaybe{
	\@ifnextchar_{\MyIntSkipScript}{%
		\@ifnextchar^{\MyIntSkipScript}{%
			\@ifnextchar\limits{\MyIntSkipTok}{%
				\@ifnextchar\nolimits{\MyIntSkipTok}{%
					\MyIntSpace}}}}%
}
\def\MyIntSkipScript#1#2{#1{#2}\MyIntSkipMaybe}
\def\MyIntSkipTok#1{#1\MyIntSkipMaybe}

\newcommand{\pushright}[1]{\ifmeasuring@#1\else\omit\hfill$\displaystyle#1$\fi\ignorespaces}
\makeatother

\usepackage[most,many,breakable]{tcolorbox}
\tcbset{shield externalize} 

\tcbuselibrary{skins}
\makeatletter
\newtcolorbox[auto counter
    ]{note}[2][Note]{enhanced,
    breakable,
    before upper={\parindent15pt\noindent}, 
    pad before break= 2mm,
    pad after break= 2mm,
    colback=white,
    colframe=blue!80!black!70,
    attach boxed title to top left={yshift*=-\tcboxedtitleheight},
    fonttitle=\bfseries \sffamily,
    label= #1,
    title={Note~\thetcbcounter:
    \mathversion{boldsans} #2},
    boxed title size=title,
    boxed title style={%
            right=-1pt,
            sharp corners,
            rounded corners=northwest,
            colback=tcbcolframe,
            boxrule=0pt,
        },
    underlay boxed title={%
            \path[fill=tcbcolframe] (title.south west)--(title.south east)
            to[out=0, in=180] ([xshift=5mm]title.east)--
            (title.center-|frame.east)
            [rounded corners=\kvtcb@arc] |-
            (frame.north) -| cycle;
        },
}
\makeatother

%% file: main.bbl
\providecommand{\href}[2]{#2}\begingroup\raggedright\begin{thebibliography}{100}

\bibitem{Feruglio:2016gvd}
F.~Feruglio, P.~Paradisi and A.~Pattori, \emph{{Revisiting Lepton Flavor
  Universality in B Decays}},
  \href{https://doi.org/10.1103/PhysRevLett.118.011801}{\emph{Phys. Rev. Lett.}
  {\bfseries 118} (2017) 011801},
  [\href{https://arxiv.org/abs/1606.00524}{{\ttfamily 1606.00524}}].

\bibitem{Aoude:2020dwv}
R.~Aoude, T.~Hurth, S.~Renner and W.~Shepherd, \emph{{The impact of flavour
  data on global fits of the MFV SMEFT}},
  \href{https://doi.org/10.1007/JHEP12(2020)113}{\emph{JHEP} {\bfseries 12}
  (2020) 113}, [\href{https://arxiv.org/abs/2003.05432}{{\ttfamily
  2003.05432}}].

\bibitem{Machado:2022ozb}
C.~S. Machado, S.~Renner and D.~Sutherland, \emph{{Building blocks of the
  flavourful SMEFT RG}},
  \href{https://doi.org/10.1007/JHEP03(2023)226}{\emph{JHEP} {\bfseries 03}
  (2023) 226}, [\href{https://arxiv.org/abs/2210.09316}{{\ttfamily
  2210.09316}}].

\bibitem{Allwicher:2023shc}
L.~Allwicher, C.~Cornella, G.~Isidori and B.~A. Stefanek, \emph{{New physics in
  the third generation. A comprehensive SMEFT analysis and future prospects}},
  \href{https://doi.org/10.1007/JHEP03(2024)049}{\emph{JHEP} {\bfseries 03}
  (2024) 049}, [\href{https://arxiv.org/abs/2311.00020}{{\ttfamily
  2311.00020}}].

\bibitem{Greljo:2023bdy}
A.~Greljo, A.~Palavri\'c and A.~Smolkovi\v{c}, \emph{{Leading directions in the
  SMEFT: Renormalization effects}},
  \href{https://doi.org/10.1103/PhysRevD.109.075033}{\emph{Phys. Rev. D}
  {\bfseries 109} (2024) 075033},
  [\href{https://arxiv.org/abs/2312.09179}{{\ttfamily 2312.09179}}].

\bibitem{Palavric:2024gvu}
A.~Palavri\'c, \emph{{Discrete Leptonic Flavor Symmetries: UV Mediators and
  Phenomenology}},  \href{https://arxiv.org/abs/2408.16044}{{\ttfamily
  2408.16044}}.

\bibitem{Allwicher:2024sso}
L.~Allwicher, M.~McCullough and S.~Renner, \emph{{New Physics at Tera-$Z$:
  Precision Renormalised}},  \href{https://arxiv.org/abs/2408.03992}{{\ttfamily
  2408.03992}}.

\bibitem{Boughezal:2024zqa}
R.~Boughezal, Y.~Huang and F.~Petriello, \emph{{Renormalization-group running
  of dimension-8 four-fermion operators in the SMEFT}},
  \href{https://arxiv.org/abs/2408.15378}{{\ttfamily 2408.15378}}.

\bibitem{Grojean:2024tcw}
C.~Grojean, G.~Guedes, J.~Roosmale~Nepveu and G.~M. Salla, \emph{{A log story
  short: running contributions to radiative Higgs decays in the SMEFT}},
  \href{https://arxiv.org/abs/2405.20371}{{\ttfamily 2405.20371}}.

\bibitem{Ardu:2021koz}
M.~Ardu and S.~Davidson, \emph{{What is Leading Order for LFV in SMEFT?}},
  \href{https://doi.org/10.1007/JHEP08(2021)002}{\emph{JHEP} {\bfseries 08}
  (2021) 002}, [\href{https://arxiv.org/abs/2103.07212}{{\ttfamily
  2103.07212}}].

\bibitem{Kley:2021yhn}
J.~Kley, T.~Theil, E.~Venturini and A.~Weiler, \emph{{Electric dipole moments
  at one-loop in the dimension-6 SMEFT}},
  \href{https://doi.org/10.1140/epjc/s10052-022-10861-5}{\emph{Eur. Phys. J. C}
  {\bfseries 82} (2022) 926},
  [\href{https://arxiv.org/abs/2109.15085}{{\ttfamily 2109.15085}}].

\bibitem{Fajfer:2023gie}
S.~Fajfer, J.~F. Kamenik, N.~Ko\v{s}nik, A.~Smolkovi\v{c} and M.~Tammaro,
  \emph{{New Physics in CP violating and flavour changing quark dipole
  transitions}}, \href{https://doi.org/10.1007/JHEP10(2023)133}{\emph{JHEP}
  {\bfseries 10} (2023) 133},
  [\href{https://arxiv.org/abs/2306.16471}{{\ttfamily 2306.16471}}].

\bibitem{Bonnefoy:2024gca}
Q.~Bonnefoy, J.~Kley, D.~Liu, A.~N. Rossia and C.-Y. Yao, \emph{{Aligned yet
  large dipoles: a SMEFT study}},
  \href{https://doi.org/10.1007/JHEP11(2024)046}{\emph{JHEP} {\bfseries 11}
  (2024) 046}, [\href{https://arxiv.org/abs/2403.13065}{{\ttfamily
  2403.13065}}].

\bibitem{Hurth:2019ula}
T.~Hurth, S.~Renner and W.~Shepherd, \emph{{Matching for FCNC effects in the
  flavour-symmetric SMEFT}},
  \href{https://doi.org/10.1007/JHEP06(2019)029}{\emph{JHEP} {\bfseries 06}
  (2019) 029}, [\href{https://arxiv.org/abs/1903.00500}{{\ttfamily
  1903.00500}}].

\bibitem{Gherardi:2020qhc}
V.~Gherardi, D.~Marzocca and E.~Venturini, \emph{{Low-energy phenomenology of
  scalar leptoquarks at one-loop accuracy}},
  \href{https://doi.org/10.1007/JHEP01(2021)138}{\emph{JHEP} {\bfseries 01}
  (2021) 138}, [\href{https://arxiv.org/abs/2008.09548}{{\ttfamily
  2008.09548}}].

\bibitem{Mantzaropoulos:2024vpe}
K.~Mantzaropoulos, \emph{{Disentangling SMEFT and UV contributions in
  h\textrightarrow{}Z\ensuremath{\gamma} and
  h\textrightarrow{}\ensuremath{\gamma}\ensuremath{\gamma} decays}},
  \href{https://doi.org/10.1103/PhysRevD.110.055041}{\emph{Phys. Rev. D}
  {\bfseries 110} (2024) 055041},
  [\href{https://arxiv.org/abs/2407.09145}{{\ttfamily 2407.09145}}].

\bibitem{Loisa:2024xuk}
E.~Loisa and J.~Talbert, \emph{{Froggatt-Nielsen meets the SMEFT}},
  \href{https://doi.org/10.1007/JHEP10(2024)017}{\emph{JHEP} {\bfseries 10}
  (2024) 017}, [\href{https://arxiv.org/abs/2402.16940}{{\ttfamily
  2402.16940}}].

\bibitem{Jenkins:2013zja}
E.~E. Jenkins, A.~V. Manohar and M.~Trott, \emph{{Renormalization Group
  Evolution of the Standard Model Dimension Six Operators I: Formalism and
  lambda Dependence}},
  \href{https://doi.org/10.1007/JHEP10(2013)087}{\emph{JHEP} {\bfseries 10}
  (2013) 087}, [\href{https://arxiv.org/abs/1308.2627}{{\ttfamily 1308.2627}}].

\bibitem{Jenkins:2013wua}
E.~E. Jenkins, A.~V. Manohar and M.~Trott, \emph{{Renormalization Group
  Evolution of the Standard Model Dimension Six Operators II: Yukawa
  Dependence}}, \href{https://doi.org/10.1007/JHEP01(2014)035}{\emph{JHEP}
  {\bfseries 01} (2014) 035},
  [\href{https://arxiv.org/abs/1310.4838}{{\ttfamily 1310.4838}}].

\bibitem{Alonso:2013hga}
R.~Alonso, E.~E. Jenkins, A.~V. Manohar and M.~Trott, \emph{{Renormalization
  Group Evolution of the Standard Model Dimension Six Operators III: Gauge
  Coupling Dependence and Phenomenology}},
  \href{https://doi.org/10.1007/JHEP04(2014)159}{\emph{JHEP} {\bfseries 04}
  (2014) 159}, [\href{https://arxiv.org/abs/1312.2014}{{\ttfamily 1312.2014}}].

\bibitem{Jenkins:2017dyc}
E.~E. Jenkins, A.~V. Manohar and P.~Stoffer, \emph{{Low-Energy Effective Field
  Theory below the Electroweak Scale: Anomalous Dimensions}},
  \href{https://doi.org/10.1007/JHEP01(2018)084}{\emph{JHEP} {\bfseries 01}
  (2018) 084}, [\href{https://arxiv.org/abs/1711.05270}{{\ttfamily
  1711.05270}}].

\bibitem{Naterop:2023dek}
L.~Naterop and P.~Stoffer, \emph{{Low-energy effective field theory below the
  electroweak scale: one-loop renormalization in the \textquoteright{}t
  Hooft-Veltman scheme}},
  \href{https://doi.org/10.1007/JHEP02(2024)068}{\emph{JHEP} {\bfseries 02}
  (2024) 068}, [\href{https://arxiv.org/abs/2310.13051}{{\ttfamily
  2310.13051}}].

\bibitem{Dekens:2019ept}
W.~Dekens and P.~Stoffer, \emph{{Low-energy effective field theory below the
  electroweak scale: matching at one loop}},
  \href{https://doi.org/10.1007/JHEP10(2019)197}{\emph{JHEP} {\bfseries 10}
  (2019) 197}, [\href{https://arxiv.org/abs/1908.05295}{{\ttfamily
  1908.05295}}].

\bibitem{Chala:2021pll}
M.~Chala, G.~Guedes, M.~Ramos and J.~Santiago, \emph{{Towards the
  renormalisation of the Standard Model effective field theory to dimension
  eight: Bosonic interactions I}},
  \href{https://doi.org/10.21468/SciPostPhys.11.3.065}{\emph{SciPost Phys.}
  {\bfseries 11} (2021) 065},
  [\href{https://arxiv.org/abs/2106.05291}{{\ttfamily 2106.05291}}].

\bibitem{DasBakshi:2022mwk}
S.~Das~Bakshi, M.~Chala, A.~D\'\i{}az-Carmona and G.~Guedes, \emph{{Towards the
  renormalisation of the Standard Model effective field theory to dimension
  eight: bosonic interactions II}},
  \href{https://doi.org/10.1140/epjp/s13360-022-03194-5}{\emph{Eur. Phys. J.
  Plus} {\bfseries 137} (2022) 973},
  [\href{https://arxiv.org/abs/2205.03301}{{\ttfamily 2205.03301}}].

\bibitem{DasBakshi:2023htx}
S.~Das~Bakshi and A.~D\'\i{}az-Carmona, \emph{{Renormalisation of SMEFT bosonic
  interactions up to dimension eight by LNV operators}},
  \href{https://doi.org/10.1007/JHEP06(2023)123}{\emph{JHEP} {\bfseries 06}
  (2023) 123}, [\href{https://arxiv.org/abs/2301.07151}{{\ttfamily
  2301.07151}}].

\bibitem{Chala:2023xjy}
M.~Chala and X.~Li, \emph{{Positivity restrictions on the mixing of
  dimension-eight SMEFT operators}},
  \href{https://doi.org/10.1103/PhysRevD.109.065015}{\emph{Phys. Rev. D}
  {\bfseries 109} (2024) 065015},
  [\href{https://arxiv.org/abs/2309.16611}{{\ttfamily 2309.16611}}].

\bibitem{Bakshi:2024wzz}
S.~D. Bakshi, M.~Chala, A.~D\'\i{}az-Carmona, Z.~Ren and F.~Vilches,
  \emph{{Renormalization of the SMEFT to dimension eight: Fermionic
  interactions I}},  \href{https://arxiv.org/abs/2409.15408}{{\ttfamily
  2409.15408}}.

\bibitem{Liao:2024xel}
Y.~Liao, X.-D. Ma and H.-L. Wang, \emph{{Probing dimension-8 SMEFT operators
  through neutral meson mixing}},
  \href{https://arxiv.org/abs/2409.10305}{{\ttfamily 2409.10305}}.

\bibitem{deBlas:2017xtg}
J.~de~Blas, J.~C. Criado, M.~Perez-Victoria and J.~Santiago, \emph{{Effective
  description of general extensions of the Standard Model: the complete
  tree-level dictionary}},
  \href{https://doi.org/10.1007/JHEP03(2018)109}{\emph{JHEP} {\bfseries 03}
  (2018) 109}, [\href{https://arxiv.org/abs/1711.10391}{{\ttfamily
  1711.10391}}].

\bibitem{Greljo:2023adz}
A.~Greljo and A.~Palavri\'c, \emph{{Leading directions in the SMEFT}},
  \href{https://doi.org/10.1007/JHEP09(2023)009}{\emph{JHEP} {\bfseries 09}
  (2023) 009}, [\href{https://arxiv.org/abs/2305.08898}{{\ttfamily
  2305.08898}}].

\bibitem{Carmona:2021xtq}
A.~Carmona, A.~Lazopoulos, P.~Olgoso and J.~Santiago, \emph{{Matchmakereft:
  automated tree-level and one-loop matching}},
  \href{https://doi.org/10.21468/SciPostPhys.12.6.198}{\emph{SciPost Phys.}
  {\bfseries 12} (2022) 198},
  [\href{https://arxiv.org/abs/2112.10787}{{\ttfamily 2112.10787}}].

\bibitem{Fuentes-Martin:2022jrf}
J.~Fuentes-Mart\'\i{}n, M.~K\"onig, J.~Pag\`es, A.~E. Thomsen and F.~Wilsch,
  \emph{{A proof of concept for matchete: an automated tool for matching
  effective theories}},
  \href{https://doi.org/10.1140/epjc/s10052-023-11726-1}{\emph{Eur. Phys. J. C}
  {\bfseries 83} (2023) 662},
  [\href{https://arxiv.org/abs/2212.04510}{{\ttfamily 2212.04510}}].

\bibitem{Guedes:2023azv}
G.~Guedes, P.~Olgoso and J.~Santiago, \emph{{Towards the one loop IR/UV
  dictionary in the SMEFT: One loop generated operators from new scalars and
  fermions}},
  \href{https://doi.org/10.21468/SciPostPhys.15.4.143}{\emph{SciPost Phys.}
  {\bfseries 15} (2023) 143},
  [\href{https://arxiv.org/abs/2303.16965}{{\ttfamily 2303.16965}}].

\bibitem{Gargalionis:2024jaw}
J.~Gargalionis, J.~Quevillon, P.~N.~H. Vuong and T.~You, \emph{{Linear Standard
  Model extensions in the SMEFT at one loop and Tera-Z}},
  \href{https://arxiv.org/abs/2412.01759}{{\ttfamily 2412.01759}}.

\bibitem{Ciuchini:1993ks}
M.~Ciuchini, E.~Franco, G.~Martinelli, L.~Reina and L.~Silvestrini,
  \emph{{Scheme independence of the effective Hamiltonian for $b \to s \gamma$
  and $b \to s g$ decays}},
  \href{https://doi.org/10.1016/0370-2693(93)90668-8}{\emph{Phys. Lett. B}
  {\bfseries 316} (1993) 127--136},
  [\href{https://arxiv.org/abs/hep-ph/9307364}{{\ttfamily hep-ph/9307364}}].

\bibitem{Ciuchini:1993fk}
M.~Ciuchini, E.~Franco, L.~Reina and L.~Silvestrini, \emph{{Leading order QCD
  corrections to $b \to s \gamma$ and $b \to s g$ decays in three
  regularization schemes}},
  \href{https://doi.org/10.1016/0550-3213(94)90223-2}{\emph{Nucl. Phys. B}
  {\bfseries 421} (1994) 41--64},
  [\href{https://arxiv.org/abs/hep-ph/9311357}{{\ttfamily hep-ph/9311357}}].

\bibitem{Jenkins:2023bls}
E.~E. Jenkins, A.~V. Manohar, L.~Naterop and J.~Pag\`es, \emph{{Two loop
  renormalization of scalar theories using a geometric approach}},
  \href{https://doi.org/10.1007/JHEP02(2024)131}{\emph{JHEP} {\bfseries 02}
  (2024) 131}, [\href{https://arxiv.org/abs/2310.19883}{{\ttfamily
  2310.19883}}].

\bibitem{DiNoi:2024ajj}
S.~Di~Noi, R.~Gr\"ober and M.~K. Mandal, \emph{{Two-loop running effects in
  Higgs physics in Standard Model Effective Field Theory}},
  \href{https://arxiv.org/abs/2408.03252}{{\ttfamily 2408.03252}}.

\bibitem{Haisch:2024wnw}
U.~Haisch and L.~Schnell, \emph{{Precision tests of third-generation four-quark
  operators: one- and two-loop matching}},
  \href{https://doi.org/10.1007/JHEP02(2025)038}{\emph{JHEP} {\bfseries 02}
  (2025) 038}, [\href{https://arxiv.org/abs/2410.13304}{{\ttfamily
  2410.13304}}].

\bibitem{Born:2024mgz}
L.~Born, J.~Fuentes-Mart\'\i{}n, S.~Kvedarait\.{e} and A.~E. Thomsen,
  \emph{{Two-Loop Running in the Bosonic SMEFT Using Functional Methods}},
  \href{https://arxiv.org/abs/2410.07320}{{\ttfamily 2410.07320}}.

\bibitem{Aitchison:1984ys}
I.~J.~R. Aitchison and C.~M. Fraser, \emph{{Fermion Loop Contribution to
  Skyrmion Stability}},
  \href{https://doi.org/10.1016/0370-2693(84)90644-0}{\emph{Phys. Lett. B}
  {\bfseries 146} (1984) 63--66}.

\bibitem{Fraser:1984zb}
C.~M. Fraser, \emph{{Calculation of Higher Derivative Terms in the One Loop
  Effective Lagrangian}}, \href{https://doi.org/10.1007/BF01550255}{\emph{Z.
  Phys. C} {\bfseries 28} (1985) 101}.

\bibitem{Aitchison:1985pp}
I.~J.~R. Aitchison and C.~M. Fraser, \emph{{Derivative Expansions of Fermion
  Determinants: Anomaly Induced Vertices, Goldstone-Wilczek Currents and Skyrme
  Terms}}, \href{https://doi.org/10.1103/PhysRevD.31.2605}{\emph{Phys. Rev. D}
  {\bfseries 31} (1985) 2605}.

\bibitem{Gaillard:1985uh}
M.~K. Gaillard, \emph{{The Effective One Loop Lagrangian With Derivative
  Couplings}}, \href{https://doi.org/10.1016/0550-3213(86)90264-6}{\emph{Nucl.
  Phys. B} {\bfseries 268} (1986) 669--692}.

\bibitem{Cheyette:1985ue}
O.~Cheyette, \emph{{Derivative Expansion of the Effective Action}},
  \href{https://doi.org/10.1103/PhysRevLett.55.2394}{\emph{Phys. Rev. Lett.}
  {\bfseries 55} (1985) 2394}.

\bibitem{Chan:1985ny}
L.~H. Chan, \emph{{Effective Action Expansion in Perturbation Theory}},
  \href{https://doi.org/10.1103/PhysRevLett.54.1222}{\emph{Phys. Rev. Lett.}
  {\bfseries 54} (1985) 1222--1225}.

\bibitem{Chan:1986jq}
L.-H. Chan, \emph{{Derivative Expansion for the One Loop Effective Actions With
  Internal Symmetry}},
  \href{https://doi.org/10.1103/PhysRevLett.57.1199}{\emph{Phys. Rev. Lett.}
  {\bfseries 57} (1986) 1199}.

\bibitem{Cheyette:1987qz}
O.~Cheyette, \emph{{Effective Action for the Standard Model With Large Higgs
  Mass}}, \href{https://doi.org/10.1016/0550-3213(88)90205-2}{\emph{Nucl. Phys.
  B} {\bfseries 297} (1988) 183--204}.

\bibitem{Dittmaier:1995cr}
S.~Dittmaier and C.~Grosse-Knetter, \emph{{Deriving nondecoupling effects of
  heavy fields from the path integral: A Heavy Higgs field in an SU(2) gauge
  theory}}, \href{https://doi.org/10.1103/PhysRevD.52.7276}{\emph{Phys. Rev. D}
  {\bfseries 52} (1995) 7276--7293},
  [\href{https://arxiv.org/abs/hep-ph/9501285}{{\ttfamily hep-ph/9501285}}].

\bibitem{Henning:2014wua}
B.~Henning, X.~Lu and H.~Murayama, \emph{{How to use the Standard Model
  effective field theory}},
  \href{https://doi.org/10.1007/JHEP01(2016)023}{\emph{JHEP} {\bfseries 01}
  (2016) 023}, [\href{https://arxiv.org/abs/1412.1837}{{\ttfamily 1412.1837}}].

\bibitem{Drozd:2015rsp}
A.~Drozd, J.~Ellis, J.~Quevillon and T.~You, \emph{{The Universal One-Loop
  Effective Action}},
  \href{https://doi.org/10.1007/JHEP03(2016)180}{\emph{JHEP} {\bfseries 03}
  (2016) 180}, [\href{https://arxiv.org/abs/1512.03003}{{\ttfamily
  1512.03003}}].

\bibitem{delAguila:2016zcb}
F.~del Aguila, Z.~Kunszt and J.~Santiago, \emph{{One-loop effective lagrangians
  after matching}},
  \href{https://doi.org/10.1140/epjc/s10052-016-4081-1}{\emph{Eur. Phys. J. C}
  {\bfseries 76} (2016) 244},
  [\href{https://arxiv.org/abs/1602.00126}{{\ttfamily 1602.00126}}].

\bibitem{Boggia:2016asg}
M.~Boggia, R.~Gomez-Ambrosio and G.~Passarino, \emph{{Low energy behaviour of
  standard model extensions}},
  \href{https://doi.org/10.1007/JHEP05(2016)162}{\emph{JHEP} {\bfseries 05}
  (2016) 162}, [\href{https://arxiv.org/abs/1603.03660}{{\ttfamily
  1603.03660}}].

\bibitem{Ellis:2016enq}
S.~A.~R. Ellis, J.~Quevillon, T.~You and Z.~Zhang, \emph{{Mixed
  heavy\textendash{}light matching in the Universal One-Loop Effective
  Action}}, \href{https://doi.org/10.1016/j.physletb.2016.09.016}{\emph{Phys.
  Lett. B} {\bfseries 762} (2016) 166--176},
  [\href{https://arxiv.org/abs/1604.02445}{{\ttfamily 1604.02445}}].

\bibitem{Fuentes-Martin:2016uol}
J.~Fuentes-Martin, J.~Portoles and P.~Ruiz-Femenia, \emph{{Integrating out
  heavy particles with functional methods: a simplified framework}},
  \href{https://doi.org/10.1007/JHEP09(2016)156}{\emph{JHEP} {\bfseries 09}
  (2016) 156}, [\href{https://arxiv.org/abs/1607.02142}{{\ttfamily
  1607.02142}}].

\bibitem{Zhang:2016pja}
Z.~Zhang, \emph{{Covariant diagrams for one-loop matching}},
  \href{https://doi.org/10.1007/JHEP05(2017)152}{\emph{JHEP} {\bfseries 05}
  (2017) 152}, [\href{https://arxiv.org/abs/1610.00710}{{\ttfamily
  1610.00710}}].

\bibitem{Ellis:2017jns}
S.~A.~R. Ellis, J.~Quevillon, T.~You and Z.~Zhang, \emph{{Extending the
  Universal One-Loop Effective Action: Heavy-Light Coefficients}},
  \href{https://doi.org/10.1007/JHEP08(2017)054}{\emph{JHEP} {\bfseries 08}
  (2017) 054}, [\href{https://arxiv.org/abs/1706.07765}{{\ttfamily
  1706.07765}}].

\bibitem{Summ:2018oko}
B.~Summ and A.~Voigt, \emph{{Extending the Universal One-Loop Effective Action
  by Regularization Scheme Translating Operators}},
  \href{https://doi.org/10.1007/JHEP08(2018)026}{\emph{JHEP} {\bfseries 08}
  (2018) 026}, [\href{https://arxiv.org/abs/1806.05171}{{\ttfamily
  1806.05171}}].

\bibitem{Cohen:2019btp}
T.~Cohen, M.~Freytsis and X.~Lu, \emph{{Functional Methods for Heavy Quark
  Effective Theory}},
  \href{https://doi.org/10.1007/JHEP06(2020)164}{\emph{JHEP} {\bfseries 06}
  (2020) 164}, [\href{https://arxiv.org/abs/1912.08814}{{\ttfamily
  1912.08814}}].

\bibitem{Kramer:2019fwz}
M.~Kr\"amer, B.~Summ and A.~Voigt, \emph{{Completing the scalar and fermionic
  Universal One-Loop Effective Action}},
  \href{https://doi.org/10.1007/JHEP01(2020)079}{\emph{JHEP} {\bfseries 01}
  (2020) 079}, [\href{https://arxiv.org/abs/1908.04798}{{\ttfamily
  1908.04798}}].

\bibitem{Angelescu:2020yzf}
A.~Angelescu and P.~Huang, \emph{{Integrating Out New Fermions at One Loop}},
  \href{https://doi.org/10.1007/JHEP01(2021)049}{\emph{JHEP} {\bfseries 01}
  (2021) 049}, [\href{https://arxiv.org/abs/2006.16532}{{\ttfamily
  2006.16532}}].

\bibitem{Ellis:2020ivx}
S.~A.~R. Ellis, J.~Quevillon, P.~N.~H. Vuong, T.~You and Z.~Zhang, \emph{{The
  Fermionic Universal One-Loop Effective Action}},
  \href{https://doi.org/10.1007/JHEP11(2020)078}{\emph{JHEP} {\bfseries 11}
  (2020) 078}, [\href{https://arxiv.org/abs/2006.16260}{{\ttfamily
  2006.16260}}].

\bibitem{Cohen:2020fcu}
T.~Cohen, X.~Lu and Z.~Zhang, \emph{{Functional Prescription for EFT
  Matching}}, \href{https://doi.org/10.1007/JHEP02(2021)228}{\emph{JHEP}
  {\bfseries 02} (2021) 228},
  [\href{https://arxiv.org/abs/2011.02484}{{\ttfamily 2011.02484}}].

\bibitem{Dedes:2021abc}
A.~Dedes and K.~Mantzaropoulos, \emph{{Universal scalar leptoquark action for
  matching}}, \href{https://doi.org/10.1007/JHEP11(2021)166}{\emph{JHEP}
  {\bfseries 11} (2021) 166},
  [\href{https://arxiv.org/abs/2108.10055}{{\ttfamily 2108.10055}}].

\bibitem{Dittmaier:2021fls}
S.~Dittmaier, S.~Schuhmacher and M.~Stahlhofen, \emph{{Integrating out heavy
  fields in the path integral using the background-field method: general
  formalism}},
  \href{https://doi.org/10.1140/epjc/s10052-021-09587-7}{\emph{Eur. Phys. J. C}
  {\bfseries 81} (2021) 826},
  [\href{https://arxiv.org/abs/2102.12020}{{\ttfamily 2102.12020}}].

\bibitem{Larue:2023uyv}
R.~Larue and J.~Quevillon, \emph{{The universal one-loop effective action with
  gravity}}, \href{https://doi.org/10.1007/JHEP11(2023)045}{\emph{JHEP}
  {\bfseries 11} (2023) 045},
  [\href{https://arxiv.org/abs/2303.10203}{{\ttfamily 2303.10203}}].

\bibitem{Banerjee:2023xak}
U.~Banerjee, J.~Chakrabortty, S.~U. Rahaman and K.~Ramkumar, \emph{{One-loop
  effective action up to any mass-dimension for non-degenerate scalars and
  fermions including light\textendash{}heavy mixing}},
  \href{https://doi.org/10.1140/epjp/s13360-024-04966-x}{\emph{Eur. Phys. J.
  Plus} {\bfseries 139} (2024) 169},
  [\href{https://arxiv.org/abs/2311.12757}{{\ttfamily 2311.12757}}].

\bibitem{Li:2024ciy}
X.-X. Li, X.~Lu and Z.~Zhang, \emph{{The Geometric Universal One-Loop Effective
  Action}},  \href{https://arxiv.org/abs/2411.04173}{{\ttfamily 2411.04173}}.

\bibitem{Henning:2016lyp}
B.~Henning, X.~Lu and H.~Murayama, \emph{{One-loop Matching and Running with
  Covariant Derivative Expansion}},
  \href{https://doi.org/10.1007/JHEP01(2018)123}{\emph{JHEP} {\bfseries 01}
  (2018) 123}, [\href{https://arxiv.org/abs/1604.01019}{{\ttfamily
  1604.01019}}].

\bibitem{Fuentes-Martin:2022vvu}
J.~Fuentes-Mart\'\i{}n, M.~K\"onig, J.~Pag\`es, A.~E. Thomsen and F.~Wilsch,
  \emph{{Evanescent operators in one-loop matching computations}},
  \href{https://doi.org/10.1007/JHEP02(2023)031}{\emph{JHEP} {\bfseries 02}
  (2023) 031}, [\href{https://arxiv.org/abs/2211.09144}{{\ttfamily
  2211.09144}}].

\bibitem{Quevillon:2021sfz}
J.~Quevillon, C.~Smith and P.~N.~H. Vuong, \emph{{Axion effective action}},
  \href{https://doi.org/10.1007/JHEP08(2022)137}{\emph{JHEP} {\bfseries 08}
  (2022) 137}, [\href{https://arxiv.org/abs/2112.00553}{{\ttfamily
  2112.00553}}].

\bibitem{Filoche:2022dxl}
B.~Filoche, R.~Larue, J.~Quevillon and P.~N.~H. Vuong, \emph{{Anomalies from an
  effective field theory perspective}},
  \href{https://doi.org/10.1103/PhysRevD.107.025017}{\emph{Phys. Rev. D}
  {\bfseries 107} (2023) 025017},
  [\href{https://arxiv.org/abs/2205.02248}{{\ttfamily 2205.02248}}].

\bibitem{Cohen:2023gap}
T.~Cohen, X.~Lu and Z.~Zhang, \emph{{Anomaly cancellation in effective field
  theories from the covariant derivative expansion}},
  \href{https://doi.org/10.1103/PhysRevD.108.056027}{\emph{Phys. Rev. D}
  {\bfseries 108} (2023) 056027},
  [\href{https://arxiv.org/abs/2301.00827}{{\ttfamily 2301.00827}}].

\bibitem{Cohen:2023hmq}
T.~Cohen, X.~Lu and Z.~Zhang, \emph{{Anomalies from the covariant derivative
  expansion}}, \href{https://doi.org/10.1103/PhysRevD.107.116015}{\emph{Phys.
  Rev. D} {\bfseries 107} (2023) 116015},
  [\href{https://arxiv.org/abs/2301.00821}{{\ttfamily 2301.00821}}].

\bibitem{DasBakshi:2018vni}
S.~Das~Bakshi, J.~Chakrabortty and S.~K. Patra, \emph{{CoDEx: Wilson
  coefficient calculator connecting SMEFT to UV theory}},
  \href{https://doi.org/10.1140/epjc/s10052-018-6444-2}{\emph{Eur. Phys. J. C}
  {\bfseries 79} (2019) 21},
  [\href{https://arxiv.org/abs/1808.04403}{{\ttfamily 1808.04403}}].

\bibitem{Cohen:2020qvb}
T.~Cohen, X.~Lu and Z.~Zhang, \emph{{STrEAMlining EFT Matching}},
  \href{https://doi.org/10.21468/SciPostPhys.10.5.098}{\emph{SciPost Phys.}
  {\bfseries 10} (2021) 098},
  [\href{https://arxiv.org/abs/2012.07851}{{\ttfamily 2012.07851}}].

\bibitem{Fuentes-Martin:2020udw}
J.~Fuentes-Martin, M.~K\"onig, J.~Pag\`es, A.~E. Thomsen and F.~Wilsch,
  \emph{{SuperTracer: A Calculator of Functional Supertraces for One-Loop EFT
  Matching}}, \href{https://doi.org/10.1007/JHEP04(2021)281}{\emph{JHEP}
  {\bfseries 04} (2021) 281},
  [\href{https://arxiv.org/abs/2012.08506}{{\ttfamily 2012.08506}}].

\bibitem{Coleman:1973jx}
S.~R. Coleman and E.~J. Weinberg, \emph{{Radiative Corrections as the Origin of
  Spontaneous Symmetry Breaking}},
  \href{https://doi.org/10.1103/PhysRevD.7.1888}{\emph{Phys. Rev. D} {\bfseries
  7} (1973) 1888--1910}.

\bibitem{Jackiw:1974cv}
R.~Jackiw, \emph{{Functional evaluation of the effective potential}},
  \href{https://doi.org/10.1103/PhysRevD.9.1686}{\emph{Phys. Rev. D} {\bfseries
  9} (1974) 1686}.

\bibitem{Fuentes-Martin:2023ljp}
J.~Fuentes-Mart\'\i{}n, A.~Palavri\'c and A.~E. Thomsen, \emph{{Functional
  Matching and Renormalization Group Equations at Two-Loop Order}},
  \href{https://arxiv.org/abs/2311.13630}{{\ttfamily 2311.13630}}.

\bibitem{Jack:1982hf}
I.~Jack and H.~Osborn, \emph{{Two Loop Background Field Calculations for
  Arbitrary Background Fields}},
  \href{https://doi.org/10.1016/0550-3213(82)90212-7}{\emph{Nucl. Phys. B}
  {\bfseries 207} (1982) 474--504}.

\bibitem{Bijnens:1999hw}
J.~Bijnens, G.~Colangelo and G.~Ecker, \emph{{Renormalization of chiral
  perturbation theory to order p**6}},
  \href{https://doi.org/10.1006/aphy.1999.5982}{\emph{Annals Phys.} {\bfseries
  280} (2000) 100--139},
  [\href{https://arxiv.org/abs/hep-ph/9907333}{{\ttfamily hep-ph/9907333}}].

\bibitem{vonGersdorff:2022kwj}
G.~von Gersdorff and K.~Santos, \emph{{New covariant Feynman rules for
  effective field theories}},
  \href{https://doi.org/10.1007/JHEP04(2023)025}{\emph{JHEP} {\bfseries 04}
  (2023) 025}, [\href{https://arxiv.org/abs/2212.07451}{{\ttfamily
  2212.07451}}].

\bibitem{Banerjee:2024rbc}
U.~Banerjee, J.~Chakrabortty and K.~Ramkumar, \emph{{Renormalization of scalar
  and fermion interacting field theory for arbitrary loop:
  Heat\textendash{}Kernel approach}},
  \href{https://doi.org/10.1140/epjp/s13360-024-05491-7}{\emph{Eur. Phys. J.
  Plus} {\bfseries 139} (2024) 714},
  [\href{https://arxiv.org/abs/2404.02734}{{\ttfamily 2404.02734}}].

\bibitem{Ford:1992pn}
C.~Ford, I.~Jack and D.~R.~T. Jones, \emph{{The Standard model effective
  potential at two loops}},
  \href{https://doi.org/10.1016/0550-3213(92)90165-8}{\emph{Nucl. Phys. B}
  {\bfseries 387} (1992) 373--390},
  [\href{https://arxiv.org/abs/hep-ph/0111190}{{\ttfamily hep-ph/0111190}}].

\bibitem{Martin:2013gka}
S.~P. Martin, \emph{{Three-Loop Standard Model Effective Potential at Leading
  Order in Strong and Top Yukawa Couplings}},
  \href{https://doi.org/10.1103/PhysRevD.89.013003}{\emph{Phys. Rev. D}
  {\bfseries 89} (2014) 013003},
  [\href{https://arxiv.org/abs/1310.7553}{{\ttfamily 1310.7553}}].

\bibitem{Martin:2017lqn}
S.~P. Martin, \emph{{Effective potential at three loops}},
  \href{https://doi.org/10.1103/PhysRevD.96.096005}{\emph{Phys. Rev. D}
  {\bfseries 96} (2017) 096005},
  [\href{https://arxiv.org/abs/1709.02397}{{\ttfamily 1709.02397}}].

\bibitem{Manohar:1997qy}
A.~V. Manohar, \emph{{The HQET / NRQCD Lagrangian to order alpha / m-3}},
  \href{https://doi.org/10.1103/PhysRevD.56.230}{\emph{Phys. Rev. D} {\bfseries
  56} (1997) 230--237}, [\href{https://arxiv.org/abs/hep-ph/9701294}{{\ttfamily
  hep-ph/9701294}}].

\bibitem{Chala:2024llp}
M.~Chala, J.~L\'opez~Miras, J.~Santiago and F.~Vilches, \emph{{Efficient
  on-shell matching}},  \href{https://arxiv.org/abs/2411.12798}{{\ttfamily
  2411.12798}}.

\bibitem{Iliopoulos:1974ur}
J.~Iliopoulos, C.~Itzykson and A.~Martin, \emph{{Functional Methods and
  Perturbation Theory}},
  \href{https://doi.org/10.1103/RevModPhys.47.165}{\emph{Rev. Mod. Phys.}
  {\bfseries 47} (1975) 165}.

\bibitem{DeWitt:1967ub}
B.~S. DeWitt, \emph{{Quantum Theory of Gravity. 2. The Manifestly Covariant
  Theory}}, \href{https://doi.org/10.1103/PhysRev.162.1195}{\emph{Phys. Rev.}
  {\bfseries 162} (1967) 1195--1239}.

\bibitem{Abbott:1980hw}
L.~F. Abbott, \emph{{The Background Field Method Beyond One Loop}},
  \href{https://doi.org/10.1016/0550-3213(81)90371-0}{\emph{Nucl. Phys.}
  {\bfseries B185} (1981) 189--203}.

\bibitem{Abbott:1981ke}
L.~F. Abbott, \emph{{Introduction to the Background Field Method}}, {\emph{Acta
  Phys. Polon.} {\bfseries B13} (1982) 33}.

\bibitem{Thomsen:2024abg}
A.~E. Thomsen, \emph{{A Partially Fixed Background Field Gauge}},
  \href{https://arxiv.org/abs/2404.11640}{{\ttfamily 2404.11640}}.

\bibitem{Barvinsky:1985an}
A.~O. Barvinsky and G.~A. Vilkovisky, \emph{{The Generalized Schwinger-Dewitt
  Technique in Gauge Theories and Quantum Gravity}},
  \href{https://doi.org/10.1016/0370-1573(85)90148-6}{\emph{Phys. Rept.}
  {\bfseries 119} (1985) 1--74}.

\bibitem{Kuzenko:2003eb}
S.~M. Kuzenko and I.~N. McArthur, \emph{{On the background field method beyond
  one loop: A Manifestly covariant derivative expansion in superYang-Mills
  theories}}, \href{https://doi.org/10.1088/1126-6708/2003/05/015}{\emph{JHEP}
  {\bfseries 05} (2003) 015},
  [\href{https://arxiv.org/abs/hep-th/0302205}{{\ttfamily hep-th/0302205}}].

\bibitem{Chetyrkin:1997fm}
K.~G. Chetyrkin, M.~Misiak and M.~Munz, \emph{{Beta functions and anomalous
  dimensions up to three loops}},
  \href{https://doi.org/10.1016/S0550-3213(98)00122-9}{\emph{Nucl. Phys. B}
  {\bfseries 518} (1998) 473--494},
  [\href{https://arxiv.org/abs/hep-ph/9711266}{{\ttfamily hep-ph/9711266}}].

\bibitem{Martin:2016bgz}
S.~P. Martin and D.~G. Robertson, \emph{{Evaluation of the general 3-loop
  vacuum Feynman integral}},
  \href{https://doi.org/10.1103/PhysRevD.95.016008}{\emph{Phys. Rev. D}
  {\bfseries 95} (2017) 016008},
  [\href{https://arxiv.org/abs/1610.07720}{{\ttfamily 1610.07720}}].

\bibitem{Herzog:2017bjx}
F.~Herzog and B.~Ruijl, \emph{{The R$^{*}$-operation for Feynman graphs with
  generic numerators}},
  \href{https://doi.org/10.1007/JHEP05(2017)037}{\emph{JHEP} {\bfseries 05}
  (2017) 037}, [\href{https://arxiv.org/abs/1703.03776}{{\ttfamily
  1703.03776}}].

\bibitem{Beneke:1997zp}
M.~Beneke and V.~A. Smirnov, \emph{{Asymptotic expansion of Feynman integrals
  near threshold}},
  \href{https://doi.org/10.1016/S0550-3213(98)00138-2}{\emph{Nucl. Phys. B}
  {\bfseries 522} (1998) 321--344},
  [\href{https://arxiv.org/abs/hep-ph/9711391}{{\ttfamily hep-ph/9711391}}].

\bibitem{Jantzen:2011nz}
B.~Jantzen, \emph{{Foundation and generalization of the expansion by regions}},
  \href{https://doi.org/10.1007/JHEP12(2011)076}{\emph{JHEP} {\bfseries 12}
  (2011) 076}, [\href{https://arxiv.org/abs/1111.2589}{{\ttfamily 1111.2589}}].

\bibitem{Dittmaier:1995ee}
S.~Dittmaier and C.~Grosse-Knetter, \emph{{Integrating out the standard Higgs
  field in the path integral}},
  \href{https://doi.org/10.1016/0550-3213(95)00551-X}{\emph{Nucl. Phys. B}
  {\bfseries 459} (1996) 497--536},
  [\href{https://arxiv.org/abs/hep-ph/9505266}{{\ttfamily hep-ph/9505266}}].

\bibitem{Collins:1984xc}
J.~C. Collins, \emph{{Renormalization}}, vol.~26 of \emph{Cambridge Monographs
  on Mathematical Physics}.
\newblock Cambridge University Press, Cambridge, 7, 2023,
  \href{https://doi.org/10.1017/9781009401807}{10.1017/9781009401807}.

\bibitem{Heisenberg:1936nmg}
W.~Heisenberg and H.~Euler, \emph{{Consequences of Dirac's theory of
  positrons}}, \href{https://doi.org/10.1007/BF01343663}{\emph{Z. Phys.}
  {\bfseries 98} (1936) 714--732},
  [\href{https://arxiv.org/abs/physics/0605038}{{\ttfamily physics/0605038}}].

\bibitem{Thomsen:2021ncy}
A.~E. Thomsen, \emph{{Introducing RGBeta: a Mathematica package for the
  evaluation of renormalization group $ \beta $-functions}},
  \href{https://doi.org/10.1140/epjc/s10052-021-09142-4}{\emph{Eur. Phys. J. C}
  {\bfseries 81} (2021) 408},
  [\href{https://arxiv.org/abs/2101.08265}{{\ttfamily 2101.08265}}].

\bibitem{Grozin:2012ec}
A.~Grozin, \emph{{Decoupling in QED and QCD}},
  \href{https://doi.org/10.1142/S0217751X13500152}{\emph{Int. J. Mod. Phys. A}
  {\bfseries 28} (2013) 1350015},
  [\href{https://arxiv.org/abs/1212.5144}{{\ttfamily 1212.5144}}].

\bibitem{Fliegner:1997ra}
D.~Fliegner, M.~Reuter, M.~G. Schmidt and C.~Schubert, \emph{{The Two loop
  Euler-Heisenberg Lagrangian in dimensional renormalization}},
  \href{https://doi.org/10.1007/BF02634170}{\emph{Theor. Math. Phys.}
  {\bfseries 113} (1997) 1442--1451},
  [\href{https://arxiv.org/abs/hep-th/9704194}{{\ttfamily hep-th/9704194}}].

\bibitem{Kors:1998ew}
B.~Kors and M.~G. Schmidt, \emph{{The Effective two loop Euler-Heisenberg
  action for scalar and spinor QED in a general constant background field}},
  \href{https://doi.org/10.1007/s100520050331}{\emph{Eur. Phys. J. C}
  {\bfseries 6} (1999) 175--182},
  [\href{https://arxiv.org/abs/hep-th/9803144}{{\ttfamily hep-th/9803144}}].

\bibitem{Dunne:2001pp}
G.~V. Dunne and C.~Schubert, \emph{{Closed form two loop Euler-Heisenberg
  Lagrangian in a selfdual background}},
  \href{https://doi.org/10.1016/S0370-2693(01)01475-7}{\emph{Phys. Lett. B}
  {\bfseries 526} (2002) 55--60},
  [\href{https://arxiv.org/abs/hep-th/0111134}{{\ttfamily hep-th/0111134}}].

\bibitem{Huet:2010nt}
I.~Huet, D.~G.~C. McKeon and C.~Schubert, \emph{{Euler-Heisenberg lagrangians
  and asymptotic analysis in 1+1 QED, part 1: Two-loop}},
  \href{https://doi.org/10.1007/JHEP12(2010)036}{\emph{JHEP} {\bfseries 12}
  (2010) 036}, [\href{https://arxiv.org/abs/1010.5315}{{\ttfamily 1010.5315}}].

\bibitem{Huet:2011kd}
I.~Huet, M.~Rausch~de Traubenberg and C.~Schubert, \emph{{The Euler-Heisenberg
  Lagrangian Beyond One Loop}},
  \href{https://doi.org/10.1142/S2010194512007507}{\emph{Int. J. Mod. Phys.
  Conf. Ser.} {\bfseries 14} (2012) 383--393},
  [\href{https://arxiv.org/abs/1112.1049}{{\ttfamily 1112.1049}}].

\bibitem{Dunne:2012vv}
G.~V. Dunne, \emph{{The Heisenberg-Euler Effective Action: 75 years on}},
  \href{https://doi.org/10.1142/S0217751X12600044}{\emph{Int. J. Mod. Phys. A}
  {\bfseries 27} (2012) 1260004},
  [\href{https://arxiv.org/abs/1202.1557}{{\ttfamily 1202.1557}}].

\bibitem{Gies:2016yaa}
H.~Gies and F.~Karbstein, \emph{{An Addendum to the Heisenberg-Euler effective
  action beyond one loop}},  \href{https://arxiv.org/abs/1612.07251}{{\ttfamily
  1612.07251}}.

\bibitem{Witten:2012bg}
E.~Witten, \emph{{Notes On Supermanifolds and Integration}},
  \href{https://doi.org/10.4310/PAMQ.2019.v15.n1.a1}{\emph{Pure Appl. Math.
  Quart.} {\bfseries 15} (2019) 3--56},
  [\href{https://arxiv.org/abs/1209.2199}{{\ttfamily 1209.2199}}].

\bibitem{DeWitt:2012mdz}
B.~S. DeWitt, \emph{{Supermanifolds}}.
\newblock Cambridge Monographs on Mathematical Physics. Cambridge Univ. Press,
  Cambridge, UK, 5, 2012,
  \href{https://doi.org/10.1017/CBO9780511564000}{10.1017/CBO9780511564000}.

\bibitem{Wegner:2016ahw}
F.~Wegner, \emph{{Supermathematics and its Applications in Statistical
  Physics}: {Grassmann Variables and the Method of Supersymmetry}}.
\newblock Springer, 2016,
  \href{https://doi.org/10.1007/978-3-662-49170-6}{10.1007/978-3-662-49170-6}.

\end{thebibliography}\endgroup
